# A review and outlook on anionic and cationic redox in Ni-, Li- and Mn-rich layered oxides Li*Me*$O_2$ (*Me* = Li, Ni, Co, Mn)

Bixian Ying[1], Zhenjie Teng[1], Sarah Day[2], Dan Porter[2], Martin Winter[1,3], Adrian Jonas[4], Katja Frenzel[4], Lena Mathies[4], Burkhard Beckhoff[4], Peter Nagel[5,6], Stefan Schuppler[5,6], Michael Merz[5,6], Felix Pfeiffer[3], Matthias Weiling[3], Masoud Baghernejad[3], Karin Kleiner[1,*]

[1] Münster Electrochemical Energy Technology (MEET), University of Münster (WWU) Corrensstraße 46, 48149 Münster, Germany

[2] Harwell Science and Innovation Campus, Diamond Light Source, Didcot, Oxfordshire OX11 0DE, U.K.

[3] Helmholtz-Institute Münster, Forschungszentrum Jülich GmbH, 48149 Muenster, Germany

[4]Physikalisch-Technische Bundesanstalt, Abbestr. 2-12, 10587 Berlin, Germany

[5] Institute for Quantum Materials and Technologies, Karlsruhe Institute of Technology (IQMT, KIT), Hermann-von-Helmholtz-Platz 1, 76344 Eggenstein-Leopoldshafen, Germany

[6] Karlsruhe Nano Micro Facility (KNMFi), Karlsruhe Institute of Technology (KIT), Campus North, Hermann-von-Helmholtz-Platz 1, 76344 Eggenstein-Leopoldshafen, Germany

* Correspondence to: karin.kleiner@wwu.de

The present work reviews the charge compensation in Ni-, Mn-, and Li-rich layered oxides ($LiNi_{1-x}Me_xO_2$, $Li_{1+y}Mn_{0.5+x}Me_{1-x-y}O_2$ with x, y ≤ 0.2, *Me* = Ni, Co, Mn) and aims to translate empirical performance parameters such as the capacity, and voltage of the cathode materials into physically measurable quantities. Upon charge and discharge two fundamentally different redox mechanisms are observed: At low and medium states of charge (SOCs) charge compensation takes mainly place at oxygen sites while electron density is shifted from the oxygen lattice to Ni (= formation of σ-bonds). At high SOCs the shift of electron density from the transition metals to oxygen (formation of π bonds) enables an additional redox process but also oxygen release from the transition metal host structure and subsequent detrimental reactions. Depending on the Ni:Co:Mn ratio, both processes lead to characteristic features in the voltage profile of the cathode materials and performance parameters like the capacity, cycling stability and open cell voltage are a function of the *Me* composition.

# Table of Contents

# 1. Introduction

Layered oxides such as Ni-rich NCMs or NCA ($LiNi_{1-x}Me_xO_2$ with x + y ≤ 0.2, *Me* = Mn, Co or Al, space group $R\bar{3}m$) are at the forefront of research because of their potential to deliver outstanding energy densities if the materials can be stabilized at high voltages, *i.e.* at degrees of delithiation y in $Li_yNi_{1-x}Me_xO_2 \ll 0.6$.[1–10] However, at high states of charge (SOCs) the increase of anisotropic lattice broadening/contraction leads to a collapse of the crystallographic structure [8,9,11–13], causing the Li diffusion coefficient to break down [14–16], and detrimental reactions such as oxygen release [17–19], surface morphology changes [11,20–22], transition metal dissolution [23,24], as well as crack formation to set in.[25–28]

Changes in the electronic structure upon charge and discharge of Ni-rich layered oxides (= de-/intercalation of Li ions) reveal that the charge compensation at low and medium SOCs involves the formation/breakage of covalent Ni-O bonds.[5,15,29–36] Core level spectroscopy (*e.g.* near edge x-ray absorption finestructure, NEXAFS, and, electron energy loss spectroscopy, EELS) provides evidence of the reversible formation of holes L̲ in the O 2p band upon charge, showing that the extraction of electrons is accompanied by a shift of electron density from the O ligands towards the transition metals (= *Me* 3*d*-O 2*p* hybridization).[29,35,37–39] So far it has been assumed that *Me*-O hybridization facilitates oxygen release at high SOCs.[31,40–43] With the discovery and application of Li- and Na-rich layered oxides as cathode materials in Li-ion batteries [44–50], however, it is discussed that excess capacity beyond *Me*-O hybridization can be achieved by a reversible oxidation of the oxygen lattice, forming molecular oxygen in a so called 'anionic redox' process.[48,51–57] These findings are supported by resonant inelastic scattering (RIXS), providing evidence for the presence of molecular oxygen due to characteristic O=O vibrations in the inelastic spectra.[52,58–60] NEXAFS data, in turn, shows that the anionic redox is at least partially irreversible and levels off soon after the first cyle.[61] Instead, it activates new redox processes at low SOCs in the subsequent cycle which are partially reversible.[47,61] Recently, O=O RIXS signatures in the O K edge of Ni-rich layered oxides at high SOCs are found, as well.[1,48,62,63] Moreover, detrimental reactions such as oxygen release, surface morphology changes, and transition metal dissolution are observed, as well.[11,15,17–22,43,64] The charge compensation at high SOCs proceeds very similar for Ni- and Mn-/Li-rich layered oxides although O=O formation and subsequent (detrimental) reactions are more pronounced in Li-rich cathodes. Beside many approaches to suppress irreversible reactions such as introducing a concentration gradient within the particles [65–68], using cationic or anionic substitution and doping [69–77], protecting the particles with surface coatings [78–82], and optimizing the morphology of the secondary and primary particles [83–85], the poor cycling stability of Ni-, Li-, and Mn-rich layered oxides remains challenging.[1,17,62]

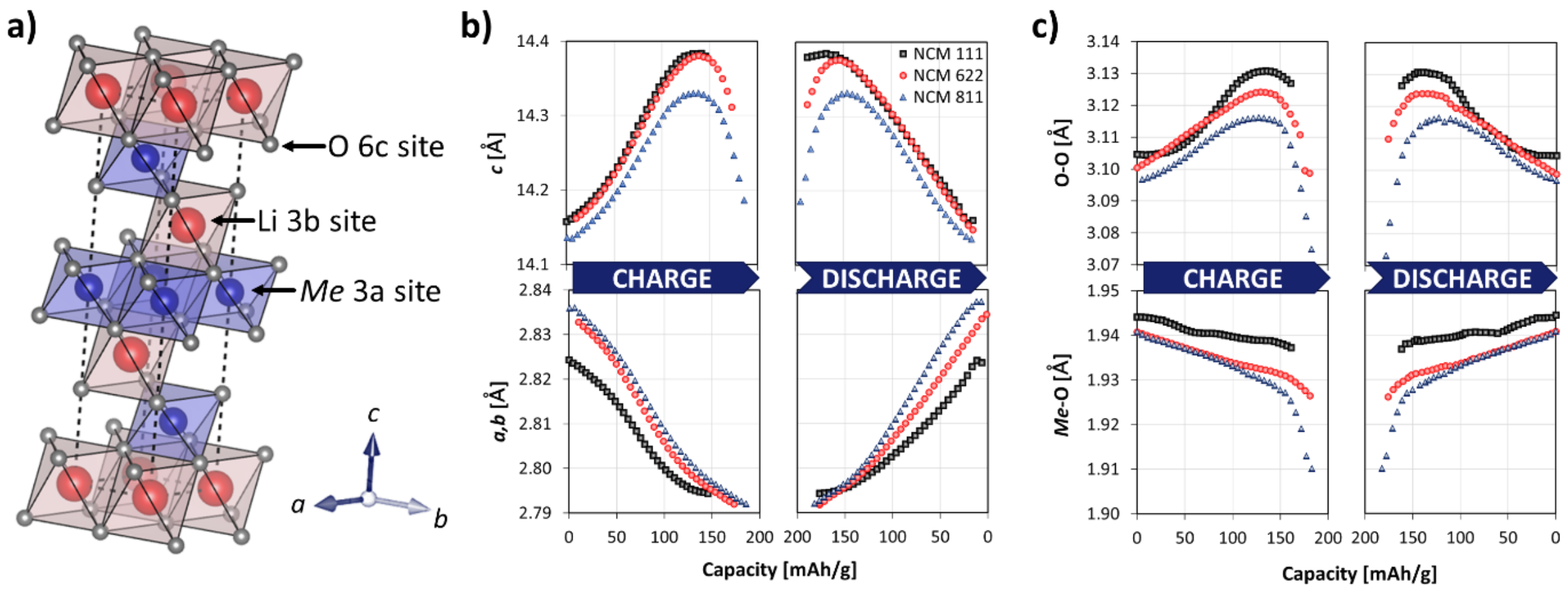

**Figure 1:** a) Unit cell with the space group $R\bar{3}m$ with which powder diffraction patterns of layered oxides $LiMeO_2$ are refined. b) Refined *c* and *a*, *b* lattice parameters of NCM111, NCM622, and NCM811 upon the first charge and discharge, and c), the O-O distance in the Li layer as well as the *Me*-O distance in the *Me* layer, obtained from operando synchrotron x-ray powder diffraction measurements (SXPD) at beamline I11 (DLS, UK). The data is reproduced from ref. [15].

In the present work the charge compensation in Ni-rich layered oxides is reviewed, providing insights into the complex electronic structure and its changes upon charge and discharge. Two fundamentally different redox mechanisms, Ni-O hybridization, and O=O dimer formation, are identified and their dependence on the Ni-, Co-, and Mn-content as well as their implication on the voltage profile is discussed. Unraveling the electronic configuration of individual transition metals in the cathodes and a discussion about formal oxidation states *vs.* the charge neutrality condition further shows that ionic $Ni^{2+}$ does not only limit the reversible capacity but also enables O=O formation.

# 2. The electronic structure of inorganic transition metal oxides

## 2.1. Crystallographic and electronic structure

Based on changes in the lattice parameters and thus changes in the *Me*-O distance upon charge and discharge of layered oxides, Tarascon *et al.* suggested early on that the reversible redox reactions include changes in the *Me* 3d-O 2p hybridization (= changes in covalence of the *Me*-O bonds).[31] *Operando* synchrotron x-ray powder diffraction (SXPD) reveals that at low and medium states of charge the *a*,*b* lattice parameter (and thus the *Me*-O distance) decreases with an increasing SOC. Shorter *Me*-O bonds in *a*,*b* direction mean a higher degree of hybridization (higher degree of covalence).[15] The *c* lattice parameter (and thus the O-O distance of tetrahedral sites in the Li layer) increases with a decreasing amount of Li ions in the host structure, **Figure 1b** and **c**, because repulsive interactions between opposing O layers increase while Li, positive charges between the O layers, are deintercalated.[5,15,31,36] At high SOCs (> 75%), in turn, a breakdown of the *c* lattice parameter (the O-O distance) is observed while oxygen release, transition metal dissolution and surface morphology changes are obsered.[5,15,31,36]

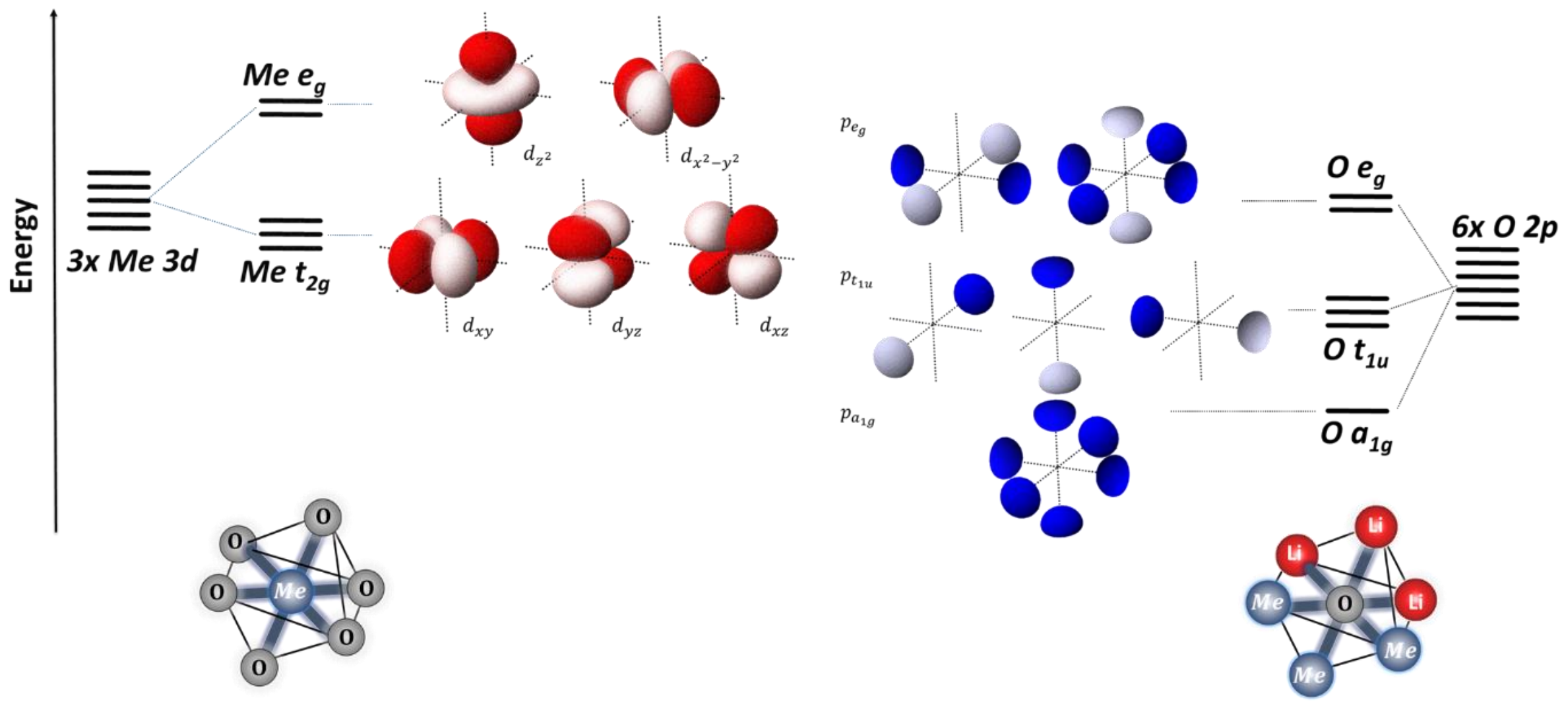

**Figure 2:** Degeneration of the *Me* 3*d*- and O 2*p*-orbitals in Li*Me*$O_2$. Thereby, the transition metal *Me* occupies an O-octahedron as.

Core level spectroscopy (e.g. near edge x-ray absorption finestructure, NEXAFS, and, electron energy loss spectroscopy, EELS) confirmed that a large portion of the charge compensation at low and medium SOCs is indeed achieved at oxygen sites upon Li de-/intercalation.[29,35,37–39] Koyama *et al.*, for example, concluded that Li deintercalation from $LiNiO_2$ reinforces the covalent bonding between O and Ni which significantly reduces the electron density of the O lattice.[29]

## 2.2. Coulomb Interactions and charge transfer

The involvement of the O lattice in the redox process upon charge and discharge goes back to the complex electronic structure of layered oxides and structural related materials.[86–89] In Li*Me*$O_2$, transition metals *Me* occupy oxygen octahedrons which are the redox active centers. Static electric fields of neighboring atoms cause a degeneration of electron orbital states as described by the ligand field (LF) theory, **Figure 2**.[88,90] The *d*-orbitals which point towards oxygen atoms in the octahedron ($d_{z^2}, d_{x^2-y^2}$ = $e_g$ orbitals) are destabilized (increase in energy) while the others ($d_{xy}, d_{yz}, d_{xz}$ = $t_{2g}$ orbitals) are stabilized (lowered in energy), **Figure 2** (left side).[91] Each O atom is surrounded by three *Me* and three Li ions (**Figure 2**, right side). Thus, the six O 2*p* orbitals are degenerated, as well, and form an orbital with $a_{1g}$ symmetry, three orbitals with $t_{1u}$ symmetry, and two orbitals with $e_g$ symmetry, respectively.[91,92] In layered oxides the redox active *Me*$O_6$ octahedrons are connected *via* their edges and ideally form an infinite crystal with an infinite amount of *Me* and O orbitals. Thus, the picture of discrete orbitals can be replaced by the band structure or the density of states (DOS), *i.e.* the number of states in the system of volume *V* as a function of the energy, **Figure 3a**.[41] If only ligand field (LF) effects are considered, transition metal oxides should be metallic as soon as the $e_g$ or $t_{2g}$ orbitals are partially filled. In reality, however, transition metal oxides reveal physical properties of semiconductors or even insulators independent of the occupancy of the 3*d* orbitals.[93–99]

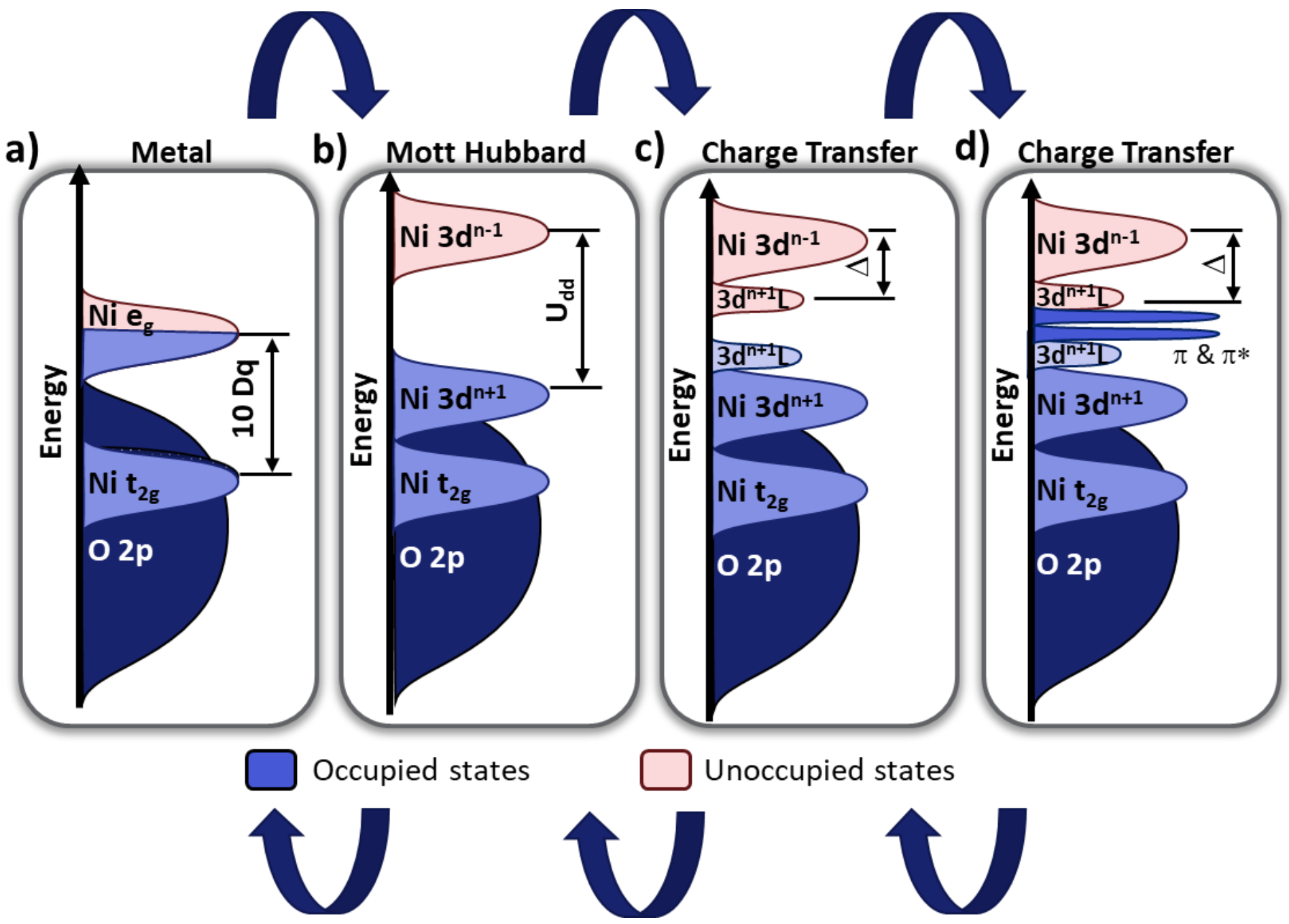

**Figure 3:** Density of states (DOS, right side) with a), only the ligand field (LF) energy 10 Dq is considered, b), Coulomb interactions $U_{dd}$, are included, and c), *Me* 3*d*, 4*s*, 4*p*-O 2*p* hybrid states are added with the charge transfer energy Δ.

Charge fluctuations between *Me* 3*d*-states ($d^n + d^n \rightleftharpoons d^{n-1} + d^{n+1}$, **Figure 4a**), which would lead to the metallic character of the solids, can be suppressed due to coulomb interactions $U_{dd}$ ($U_{dd} = E(d^{n-1}) + E(d^{n+1}) - 2 \cdot E(d^n)$), *i.e.* the energy which is necessary to place two electrons into the same *d* site, introducing a band gap between occupied (higher Hubbard) and unoccupied (lower Hubbard) states, **Figure 3b**.[89,94,96,97,99,100] Indeed, $U_{dd}$ values of up to 10 eV are reported for late transition metal compounds due to the relatively low extension of the radial distribution function of 3*d* orbitals (Mott-Hubbard picture).[93,96] The radial contribution to the wavefunction decreases with an increasing number in the group of 3*d* metals and thus, $U_{dd}$ increases. However, the "Mott-Hubbard" picture breaks down when trying to explain the influence of the ligands on the band gap – the band gap is proportional to the electronegativity of the ligands.[89,95,98,101–104] This linear relationship indicates the presence of charge fluctuations from the ligands towards the metals ($d^n \rightleftharpoons d^{n+1}$ L, L denotes an electron hole in the *p* states of the ligands) which is called ligand to metal charge transfer (LMCT), **Figure 4b**. If the LMCT energy Δ is lower than $U_{dd}$, electron density is shifted from O 2*p* to *Me* 3*d* sites (= hybridization or formation of covalent bonds) which results in the reduction of *Me* and an oxidation of the O lattice (from an oxidation state of -2 towards 0).[89,95,98,101–103,105] This introduces additional hybrid states between the upper and lower Hubbard band, **Figure 3c**, and the materials turn into charge transfer type semiconductors.

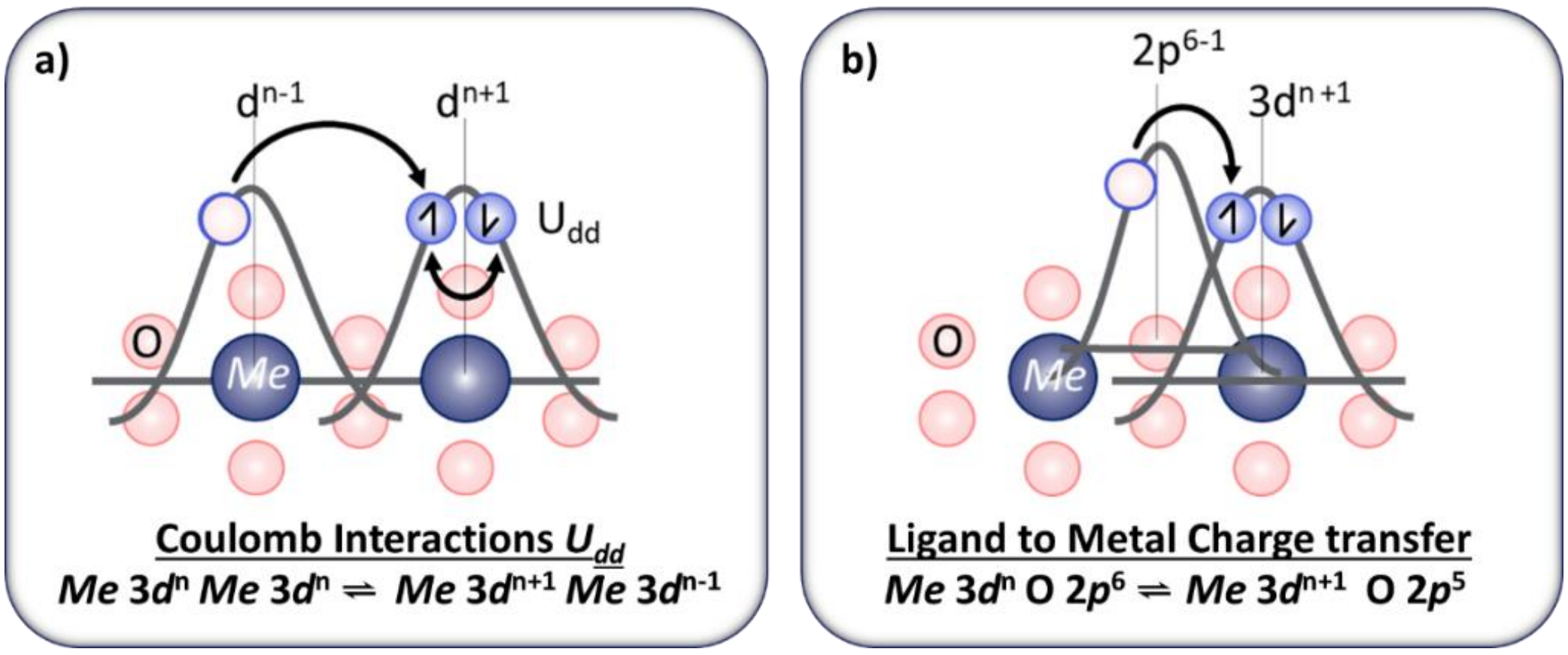

**Figure 4:** Depiction of charge fluctuations between two *Me*-sites a) and between *Me* and O b).

## 2.3. Coordination chemistry and Electrochemical performance

Depending on the coordination number of O and the arrangement of *Me* around, up to six *Me*-O σ-bonds between *Me* 3*d* ($e_g$)-, *Me* 4*s*-, *Me* 4*p*- and O 2*p*-orbitals are formed (**Figure 5a**).[87,91,92] The formation of σ-bonds leaves holes in the O 2p band (see LMCT in **Section 2.2**) which can be probed with core level spectroscopy at the O K edge.[15,29,35,106,107] Moreover, the finestructure of the *Me* $L_{2,3}$ edge is changing with a change in hybridization. Pioneering work from Thole and co-workers showed that multiplet structures of the *Me L* edge are washed out with an increasing degree of hybridization.[108] If electrons are shared between O and *Me* the electronic configuration of *Me* becomes a mixture of a *Me* $d^n$ and a *Me* $d^{n+1}$ (+ L) configuration which means that the peaks in the *Me L* edge, *i.e.* the finestructure or multiplets of the spectra, become a convolution of both configurations with characteristic satellites.[29,86,104,109–113]

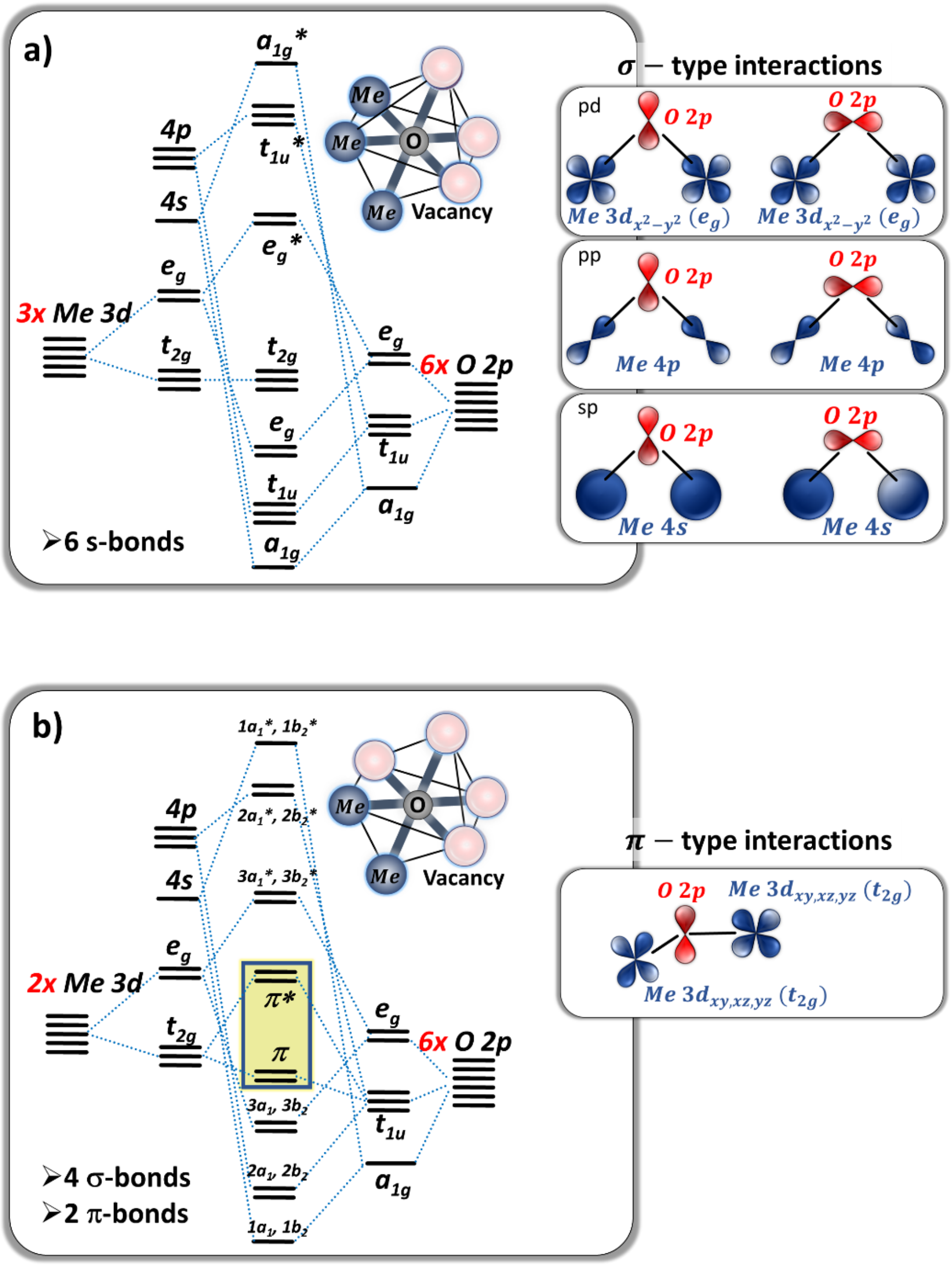

**Figure 5**: The formation of *Me*-O (*Me* = transition metal) molecular orbitals (MOs) in dependence on the coordination number of O in an octahedral environment. a) shows O with three *Me* neighbours which leads to 6 σ-type MOs. In b) only two *Me* sites are present in the neighbourhood of O which leads to 4 s-type and 2 p-type MOs. The figure is reproduced from [92,114].

If O is coordinated by less than three transition metals or if the symmetry around the *Me* is unfavorable for the formation of σ-bonds, only four or even less O 2*p* orbitals form σ-bonds with neighboring transition metals. The remaining O 2*p* orbitals can undergo weak π-type interactions with *Me* $t_{2g}$ orbitals parallel to the existing σ-bonds (**Figure 3d** and **Figure 5b**).[54,92,114–116] The underlying π-type interactions are considered as a metal to ligand charge transfer (MLCT), *i.e.* an oxidation of *Me* and a reduction of the O sites, because the *Me* $t_{2g}$ orbitals are completely occupied while the unhybridized O 2*p* orbitals have holes in the *p*-band.[91,92] The weak MLCT interactions leave almost orphaned O 2*p* orbitals near the Fermi level which become redox active at high states of charge (see **Paragraph 4**).[92,114]

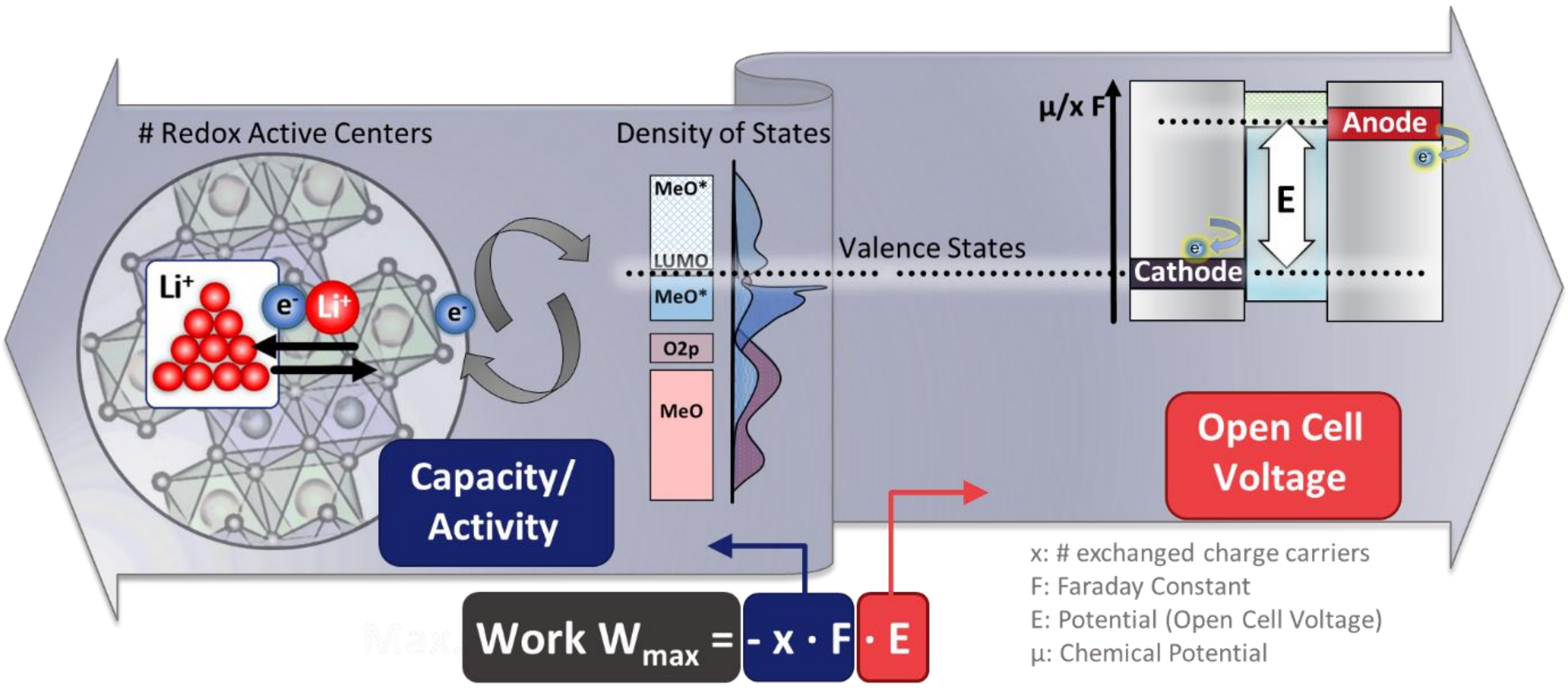

**Figure 6:** Schematic depiction of charge exchange in layered oxides: The number of electrons/holes in the valence states determine the capacity $x \cdot F$ while its relative energetic position versus the anode material is proportional to the open cell voltage $E$.

# 3. Charge compensation at low and medium states of charges

## 3.1. Cationic redox

The formal redox reaction for the charge and discharge of layered oxides is given in **Eq. 1**. According to this, the transition metals, often Ni, Co and/or Mn, are oxidized from a mean oxidation state of +3 to +4 in a so called ‘cationic redox` process.[117] In discharged Li*Me*$O_2$, Ni can reveal a +2 oxidation state if Mn is present because Mn has a +4 oxidation state.[15,32,115,118]

$$\mathrm{Li}\boldsymbol{Me(III)}\mathrm{O_2} \underset{\text{Discharge}}{\overset{\text{Charge}}{\rightleftharpoons}} (1\text{-}x)\mathrm{Li^+} + x\mathrm{e^-} + \mathrm{Li_x}\boldsymbol{Me(III+x)}\mathrm{O_2} \qquad \textbf{Eq. 1}$$

If one Li ion is extracted per formula unit of Li*Me*$O_2$, the theoretical capacity $Q_{theo}$ is given by $\frac{z \cdot F}{M}$, where $z$ is the number of exchanged charges, $F$, the Faraday constant, and $M$, the molar mass of Li*Me*$O_2$. This leads to theoretical capacities of ~270 mAh/g - ~275 mAh/g. In reality, however, less than 200 mAh/g can be achieved before the *Me*-O host structures collapse.[5,15,36,119] The point, at which the structures start to crumble is often determined with powder diffraction [1,5,16,31,36,119] and thus, the real capacities of layered oxides are so far empirical determined parameters. This also holds for the mean charge and discharge voltages relative to graphite or metallic lithium (or the empirically chosen cutoff voltages), although trends like a decreasing stability at high voltages with an increasing Ni-content are well documented.[5,7,17,24,120–124] In order to understand the stability limitations of Ni-rich layered oxides, many studies and reviews point out that oxygen plays a major role in the charge compensation (see also **Paragraph 4**).[1,32,41,115,125–131] Upon charge, electrons are taken out from the valence band of the cathode $\mu_{cathode}$ (herein $t_{2g}$, $e_g$ and/or O 2*p* orbitals as depicted in **Figure 4a**, LF picture) and are filled into the valence band of the anode, **Figure 6**.[41,125,129–131] Upon discharge, the electron exchange is reversed. According to **Eq. *2***, the difference in chemical potential $\mu_i$ between the valence band of the cathode and the anode determines the difference in Gibbs free energy $\Delta G$ and consequently, the cell voltage E.

$$\Delta G = \sum_i \mu_i = -x \cdot F \cdot E \qquad \textbf{Eq. 2}$$

Essentially, this approach describes that charge compensation takes place at *Me* and O sites and the relative energetic position of the O 2*p* and *Me* 3*d* states determines, to which extent the elements are oxidized/reduced, **Figure 7**. The chemical potentials of different transition metal oxides are tabulated in [132].

Caution must be taken when discussing the relative oxidation/reduction potential of correlated transition metals, *e.g.* if they occupy the same sites in layered oxides: The LF depiction of the DOS (**Figure 4a** and **Figure 7**) implies that all transition metals become redox active (reduction/oxidation) once the higher/lower lying valence states are completely empty/full.[125,129–131] However, x-ray absorption spectroscopy results reveal, for example, that transition metals like Co and Mn are hardly redox active in the presence of Ni [1,15,34], findings which are the basis of controversial discussions in many studies.[5,15,34–36,110,134] Before Mn and Co show a significant contribution to the charge compensation, the structures collapse leading to a breakdown in Lis diffusion kinetics [14], subsequent oxygen release [17,18,120], electrolyte decomposition reactions [17,135,136], surface morphology changes [34,84,122,137], and transition metal dissolution.[15,17,18,115,138] In case of NCM111, NCM622 and NCM811 ($LiNi_{1/3}Co_{1/3}Mn_{1/3}O_2$, $LiNi_{0.6}Co_{0.2}Mn_{0.2}O_2$, and $LiNi_{0.8}Co_{0.1}Mn_{0.1}O_2$), for example, significant changes between the charged and the discharged state are only observed at the Ni $L_{2,3}$ and O K edge, while the Mn $L_{2,3}$ and Co $L_{2,3}$ do not show a difference, **Figure 9**.

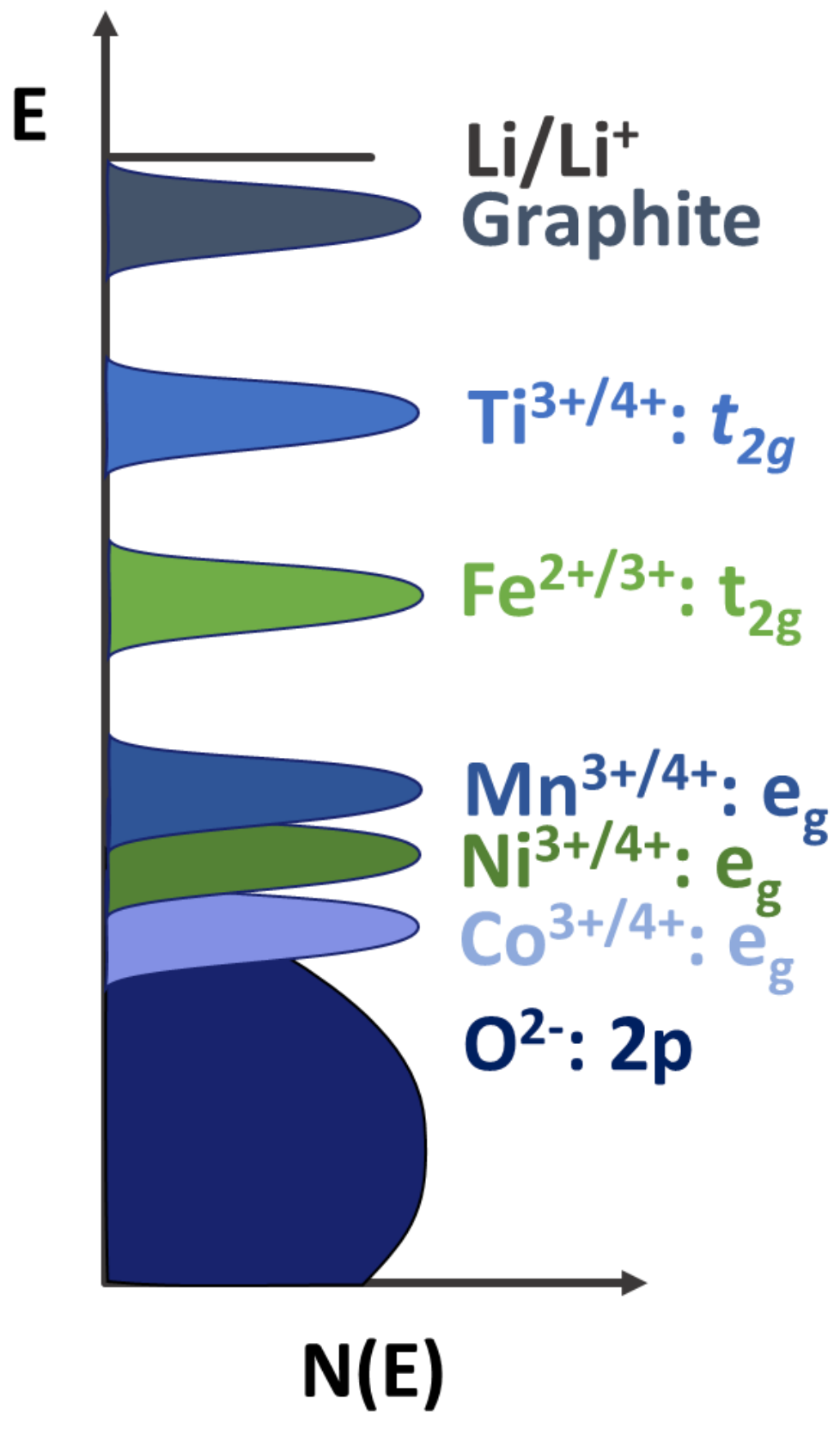

**Figure 7:** Relative positions of *Me* 3d valence states with respect to the of the O:2p band as reproduced from [129,130,133].

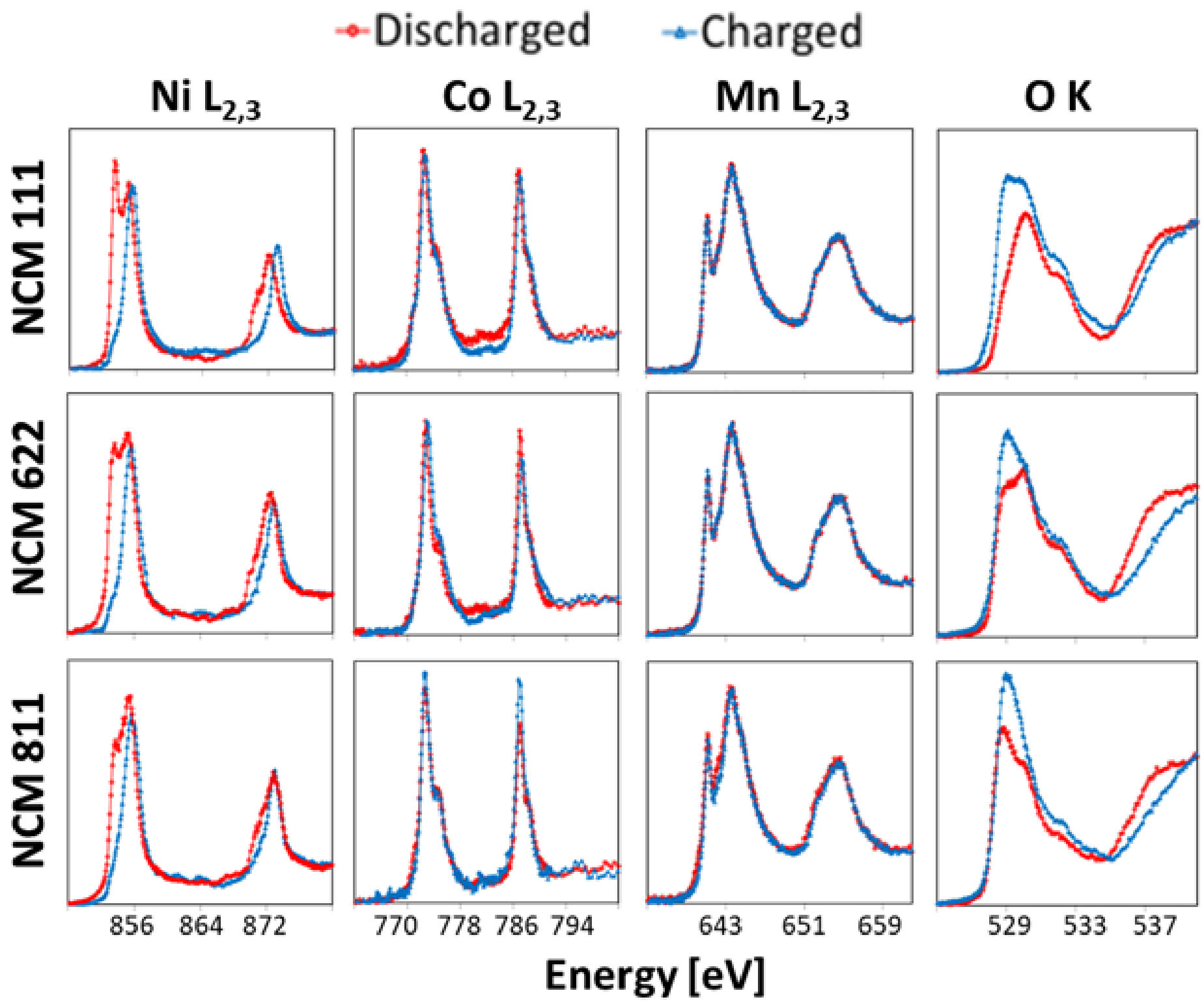

**Figure 9:** Ni, Co, Mn $L_{2,3}$ and O K edge, determined with NEXAFS spectroscopy in fluorescence yield mode in the charged and discharged state of NCM111 ($LiNi_{1/3}Co_{1/3}Mn_{1/3}O_2$), NCM622 ($LiNi_{0.6}Co_{0.2}Mn_{0.2}O_2$) and NCM811 ($LiNi_{0.8}Co_{0.1}Mn_{0.1}O_2$). The data is reproduced from reference [15].

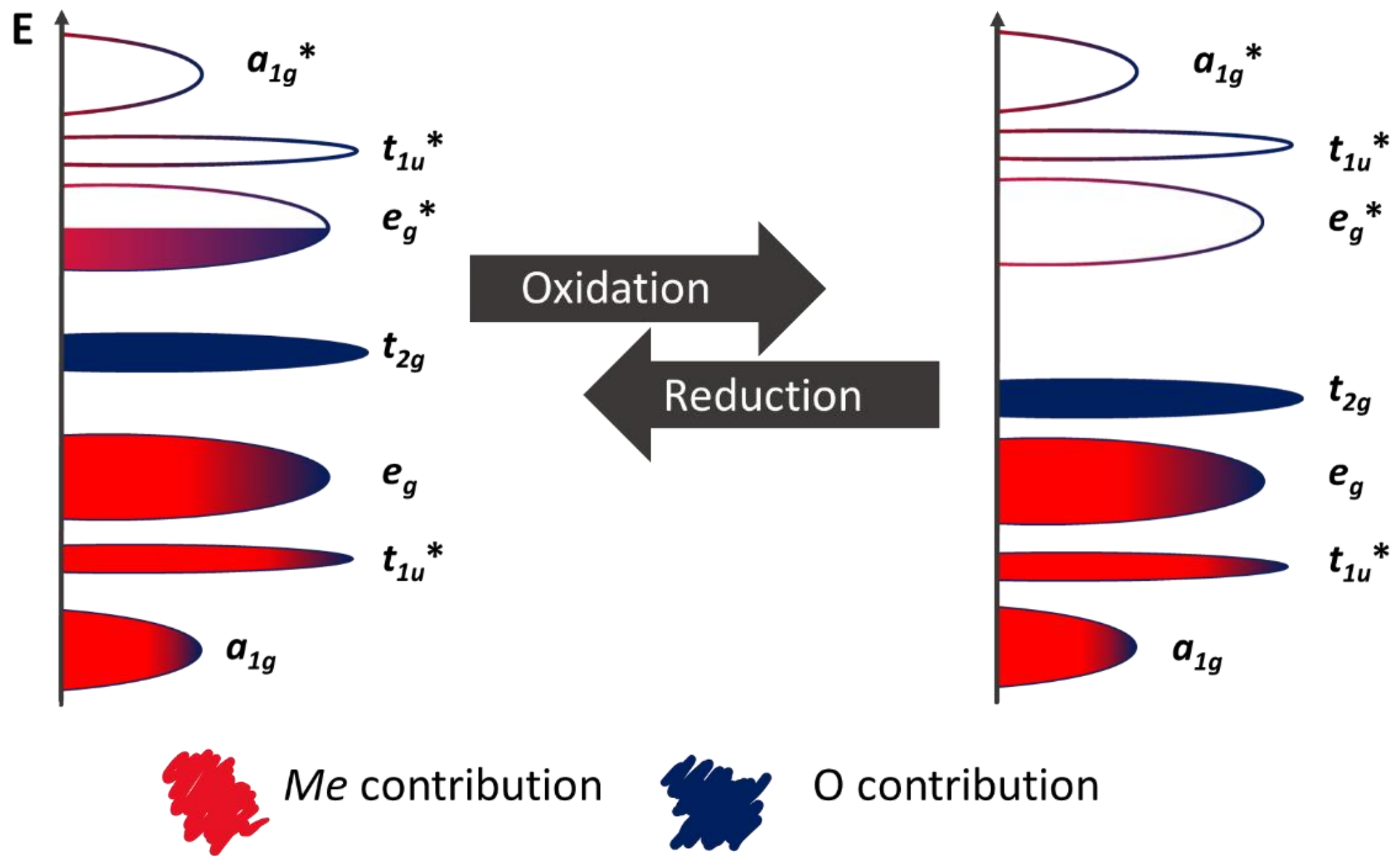

**Figure 8:** Depiction of the DOS of $LiNi_{1-x}Me_xO_2$ (left, charged layered oxides) and $Ni_{1-x}Me_xO_2$ (right, discharged layered oxides) deduced from the MO scheme in **Figure 5**a and reproduced from [92].

## 3.2. *Me*-O hybridization

### 3.2.1. Charge compensation from the perspective of O

The LF picture (**Figure 4a**) allows transition metal oxidation/reduction without the involvement of the O lattice or in other words, the independent oxidation of oxygen and transition metals.[15,29,34,35] However, oxygen does undergo σ and π type charge transfer interactions with its transition metal neighbors and thus, the oxidation/reduction of oxygen cannot be considered independent from the redox process at *Me* sites. A more advanced picture includes *Me* 3*d*-O 2*p* hybridizations (**Figure 4c**) assuming that the band gap in lsayered transition metal oxides is of charge transfer type.[96,97,139–142] In Ni-rich layered oxides, for example, the highest occupied states are Me $3d^{n+x}\,\underline{L}^{x}$ ($0 \leq x \leq 1$) states, where $\underline{L}$ denotes a hole in the O 2*p*-band. The lowest, unoccupied states are of the type Me $3d^{n+1}$, respectively (**Figure 8**).[97,100,143]

Kuiper *et al.* showed with O K NEXAFS on $Li_xNi_{1-x}O$ ($0 < x < 0.5$) that Li doping of NiO leads to the formation of holes in the O 2*p* band while the number of *d* electrons around Ni (starting with a $3d^8$ configuration at x = 0) does hardly change.[144,145] The charge imbalance is compensated by a charge transfer from O 2*p* towards Ni 3*d* states, *i.e.* by the formation of covalent Ni-O bonds as evident from upcoming peaks at 529 eV (Ni $e_g$-O 2*p*) and 530 eV (Ni $t_{2g}$-O 2*p*) in the O K spectra which represent holes in the O 2*p* band.[29,100,144–146] The underlaying charge compensation in Ni-containing layered oxides ($Li_yNi_{1-x}Me_xO_2$ with $0.6 < y < 1$, $0 < x \leq 1$) is based on the same principle: With an increasing SOC (= decreasing Li content, *i.e.* upon charge) peaks at 529 eV and 530 eV in the O K spectra increase until the lattice starts to collapse, **Figure 10**.[15,22,34,38,118] This means, electron density is shifted from Ni towards O sites (LMCT), the *Me*-O bonds become more covalent (formation of σ bonds), and the gap between bonding and antibonding states increases, **Figure 8**.[92,144,145] A higher degree of covalence also means that the *Me*-O bonds become shorter, **Figure 1c**.[1,15,31,34,35,92,110,147] A major conclusion from these findings is that the number of electrons around Ni in $Li_xNi_{1-x}O$ ($0 < x < 1$) does hardly change – instead, electron holes $\underline{L}$ at O sites evolve indicating *Me*-O hybridization to compensate for the charge imbalance.

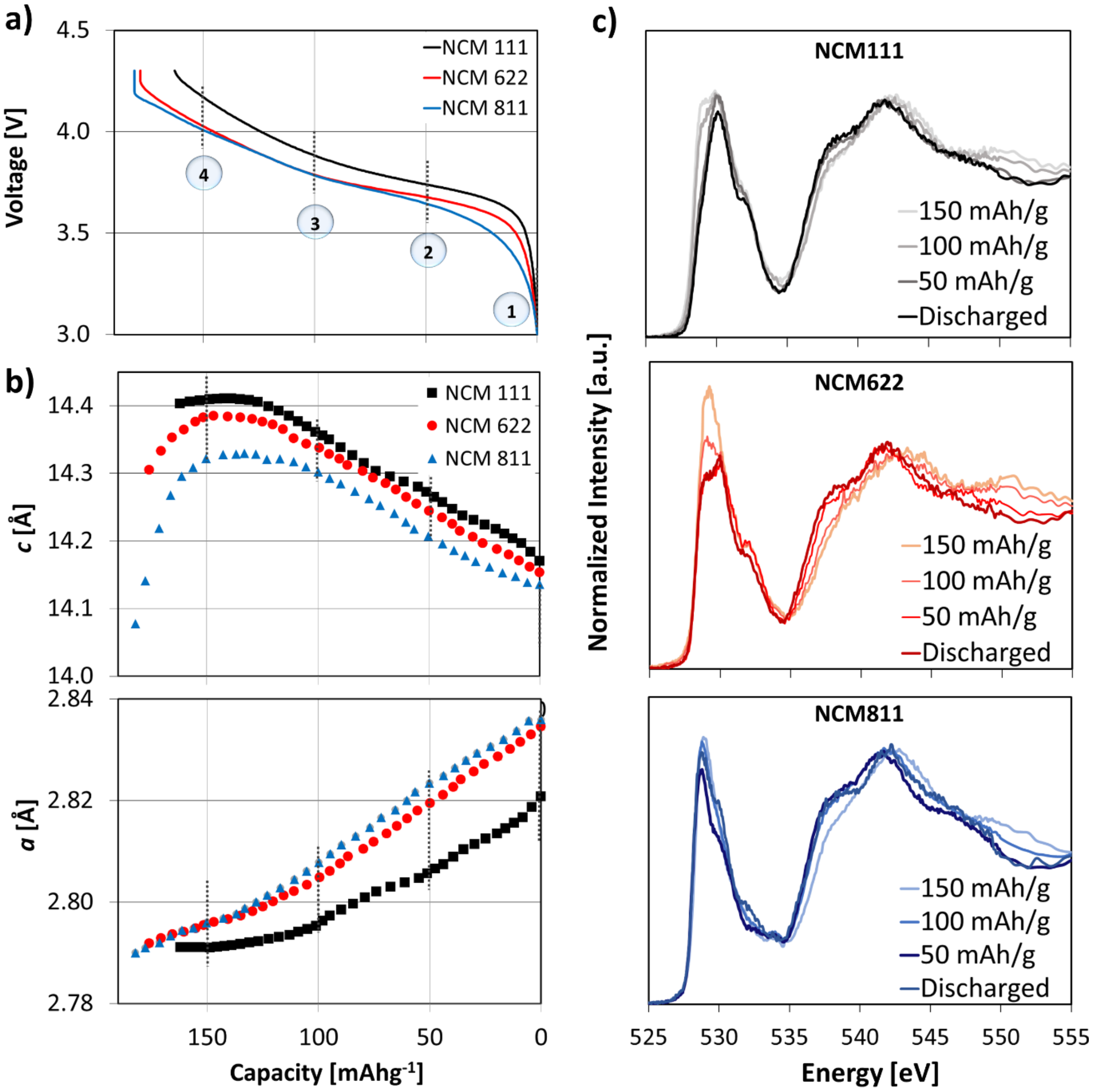

**Figure 10:** a) 1st discharge of NCM111, NCM622 and NCM811, b) the corresponding *c* and *a*, *b* lattice parameters, and c), the O K edge at 150 mAh/g (4), at 100 mAh/g (3), at 50 mAh/g (2) and in the discharged state (1). The data is reproduced from reference [15].

The comparison of the O K pre-peaks with reference materials such as $LiNiO_2$ (LNO), $LiCoO_2$ (LCO) and $Li[Li_{1/3}Mn_{2/3}]O_2$ (LMO) proves that only Ni-O hybrid states show a significant redox activity in NCMs, **Figure 11**. The position and changes in the intensity of the peaks at 528 eV and 531 eV are similar to the changes in the O K spectra of $LiNiO_2$. For the choice of the reference materials, the electronic configuration of the transition metal *Me* needs to be the same as in the NCMs, and the references should be isostructural or structural related to the $R\bar{3}m$ space group. Therefore, rhombohedral LCO ($LiCoO_2$, $Co^{3+}$ low spin) and LNO ($LiNiO_2$, $Ni^{2+/3+}$ low spin) is used. Mn in $LiMnO_2$ has a $t_{2g}^4$ $e_g^0$ (formal $Mn^{3+}$) configuration and is therefore not appropriate for a comparison to a $t_{2g}^3$ $e_g^0$ (formal $Mn^{4+}$) configuration as present in the NCMs. Instead, LLO ($Li_2MnO_3$ = $Li(Li_{1/3}Mn_{2/3}O_2)$ with a formal +4 oxidation state of Mn and a higher ordered rhombohedral structure is used.[15,148]

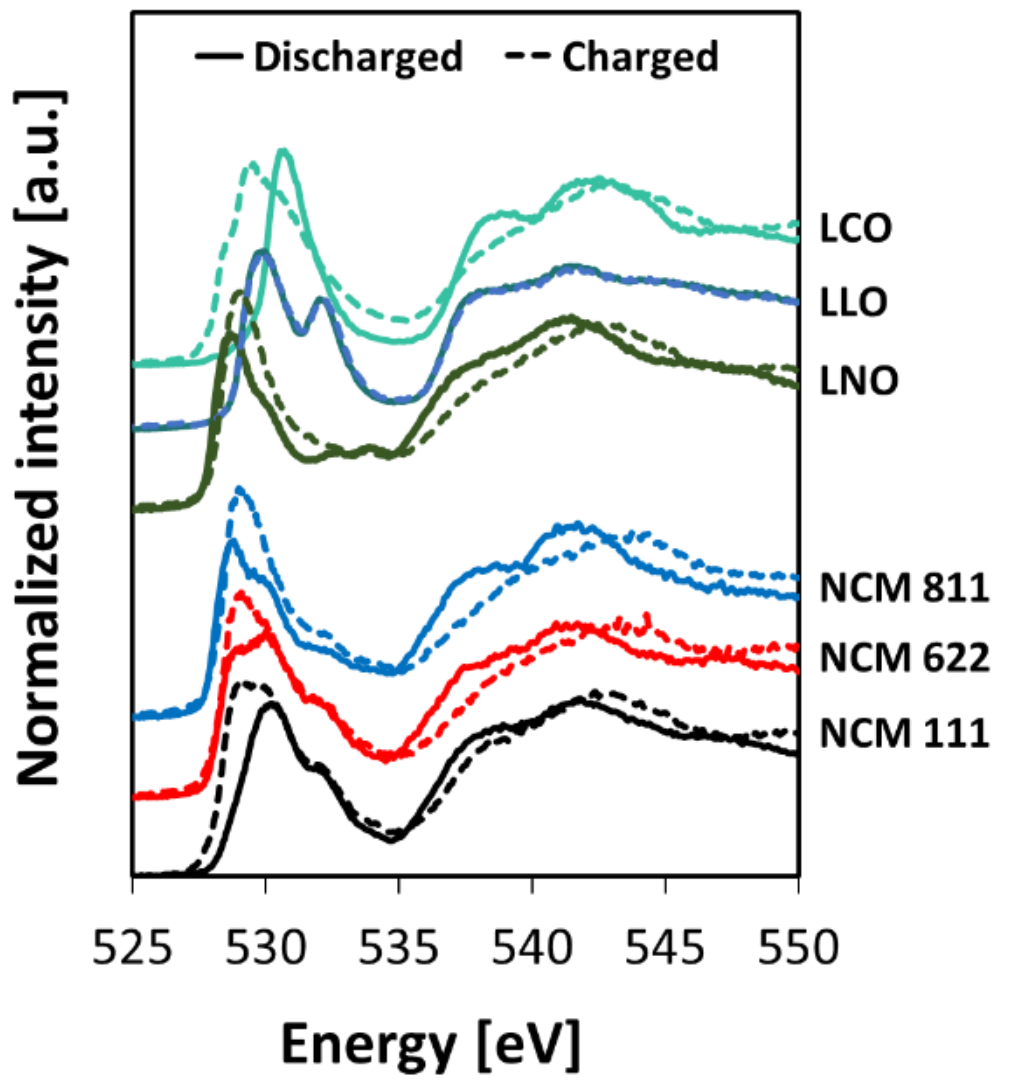

**Figure 11:** O K NEXAFS spectra of NCM111 ($LiNi_{1/3}Co_{1/3}Mn_{1/3}O_2$), NCM622 ($LiNi_{0.6}Co_{0.2}Mn_{0.2}O_2$), NCM811 ($LiNi_{0.8}Co_{0.1}Mn_{0.1}O_2$), LNO ($LiNiO_2$), LLO ($Li[Li_{1/3}Mn_{2/3}]O_2$) and LCO ($LiCoO_2$) in the charged (solid line) and discharged states (dashed lines). The Figure is reprinted from reference [15].

### 3.2.2. Charge compensation from the perspective of Ni

More insights into the charge compensation can be revealed from the changes in the Ni $L_{2,3}$, **Figure 12**).[15,34,35,109,110] Pioneering work on multiplet calculations allows to simulate and interpret the finestructure in *Me* $L_{2,3}$ edges including ligand field effects, charge transfer, spin orbit coupling, coulomb interactions and many more interactions, excitations and transitions.[86,87,108,111,149–153] Among the first materials which were extensively studied with regards to the Ni $L_{2,3}$ edge is NiO.[22,29,87,99,137,144,153–157] Herein, Ni reveals a +2 configuration which is very similar to one of the configurations observed in the Ni $L_{2,3}$ edges of NCMs, **Figure 12**. This configuration (yellow CTM calculation, **Figure 12**) with main peaks at 853 eV ($L_3$ edge) and 873 eV ($L_2$ edge) is very ionic (Ni $3d^{8.04}$ $\underline{L}^{0.04}$).[15] With an increasing atomic number in the periodic table, Coulomb interactions between *d*-electrons ($U_{dd}$) increase and thus, pairing electrons in $e_g$ orbitals becomes energetically less favorable. $Ni^{2+}$ has two half-filled $e_g$ orbitals and an electron transfer from O towards Ni (formation of σ-bonds) would require overcoming $U_{dd}$. Consequently, the $Ni^{2+}$ configuration remains very ionic.[15,34] Upon charge, the $Ni^{2+}$ peaks in **Figure 12** vanish (from the top to the bottom) and its relative intensity is decreasing with an increasing Ni content (from NCM111 to NCM811, **Figure 12**, top row). But looking at the Ni stoichiometry (1-*x* in $LiNi_{1-x}Co_{x/2}Mn_{x/2}O_2$) which is increasing from 1/3 (NCM111) *via* 0.6 (NCM622) to 0.8 (NCM811), the absolute amount of $Ni^{2+}$ remains close to 0.25 almost independent of the Ni content.[15] While the ionic $Ni^{2+}$ configuration is well known and studied in literature [22,29,87,144,153–155,157] and thus, indisputable, the oxidized Ni configuration, *i.e.* the almost solely remaining configuration at ≥ 150 mAh/g (**Figure 12**, last row, blue calculation), requires a careful discussion. At high SOCs, in turn, layered oxides reveal the highest degree of covalence because the number of holes in the O K band reach a maximum at ~ 150 mAh/g (**Figure 10**, for references see also [29,100,144–146]). In agreement with that, the oxidized Ni configuration (**Figure 13a**) shows blurred peaks which indicate that the increase in covalence stems from increasing covalent interactions between Ni and O (see also **Paragraph 2**).[108] Charge transfer (CT) effects, *i.e.* a convolution of more than one configuration such as Ni $d^n \rightleftharpoons$ Ni $d^{n+1}\underline{L}$ (**Figure 13a**), lead to the observed peak broadening.[108] The oldest calculations of oxidized Ni species such as Ni in $NiO_2$ reveal a 4+ configuration, **Figure 13b**. This is the oxidation state expected from stoichiometric considerations based on **Eq. 1**.[111,112] $Ni^{4+}$ is a $d^6$ low spin configuration which leads to a sharp peak in the Ni $L_3$ (peak A) and $L_2$ edge (peak B), respectively, because the dipole selection rule only allows electron transitions of the type $2p_{1/2}^2\ 2p_{3/2}^4\ t_{2g}^6\ e_g^0 \rightarrow 2p_{1/2}^2\ 2p_{3/2}^3\ t_{2g}^6\ e_g^1$ ($L_3$ edge) and $2p_{1/2}^2\ 2p_{3/2}^4\ t_{2g}^6\ e_g^0 \rightarrow 2p_{1/2}^1\ 2p_{3/2}^4\ t_{2g}^6\ e_g^1$ ($L_2$ edge). The satellites at lower and higher energies (peaks C-E in the calculation, **Figure 13b**) are assigned to admixing high spin configurations (the excited states are $2p_{1/2}^2\ 2p_{3/2}^3\ t_{2g}^5\ e_g^2$, $2p_{1/2}^1\ 2p_{3/2}^4\ t_{2g}^4\ e_g^3$, $2p_{1/2}^2\ 2p_{3/2}^3\ t_{2g}^4\ e_g^3$, $2p_{1/2}^1\ 2p_{3/2}^5\ t_{2g}^4\ e_g^2$).[29,112] For the calculations, a trigonal distortion of the Ni-octahedron resulting in a $D_{4h}$ symmetry was implemented as observed upon delithiation of $LiNiO_2$ with *operando* powder diffraction (PD).[31,137,158,159] Note that *operando* PD of conventional layered oxides of the type $Li_yNi_{1-x}Me_xO_2$ ($1 < y < 0.4$), however, reveal a rhombohedral structure at all SOCs.[5,12,15,36,123,124,159] The deviation between $Li_yNi_{1-x}Me_xO_2$ and $Li_yNiO_2$ goes either back to a lower degree of delithiation at the end of charge (usually only 50%-70% of the Li ions are extracted from $Li_yNi_{1-x}Me_xO_2$ before the structures crumble [119,160]) or other metals such as Mn, Co, Al or Mg stabilize the layered structures due to pillar effects.[7,161,162] Even though the CTM calculations of $Ni^{4+}$ were another advance in the field, simulating the asymmetric shape and the peak broadening observed in the spectra of oxidized Ni (**Figure 13a**) became only possible with the consideration of ligand to metal charge transfer (LMCT) interactions.[111,153,163] Building on this, the work from Montoro *et al.* and Kleiner *et al.* reveal a covalent $Ni^{3+}$ configuration in charged NCMs using charge transfer multiplet (CTM) calculations (**Figure 13c**), although the experimentally measured Ni $L_{2,3}$ edge looks very similar to that obtained earlier.[15,34,109,110] In these studies ligand to metal charge transfer (LMCT) is included by admixing excitations of the type $Me\ 2p^6\ 3d^{n+1}\ O\ 2p^5 \rightarrow Me\ 2p^5\ 3d^{n+2}\ \underline{L}$. Herein, an $O_h$ symmetry was used to account for the ligand field around Ni.[5,12,15,36,123,124,159] Dipole allowed transitions are $2p_{1/2}^2\ 2p_{3/2}^4\ t_{2g}^6\ e_g^1 \rightarrow 2p_{1/2}^2\ 2p_{3/2}^3\ t_{2g}^6\ e_g^2$ ($L_3$ edge), $2p_{1/2}^2\ 2p_{3/2}^4\ t_{2g}^6\ e_g^1 \rightarrow 2p_{1/2}^1\ 2p_{3/2}^4\ t_{2g}^6\ e_g^2$ ($L_2$ edge), $2p_{1/2}^2\ 2p_{3/2}^4\ t_{2g}^6\ e_g^2\ \underline{L} \rightarrow 2p_{1/2}^2\ 2p_{3/2}^3\ t_{2g}^6\ e_g^3\ \underline{L}$ ($L_3$ edge) and $2p_{1/2}^2\ 2p_{3/2}^4\ t_{2g}^6\ e_g^2\ \underline{L} \rightarrow 2p_{1/2}^1\ 2p_{3/2}^4\ t_{2g}^6\ e_g^3\ \underline{L}$ ($L_2$ edge), respectively, which lead to main peaks at 855 eV, 857 eV, 873 eV and 874 eV (peaks A-D, **Figure 13c**). The satellites (peaks F and G) do not correspond to an excitation in any unoccupied state but are peaks which arise from many-electron effects.[86,89,91] Screening of a relatively large parameter field (10 Dq = 1 eV – 4 eV, CT = -15 eV – 15 eV, $d^n$ with n = 8, 7, 6) revealed that the best match between the simulation and the experimental spectra is obtained for Ni $3d^{7.7}$ O $2p^{5.3}$.[15,109]

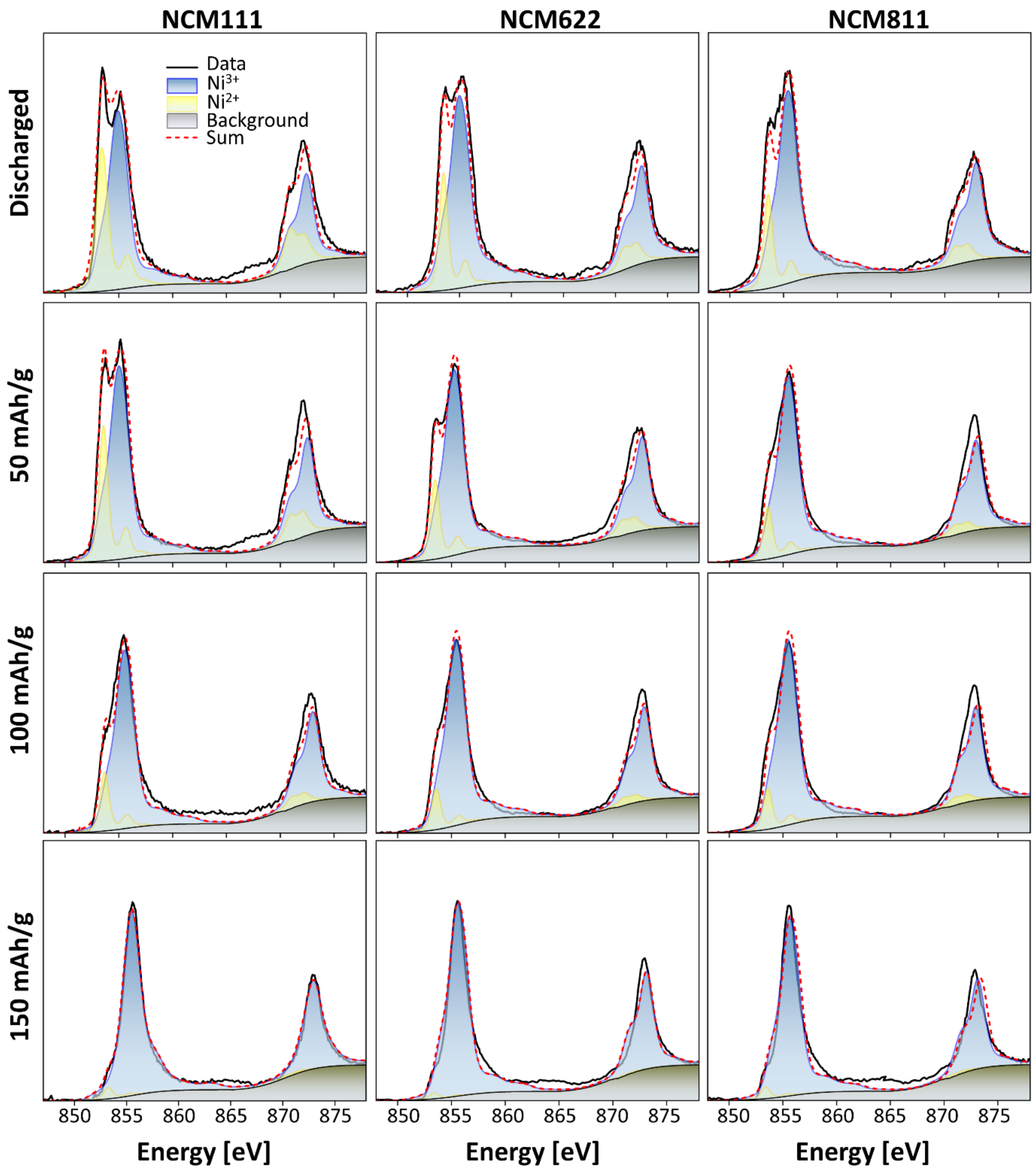

**Figure 12:** Ni $L_{2,3}$ edge of NCM111, NCM622 and NCM822, determined upon the 1st C/20 discharge at 150 mAh/g, at 100 mAh/g, at 50 mAh/g and in the discharged state, respectively. The blue and yellow areas represent charge transfer multiplet calculations for a covalent $Ni^{3+}$ (Ni $3d^{7.7}$ O $2p^{5.3}$) and an ionic $Ni^{2+}$ (Ni $3d^{8.04}$ O $2p^{5.96}$) configuration as reproduced from reference [15].

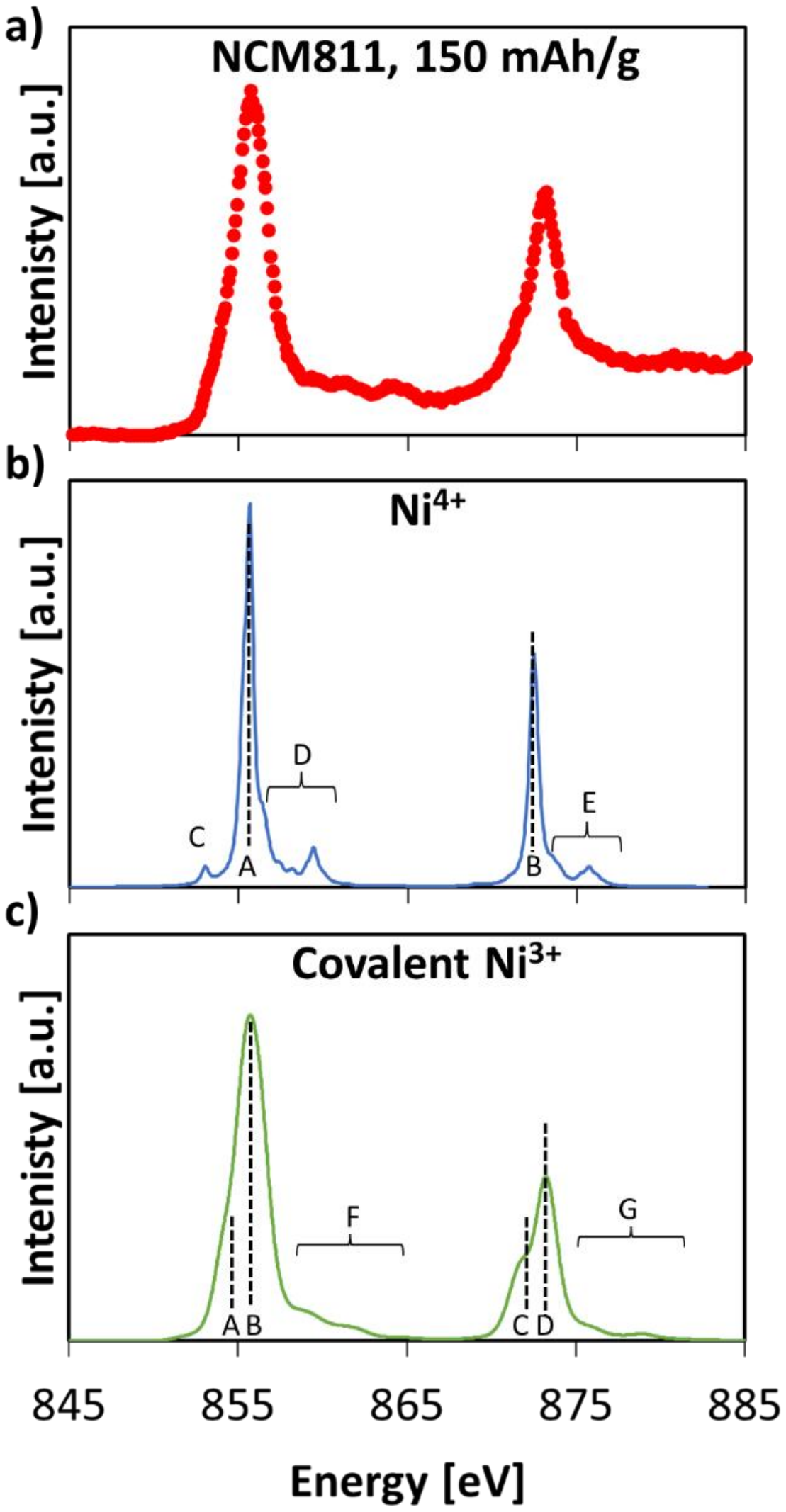

**Figure 13:** (a), Measured Ni $L_{2,3}$ edge of NCM811 discharged to 150 mAh/g, (b), Ni $L_{2,3}$ edge of a $Ni^{4+}$ configuration determined with CTM calculations reproduced from [29,112], and, (c), CTM calculation of covalent $Ni^{3+}$ reproduced from [15,34,109,110].

The covalent $Ni^{3+}$ describes the finestructure of the measured spectrum better compared to $Ni^{4+}$ (**Figure 13a-c**, *e.g.* the low energetic shoulders cannot be described by the 4+ configuration).[15] This even holds if LMCT effects are included in the $Ni^{4+}$ calculations, **Figure 15**.[15] These findings have recently been confirmed by density functional theory and dynamical mean-field theory calculations based on maximally localized Wannier functions.[164] However, if covalent $Ni^{3+}$ is considered as the correct configuration, a charge imbalance as described by Galakhov *et al.* is the consequence, because the formal oxidation state of Ni in $NiO_2$ (= charged $LiNiO_2$) is +4.[100] The origin of this charge imbalance is discussed in more detail in **Paragraph 4**. Besides the discrepancies in the symmetry and charge transfer between the 4+ and 3+ simulation (**Figure 13b** and **c**), it also needs to be mentioned that CTM calculations describe the finestructure of a *Me* $L_{2,3}$ spectra based on the number of electrons in the 3*d* states. LMCT of more than one electron are not included in the present calculations - a Ni $3d^{6+2}$ ($Ni^{4+}$ $\underline{L}^2$) and a Ni $3d^{7+1}$ ($Ni^{4+}$ $\underline{L}^1$), however, would reveal very similar spectra. Thus, the real oxidation state of Ni remains undefined since CTM calculations only provide an estimation of the numbers of *d* electrons while holes in the O 2*p* band cannot be absolutely quantified.

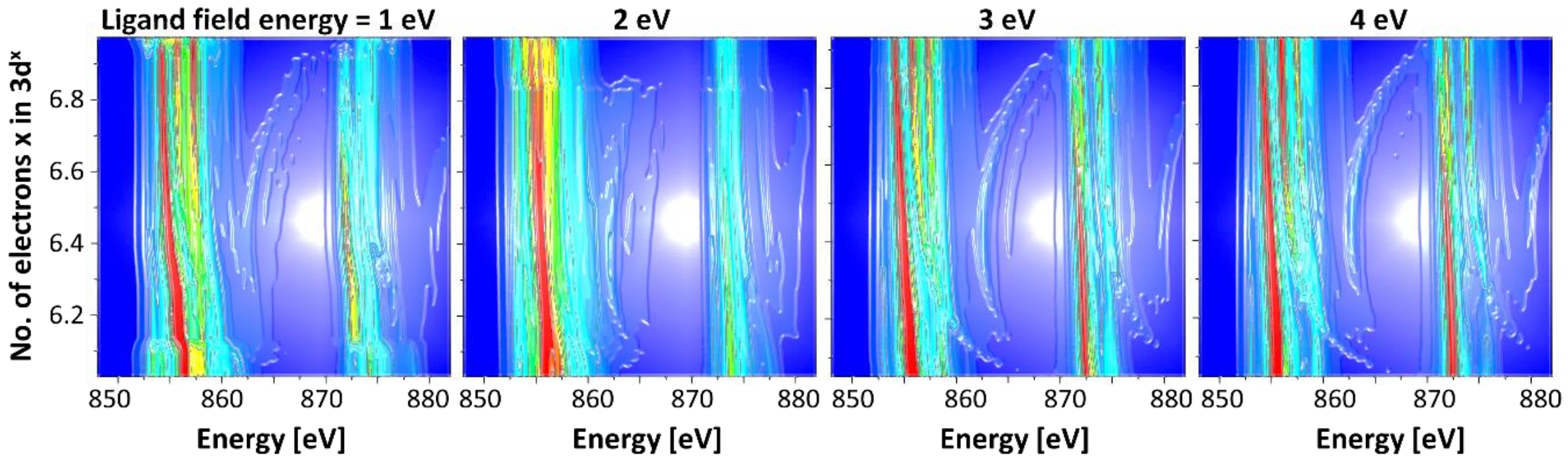

**Figure 15:** Charge transfer multiplet calculations of Ni $3d^{6+y}$ $\underline{L}^y$ ($0 \leq y \leq 1$), performed with a ligand field energy of 1 eV (A), 2 eV (B), 3 eV (C) and 4 eV (D), respectively. Below the calculations the charged Ni $L_{2,3}$ edges of NCM111, NCM622 and NCM811 are shown. The data is reproduced from reference [15].

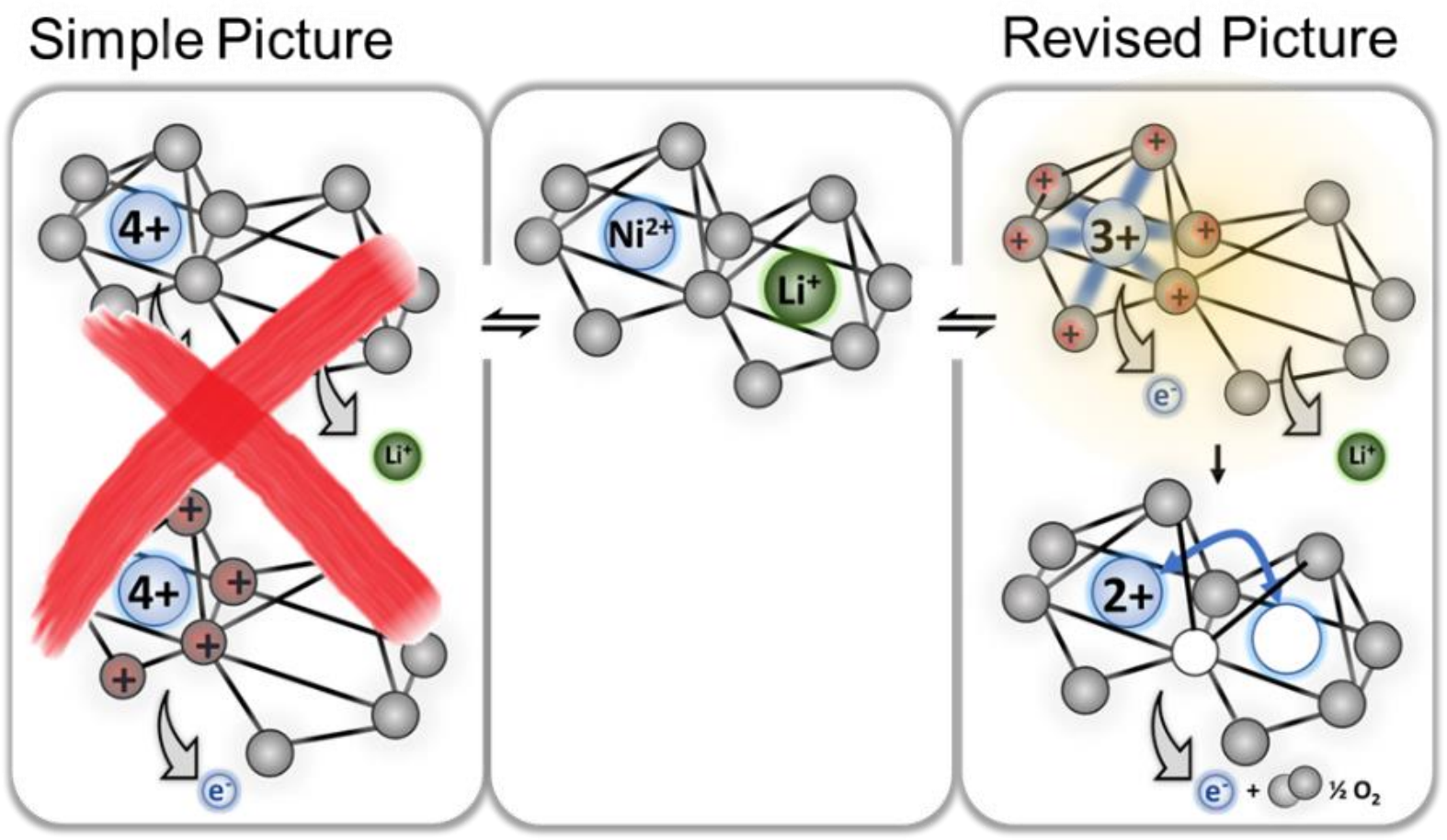

**Figure 14:** Schematic depiction of the charge compensation in Ni-containing layered oxides which involves the oxidation of ionic $Ni^{2+}$ to covalent $Ni^{3+}$. The schematic illustrations are reproduced from reference [15].

From the discussion above it can be concluded that at low and medium SOCs electron density is shifted from O 2*p* towards Ni 3*d* sites upon charge and vice versa upon discharge, forming/breaking covalent bonds between Ni and O, **Figure 12**.[15,29,127] Ni is oxidized/reduced from/to ionic $Ni^{2+}$ to/from covalent $Ni^{3+}$, **Figure 14**.[15,34,109] Thus, $Ni^{2+}$ is the redox active configuration in Ni-containing layered oxides and recent advances in the field aim at maximizing the Ni-content.[4–6,165–169]

## 3.3. Physical limitations in energy density

### 3.3.1. Onset of irreversible reactions

Ni-rich layered oxides show outstanding energy densities which come at the expense of cycling stability.[1–10,14,15,17,22,124,169–171] Despite many efforts to stabilize the structures by introducing concentration gradients within the particles [65–68], using cationic or anionic substitution and doping [69–77], protecting the particles with surface coatings [78–82], and optimizing the morphology of the secondary and primary particles [83–85], the poor cycling stability of Ni-rich layered oxides remains a major challenge.[1,17,62] At high SOCs (> 150 mAh/g or x in $Li_x[Ni,Co,Mn]O_2$ < 0.55) the *c* lattice parameters of the layered oxides collapse as evident from the strong shift of the 003 reflection towards higher 2θ values, **Figure 16** (yellow marked areas).[8,9,11–13,119]

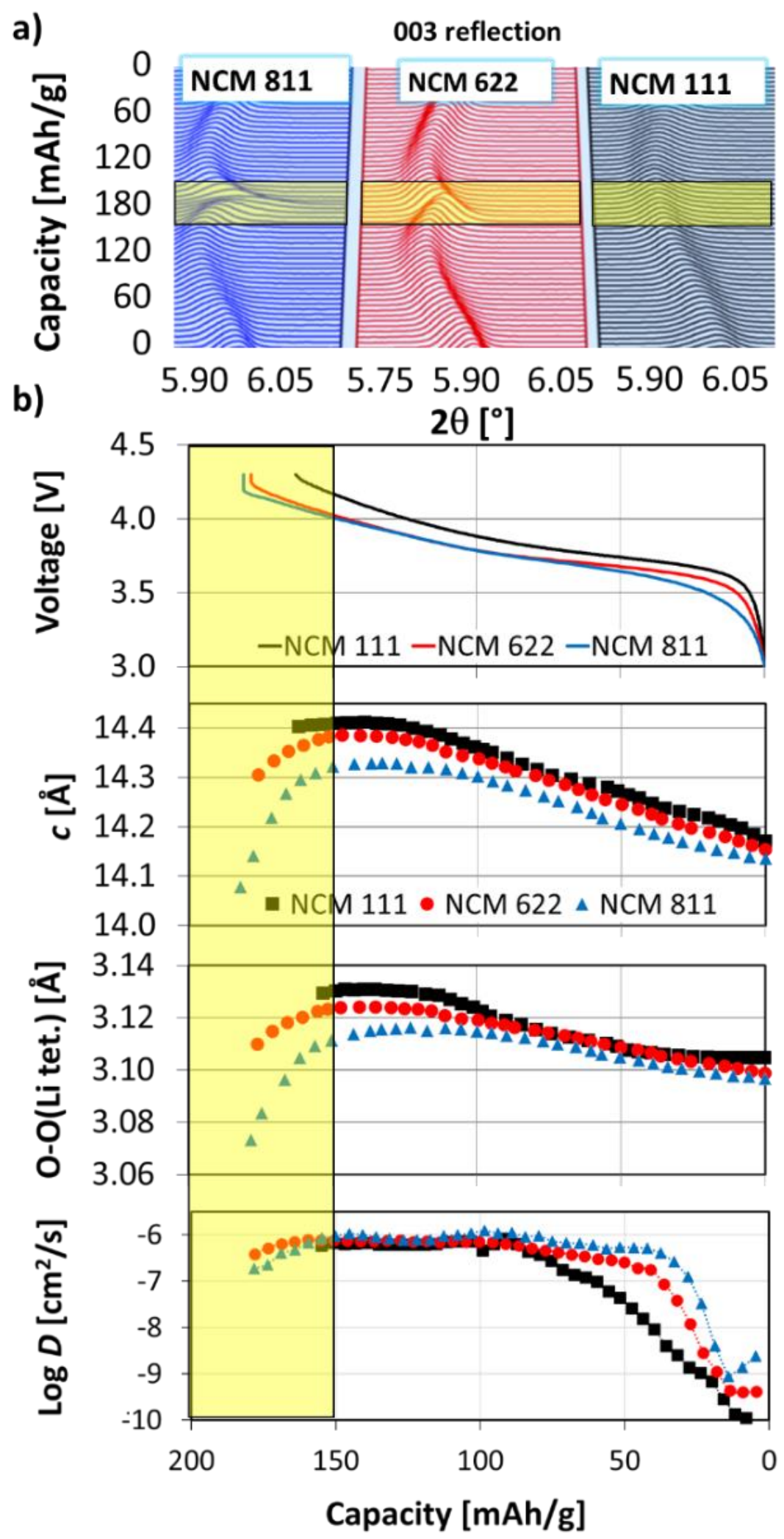

**Figure 16:** (a) 003 reflection of NCM811, NCM622 and NCM111 using operando SXPD at 25 eV. (b) shows the voltage profile, the *c* lattice parameter, the O-O distance of the Li tetrahedron, and the diffusion coefficient, determined with GITT. The data was collected upon the first discharge of the NCMs and is reproduced from [15].

The *c* lattice parameter, in turn, determines the O-O distance of the tetrahedral sites in the Li layer (**Figure 16b,** 2nd and 3rd panel), the intermediate sites upon Li diffusion (**Figure 17**). With the breakdown of the *c* lattice parameter the Li tetrahedron shrink, Li must pass through much smaller O-triangles while hopping into a neighboring Li site, and consequently, the activation barrier to overcome the attractive interactions between O and $Li^+$ increases. This slows down the Li diffusion as revealed with GITT (**Figure 16b,** 4th panel) [14,15,34] and $^7$Li NMR.[14–16,124]

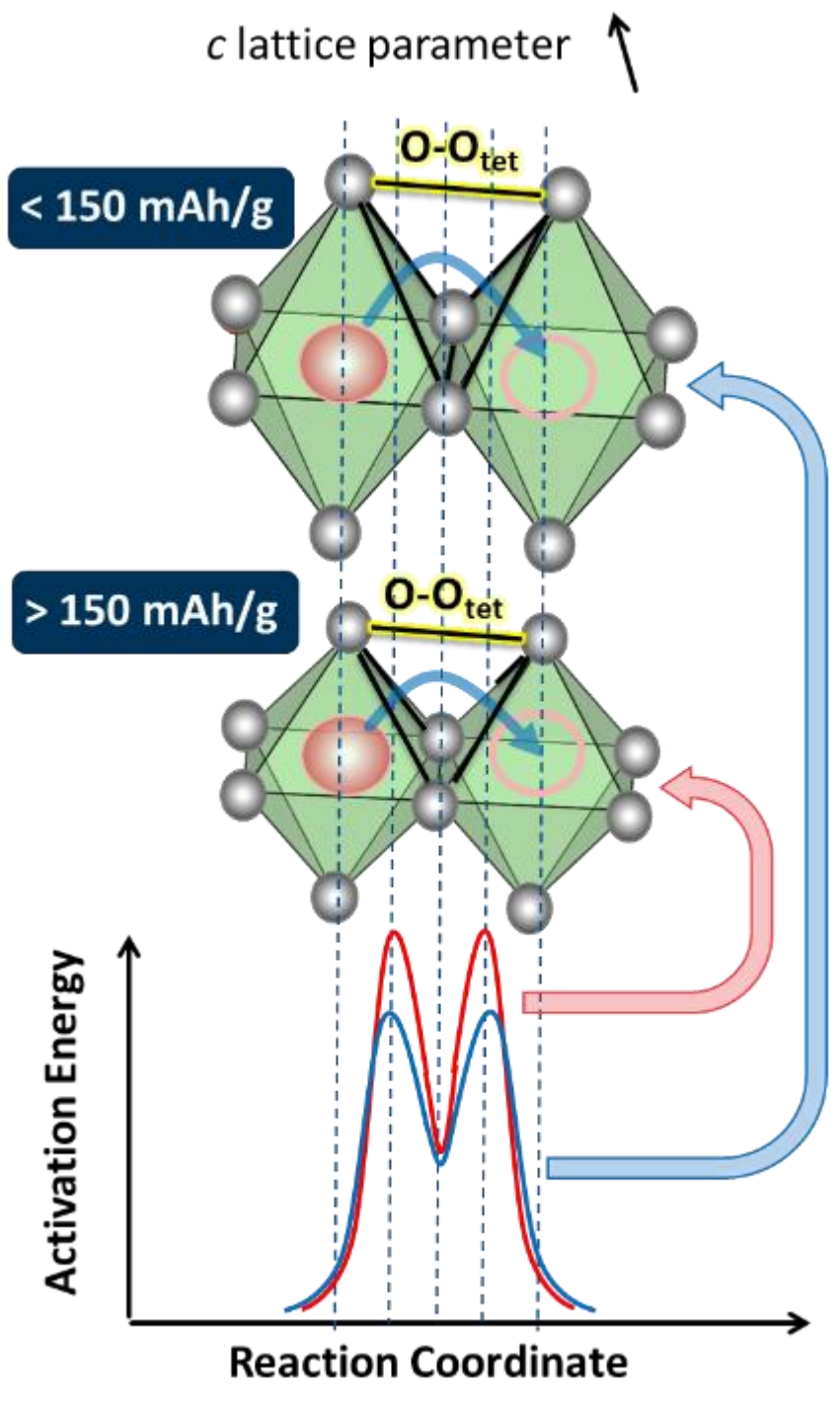

**Figure 17:** Schematic depiction of the Li diffusion pathway in layered oxides at low to medium SOCs (< 150 mAh/g), and at high SOCs (> 150 mAh/g). The lower graph shows the energy profile along the Li diffusion pathway.

Upon operation of a battery, the consumer usually does not know about kinetic limitations and thus, does not adjust the applied load at high SOCs. To deliver the same current with a reduced Li diffusion kinetics and to compensate for the abrupt volumetric changes which come along with the breakdown of the lattice, irreversible reactions such as oxygen release [17–19,138], surface morphology changes [11,20–22,28] (**Figure 18a**), transition metal migration [148,162,172,173] (**Figure 18b**), transition metal dissolution [23,24,61] (**Figure 18c**), and crack formation [25–28] (**Figure 18d**) set in.

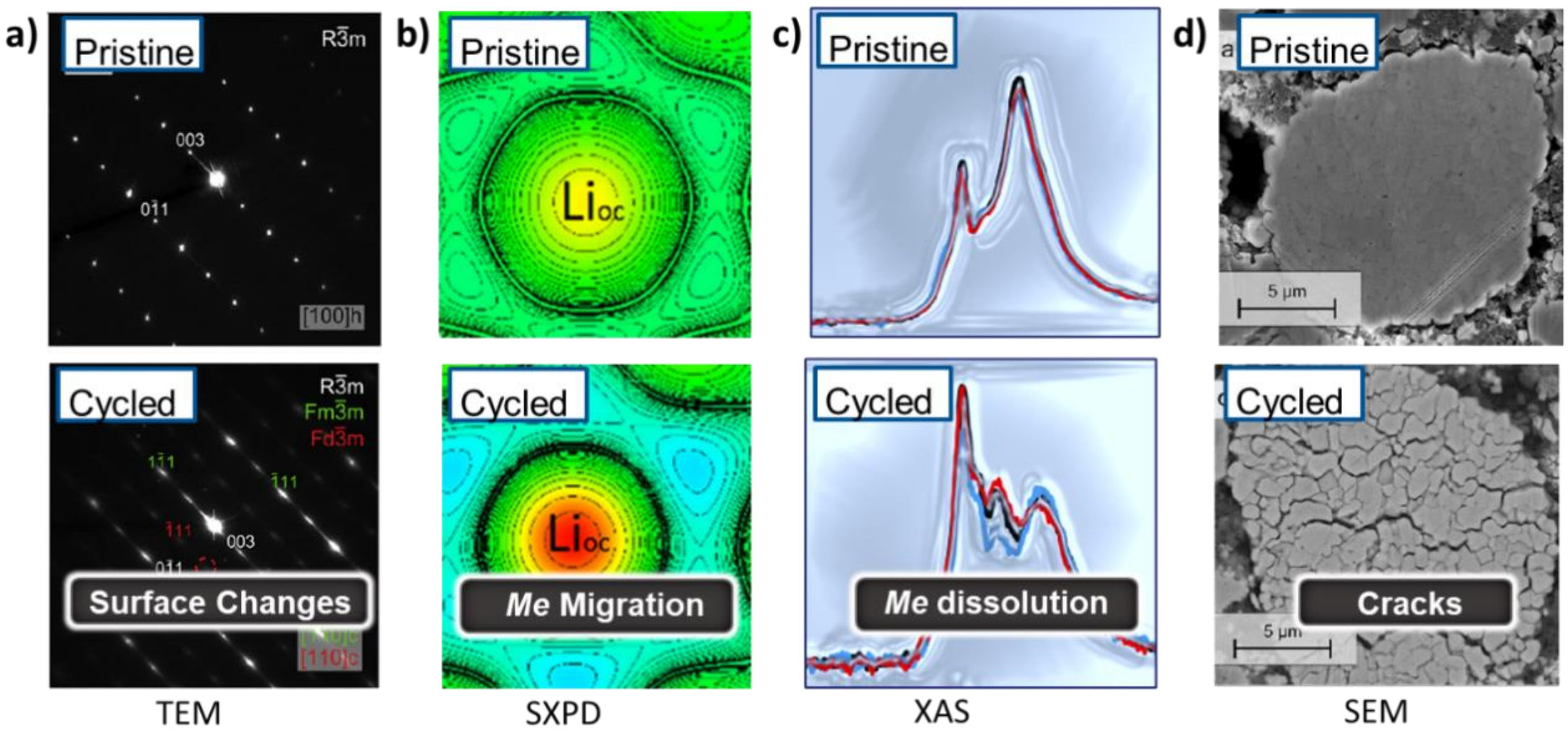

**Figure 18:** Cathode material degradation of layered oxides and structural related materials upon cycling. The plots highlight (a), surface morphology changes determined with transmission electron microscopy (TEM) [28] (b), transition metal migration measured with SXPD [148], (c), transition metal dissolution evident from core level spectroscopy (e.g. soft XAS/NEXAFS) [61], and (d), crack formation visible in scanning electron microscopy (SEM) images of cross sections of the active particles [28].

The lattice parameters start to collapse at the point, where the $Ni^{2+}$ content reaches zero (**Figure 19**).[15] The higher the Ni content in the layered oxides, the more capacity can be achieved after the $Ni^{2+}$ content decreases to zero, but the breakdown of the lattice is stronger.[5,15,17,36] Thus it can be concluded, that $Ni^{2+}$ is the redox active center in Ni-containing layered oxides at low and medium SOCs.

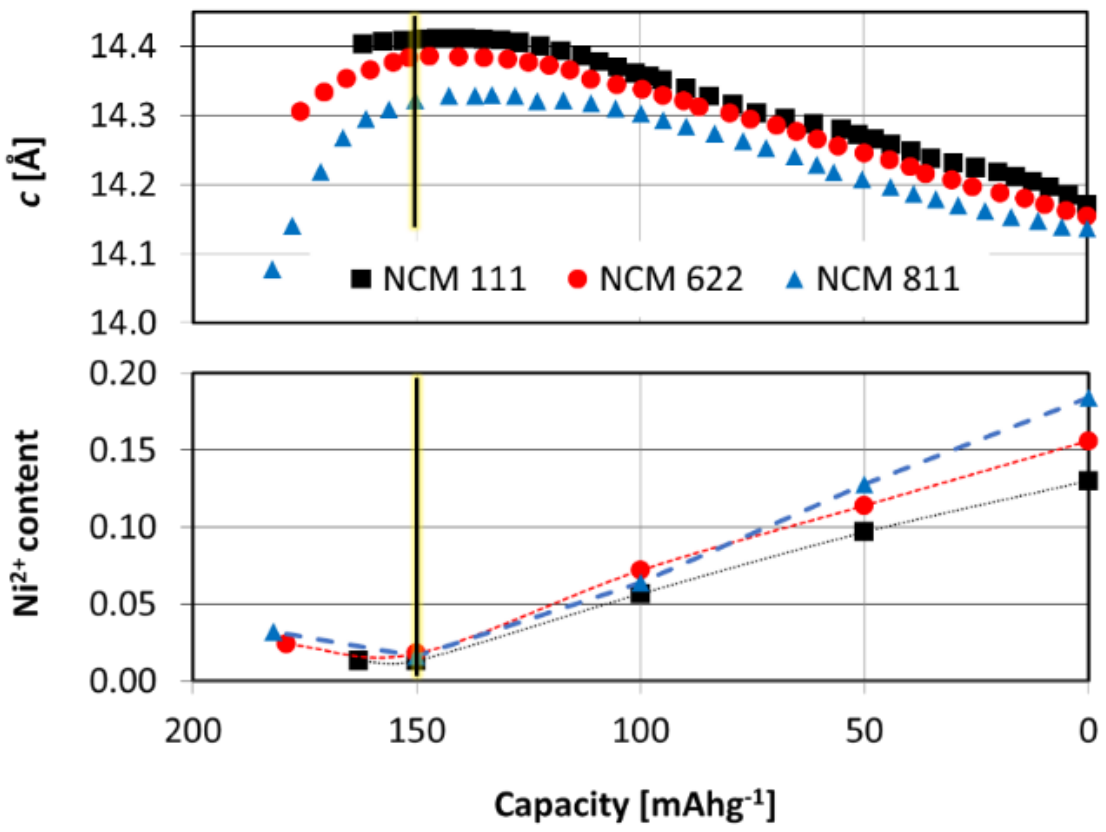

**Figure 19:** *C* lattice parameter in comparison to the Ni2+ content in NCM111, NCM622 and NCM811 upon the first discharge. The relative $Ni^{2+}$ content is determined from the CTM calculations depicted in **Figure 12**.

### 3.3.2. Voltage limitations

Increasing the amount of Ni in the rhombohedral structures mainly leads to an increase of covalent $Ni^{3+}$; the amount of redox active $Ni^{2+}$ hardly increases (**Figure 20**, left side). Due to the relatively little increase of the $Ni^{2+}$ content, the reversible capacity, herein defined at the point, where the lattice starts to collapse (**Figure 16a**), remains almost constant.[15] The increasing amount of covalent $Ni^{3+}$ with an increasing Ni content, however, shifts up the chemical potential of the antibonding *Me*-O hybrid states [15,174–178], *i.e.*, the redox active valence states in Ni-containing layered oxides (see **Paragraph 3.1** and **3.2**). The chemical potential of the cathodes relative to the chemical potential of the anodes, in turn, defines the open cell voltage of the battery (**Figure 6**).[15,174,178,179] Due to the disproportionately high increase of covalent $Ni^{3+}$ with an increasing Ni content, the covalent character of the materials increases, and along with this, the mean discharge voltage decreases. From NCM111 *via* NCM622 to NCM811, for example, the mean discharge voltage decreases from 3.85 V via 3.79 to 3,75 V, **Figure 20** (right side).[15]

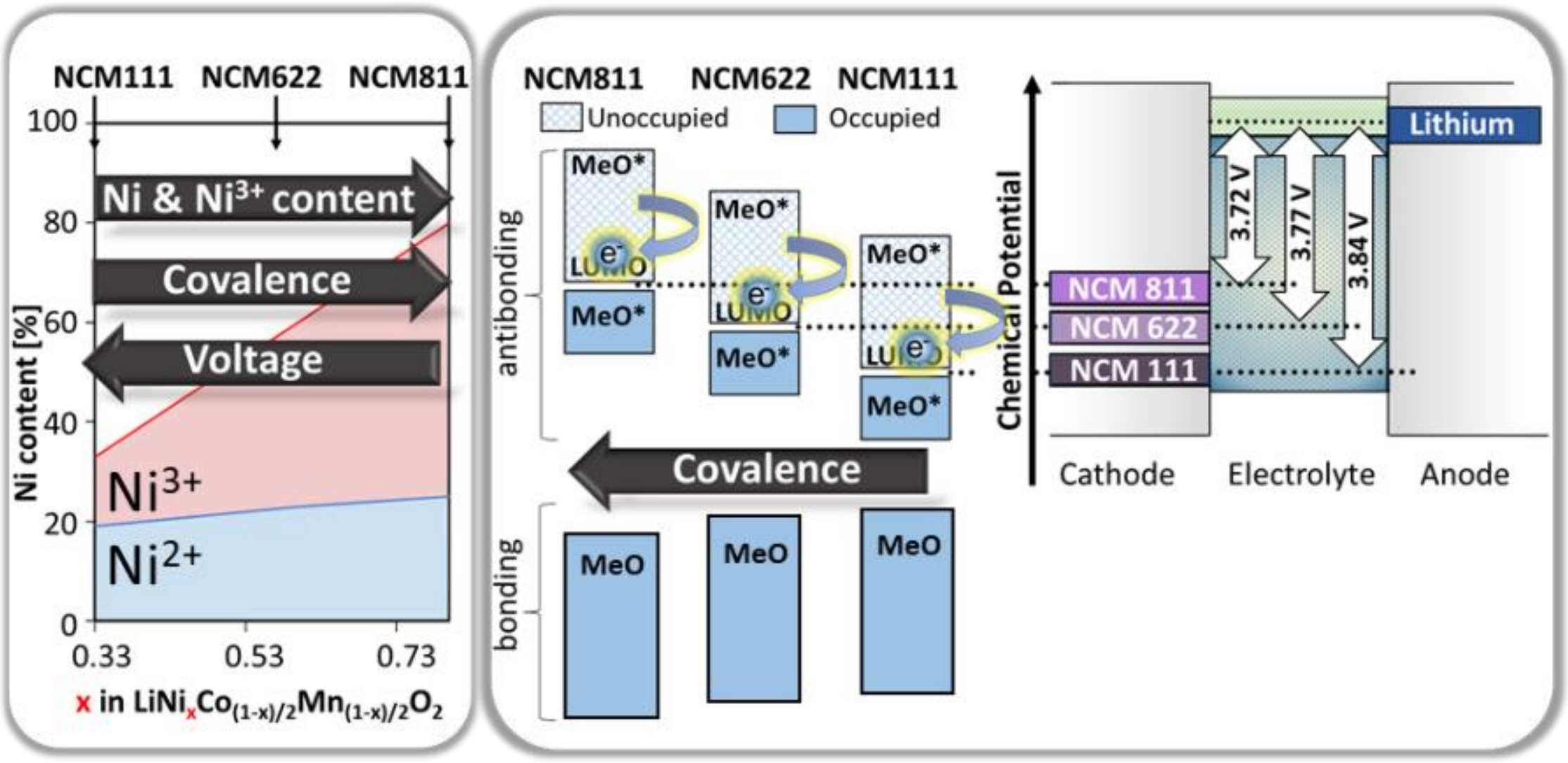

**Figure 20:** $Ni^{2+}$ and $Ni^{3+}$ content in NCMs depending on the overall Ni-content (left), a schematic depiction of the electronic structure (middle) and the corresponding cell voltages of the materials versus metallic lithium. The illustrations are reproduced from reference [15].

The complex relation between the electronic structure (the DOS), the $Ni^{2+}$ and $Ni^{3+}$ content, and the cell voltage can also be studied by a systematic substitution of Ni in $LiNiO_2$ with Co and/or Mn: The differential capacity (dq/dV) as a function of the voltage (**Figure 21**) is closely related to the DOS, because it shows the energetic position of the valence band of a cathode relative to that of the anode (see **Eq. *2***), the states from which electrons are taken out or into which electrons are filled in upon operation, **Figure 6**.[130,174,180] From the LNMOs ($LiNi_{1-x}Mn_xO_2$ with x=0.2, 0.4 and 0.5) *via* the NCMs ($LiNi_{1-x}Co_{x/2}\,Mn_{x/2}O_2$ with x=0.5, 0.4 and 0.2) to the LNCOs ($LiNi_{1-x}Co_xO_2$ with x=0.2, 0.4 and 0.5) the mean voltage, *i.e.* the position of the dq/dV peaks in **Figure 21**, shift to lower values. The electrochemical and analytical data demonstrates that Mn and Co crucially alter the electronic structure of layered oxides although the redox active configuration is, in all materials, ionic $Ni^{2+}$ (**Paragraph 3.1** and **3.2**). To understand the influence of Mn and Co on the DOS, their electronic configurations need to be discussed, first.

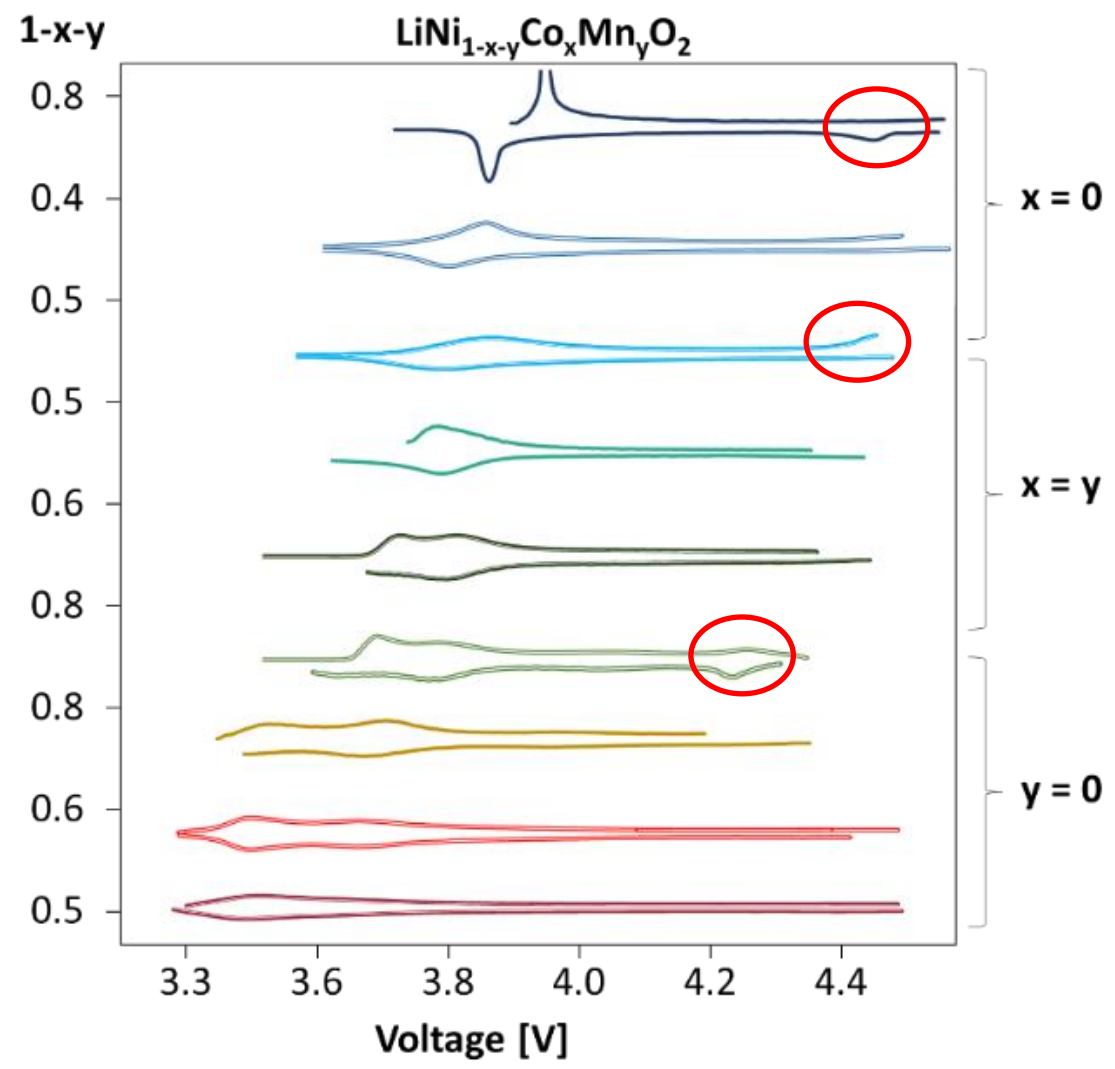

**Figure 21:** (a) Differential capacity of LNMOs ($LiNi_{1-x}Mn_xO_2$), NCMs ($LiNi_{1-x}Co_{x/2}$ $Mn_{x/2}O_2$), and LNCOs ($LiNi_{1-x}Co_xO_2$) with x = 0.2, 0.4 and 0.5. (b) shows a 2D map of the differential capacity upon discharge as a function of the Ni stoichiometry in $LiNi_{1-x}Me_xO_2$ (*Me* = Co, Mn). The differential capacity was determined using GITT pulses as described in [15] – the voltage was taken at the end of the rest step of the GITT pulses to ensure, that kinetic effects do not alter the findings.

### 3.3.2.1. Influence of Mn on the characteristic voltage profile

In layered oxides, Mn reveals a 4+ oxidation state as deduced from CTM calculations of the Mn $L_{2,3}$ edge, **Figure 22a-c**.[15,104,113,163,181–187] In case of such high valence transition metals the ligand to metal charge transfer is very pronounced leading to a population of the $e_g$ orbitals of > 15%.[163] Thus, Mn $3d^{3+1}$ $\underline{L}$ and higher order configurations such as Mn $3d^{3+2}$ $\underline{L}^2$ need to be included in the calculations (**Figure 22c**).[15] Moreover, Jahn Teller distortions of the $MnO_6$ octahedra are frequently reported which alter the valence states (the finestructure in the Mn $L_{2,3}$), in addition.[188–191] The high oxidation state of Mn reduces Ni from a formal 3+ to a 2+ configuration, *i.e.* from a very covalent to a very ionic species (see **Paragraph 3.2.2**). Consequently, the ionic character of the redox active valence states increases by admixing Mn and the dq/dV peaks are shifted to higher values, **Figure 21**.

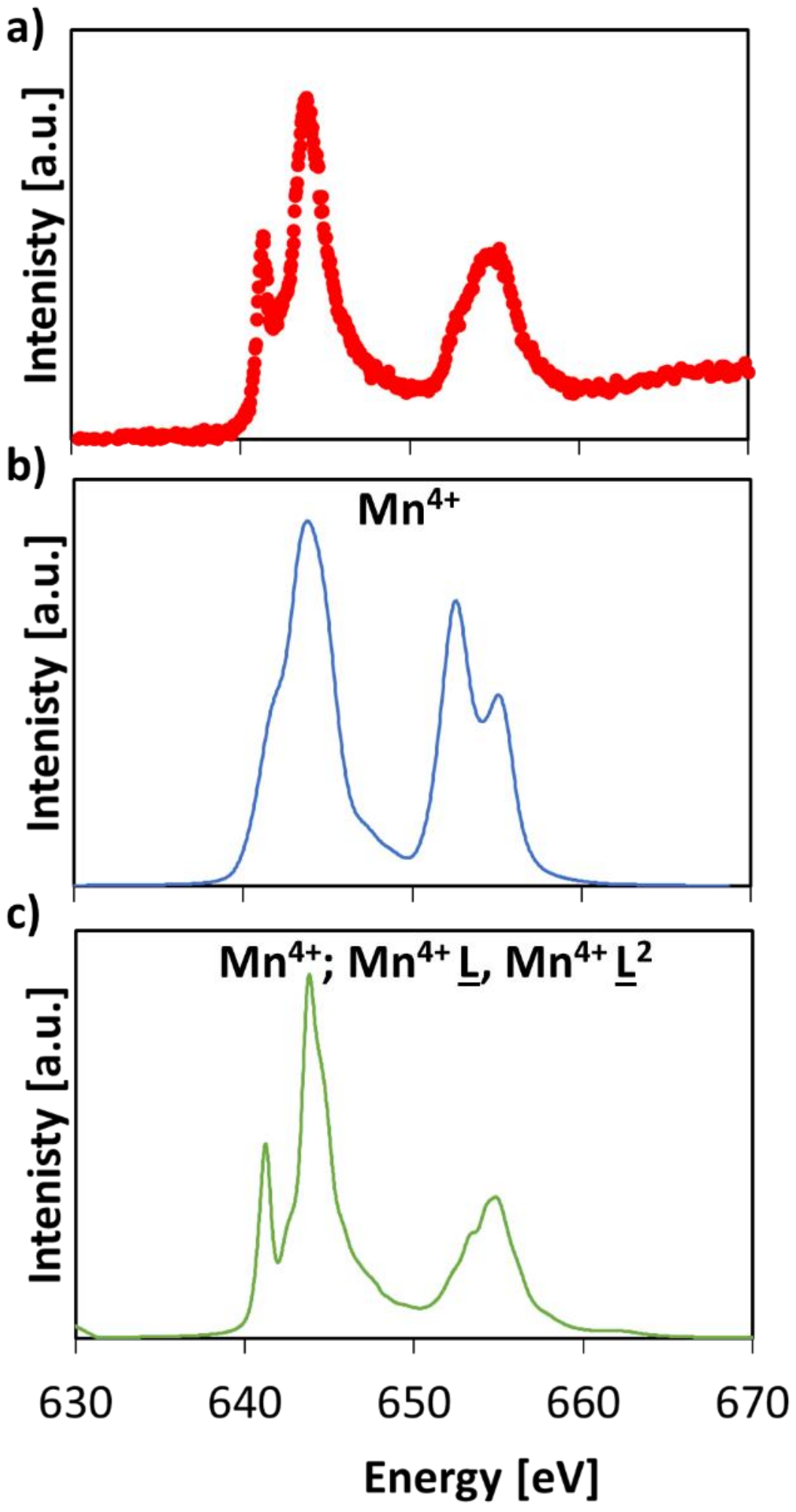

**Figure 22:** (a) Mn $L_{2,3}$ edge, taken from NCM111 as reported in reference [15], (b), CTM calculation of a Mn $3d^{3+}$ configuration and, (c), the configuration including the higher order Mn $3d^{3+1}$ $\underline{L}$ and Mn $3d^{3+2}$ $\underline{L}^2$ configurations.

#### 3.3.2.2. Influence of Co on the characteristic voltage profile

The NEXAFS spectra of the Co $L_{2,3}$ edges reveal a covalent 3+ configuration ($3d^{6.15}\,\underline{L}^{0.15}$), **Figure 23**.[15,104,192–195] Thus, the formal oxidation state of Ni in LNCOs ($LiNi_{1-x}Co_xO_2$) is +3, too. As discussed in **Paragraph 3.2.2**, this means that $Ni^{3+}$ is stabilized by a strong LMCT from the oxygen lattice towards Ni leading to a highly covalent Ni configuration. Consequently, admixing Co increases the covalent character of layered oxides and the dq/dV peaks appear at lower voltages, **Figure 21**.

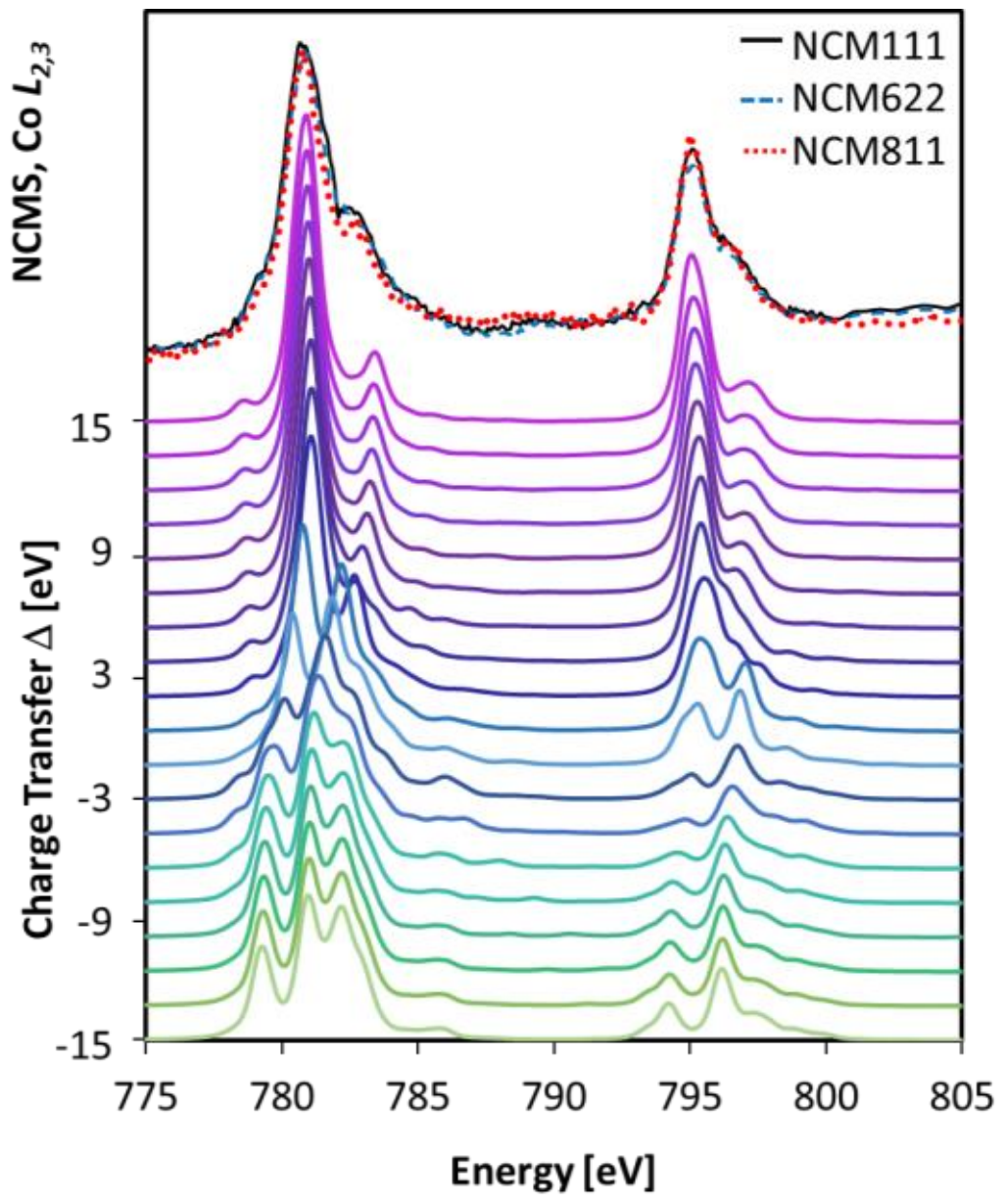

**Figure 23:** Co $L_{2,3}$ edge of NCM111, NCM622 and NCM811 (top) and calculated $d^6$ configurations varying the charge transfer energy Δ from -15 eV to 15 eV. For the simulations, $U_{dd}$ minus $U_{pd}$ was kept at 1 eV, and 10 Dq = 3 eV. The Slater integrals were reduced to 80% to account for configuration interactions.

#### 3.3.2.3. Influence of Ni on the characteristic voltage profile

Ni in layered oxides is a convolution of a covalent $Ni^{3+}$ and an ionic $Ni^{2+}$ configurations (see **Paragraphs 3.2.2**). The $Ni^{2+}/Ni^{3+}$ ratio can be estimated using core level spectroscopy such as XPS (**Figure 24** and **S2**). The spectra are simulated with eight gaussian functions using CTM calculations, a NiO reference with prominent $Ni^{2+}$ features, and constraints for the energetic positions and the full width at half maxima (FWHM) of the peaks (see **S2**). Thereby, the peaks A, C, E, and G are attributed to the ionic $Ni^{2+}$ configuration while the peaks B, D, F, and H are assigned to covalent $Ni^{3+}$ (**Figure 24**). The results are summarized in **Figure 25a**. With a decreasing Ni content in the LNMOs, the spectral weight in the Ni $L_{2,3}$ edge shifts towards lower energies, *i.e.*, towards more $Ni^{2+}$. Less Ni but a higher $Ni^{2+}/Ni^{3+}$ ratio means that the Ni content increases at the expense of $Ni^{3+}$ while the $Ni^{2+}$ content remains constant. The Ni $L_{2,3}$ XPS spectra of the LNCOs are independent of the Ni stoichiometry x in $LiNi_{1-x}Co_xO_2$ (bottom panel in **Figure 24**). The $Ni^{2+}/Ni^{3+}$ ratio remains constant which means that both the $Ni^{3+}$ and $Ni^{2+}$ content decrease with a decreasing x. The NCM XPS spectra reveal trends between the spectra of LNMO and LNCO.

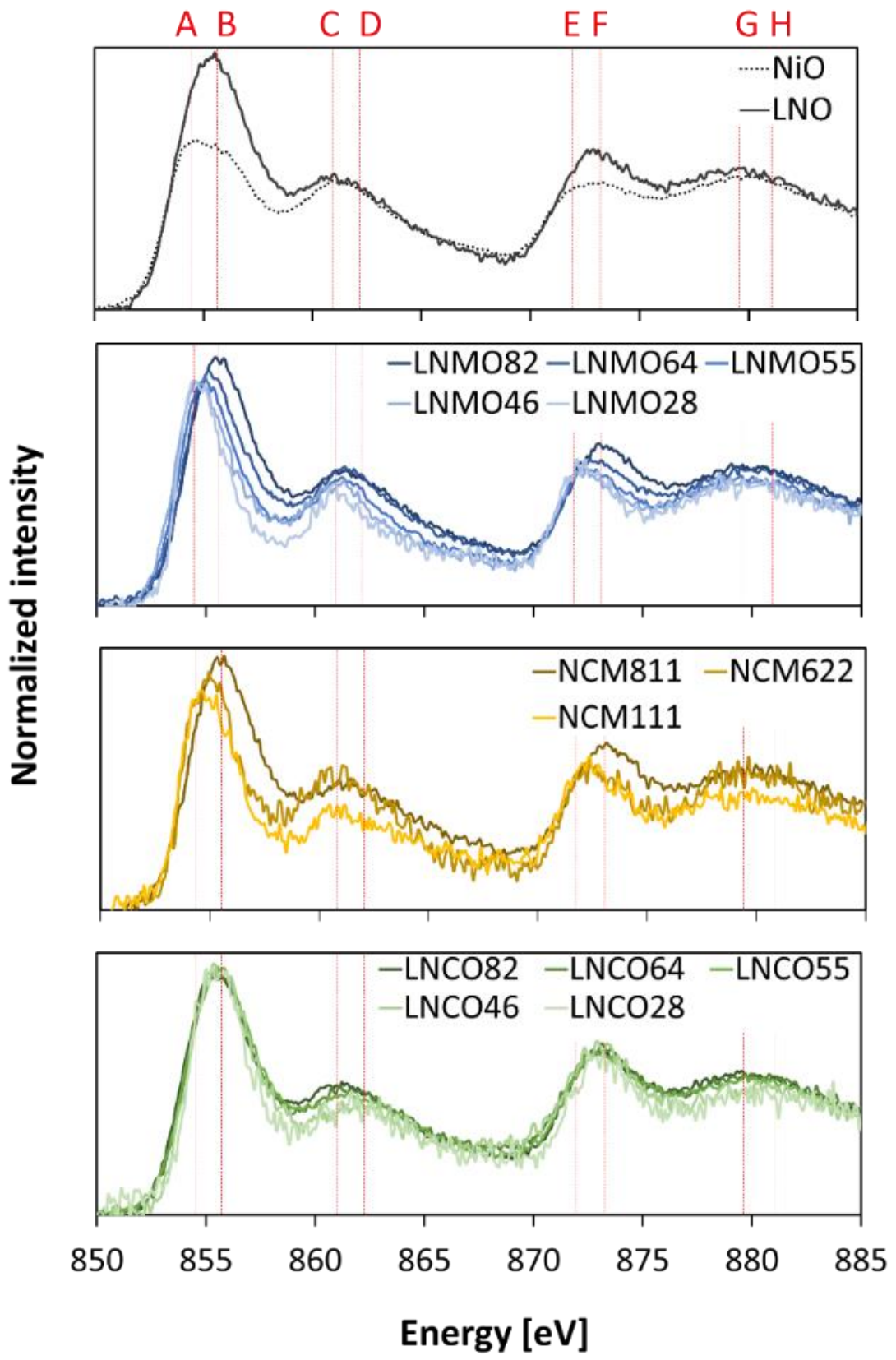

**Figure 24:** X-ray photoelectron spectroscopy (XPS) spectra of LNO ($LiNiO_2$) and NiO (1st panel), LNMOs (2nd panel, LNMO82: $LiNi_{0.8}Mn_{0.2}O_2$, LNMO64: $LiNi_{0.6}Mn_{0.4}O_2$, LNMO55: $LiNi_{0.5}Mn_{0.5}O_2$, LNMO46: $LiNi_{0.4}Mn_{0.6}O_2$, LNMO28: $LiNi_{0.2}Mn_{0.8}O_2$), NCMs (3rd panel, LNO: $LiNiO_2$, NCM811: $LiNi_{0.8}Co_{0.1}Mn_{0.1}O_2$, NCM622: $LiNi_{0.6}Co_{0.2}Mn_{0.2}O_2$, NCM111: $LiNi_{1/3}Co_{1/3}Mn_{1/3}O_2$), and LNCOs (4th panel, LNCO82: $LiNi_{0.8}Co_{0.2}O_2$, LNCO64: $LiNi_{0.6}Co_{0.4}O_2$, LNCO55: $LiNi_{0.5}Co_{0.5}O_2$, LNCO46: $LiNi_{0.4}Co_{0.6}O_2$, LNCO28: $LiNi_{0.2}Co_{0.8}O_2$), and. All spectra are energy calibrated, background corrected and normalized.

A maximum of ~ 33% $Ni^{2+}$ is reached for $LiNiO_2$ (LNO), **Figure 26a**. Note that this is incompatible with the charge neutrality condition: According to the formal oxidation states of the elements in LNO Ni should reveal a pure $Ni^{3+}$ configuration. The lack of σ-type hybridizations and the presence of $Ni^{2+}$, in turn, leads to the formation of π-states (**Figure 5b**). π-interactions are metal to ligand charge transfer (MLCT) interactions and thus lead to an oxidation state of oxygen < -2 (closer to zero). This enables a lower, mean oxidation state of the transition metals [86,92,114,196,197] and thus allows Ni to stay in a +2-oxidation state. With CTM calculations of NCM811 NEXAFS spectra, for example, it was estimated that ~ 25% ionic $Ni^{2+}$ ($d^8$ configuration,), ~ 55% covalent $Ni^{3+}$ ($d^7$ low spin configuration), ~ 10% covalent $Co^{3+}$ ($d^6$ low spin configuration) and ~ 10% covalent $Mn^{4+}$ ($d^3$ configuration) is present in the *Me*-O host structure.[15] This leads to a mean oxidation state of the transition metals *Me* of ~ +2.85.[15] The charge neutrality condition therefore requires a mean oxidation state for oxygen of ~ -1.9 due to MLCT interactions. The lower oxidation state of oxygen may also be (partially) caused by an oxygen deficiency. However, to explain an oxidation state of -1.9 instead of -2 requires 5% oxygen vacancies which is excluded by the refinement of the SXPD data (**Figure 1**).

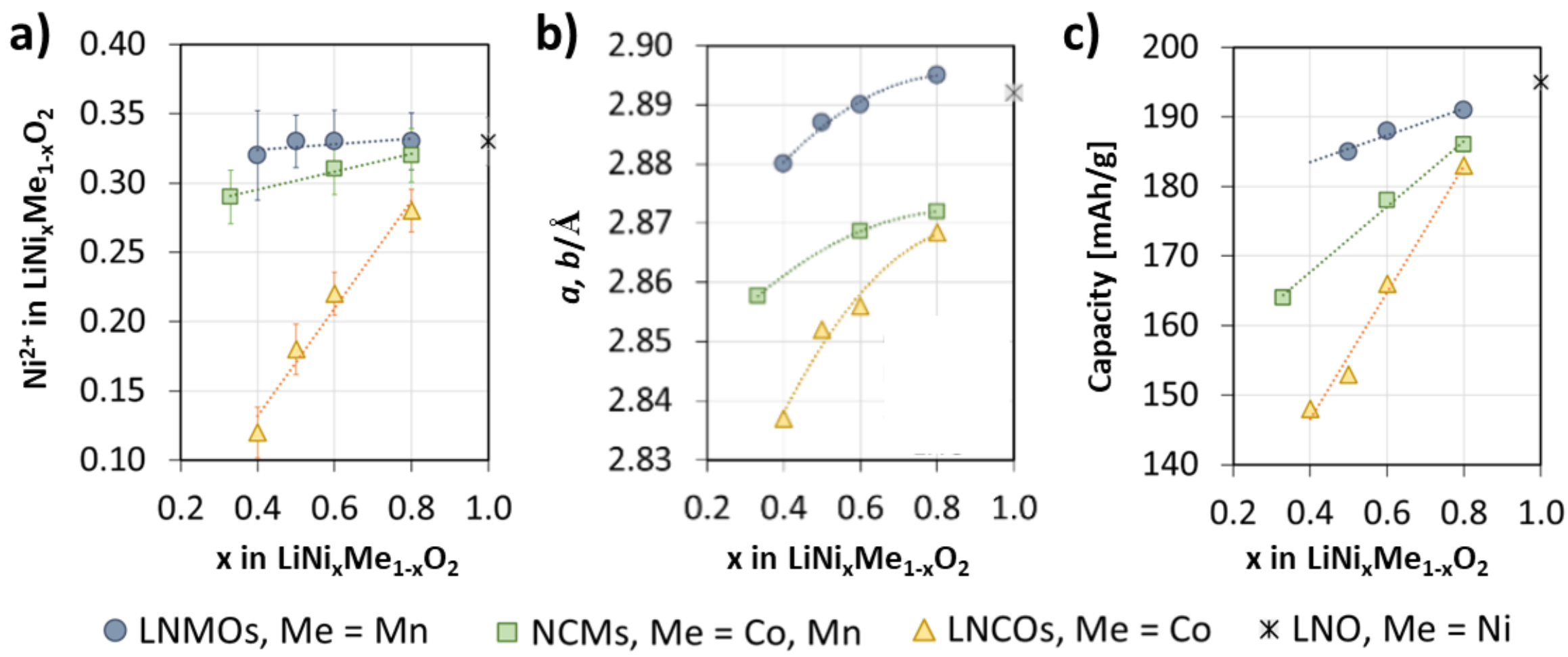

**Figure 25:** (a) $Ni^{2+}$-content determined with XPS, (b) *a,b* lattice parameter determined with powder diffraction (Cu Ka source) of the pristine layered oxides, and (c), 1st discharge capacity measured with GITT [15] of LNMOs ($LiNi_{1-x}Mn_xO_2$), NCMs ($LiNi_{1-x}Co_{x/2}Mn_{x/2}O_2$), and LNCOs ($LiNi_{1-x}Co_xO_2$),.

In this context it is proposed that $Ni^{2+}$ can also be treated like *Me* vacancies in LLO because $Ni^{2+}$ does hardly undergo σ-type interactions with O-neighbors (**Figure 5** and **Paragraph 3.3**).[15] These excess Li ions (*Me* vacancies) form a honeycomb structure in the transition metal layer as depicted in **Figure 26**. In such a superstructure a maximum of 1/3 of the *Me* (3a) sites can be occupied by $Ni^{2+}$ which is the amount found for LNO. In **Paragraph 3.3.3.4** advances in analytics are discussed to further proof/evaluate this hypothesis. In LNMOs, the $Ni^{2+}$ content remains close to 1/3 of the transition metals and is almost independent of the overall Ni content. This goes back to the presence of $Mn^{4+}$ which introduces a stoichiometric amount of $Ni^{2+}$. Co, in turn, is a +3 configuration which forces Ni into a +3-oxidation state, too. Thus, the amount of $Ni^{2+}$ is a function of the overall Ni content – it is roughly 1/3 of the overall Ni content and decreases significantly with a decreasing Ni stoichiometry, **Figure 25a** (*x* in $LiNi_xCo_{1-x}O_2$ = 0.8 → $Ni^{2+}$ stoichiometry = 0.8·1/3 = 0.26; *x=0.6* → $Ni^{2+}$ stoichiometry = 0.6·1/3 = 0.2; *x=0.33* → $Ni^{2+}$ stoichiometry = 0.33·1/3 = 0.11). The $Ni^{2+}$ content in NCMs lies between the LNMO and LNCO line showing that both, Co, and Mn have an influence on the Ni configuration.

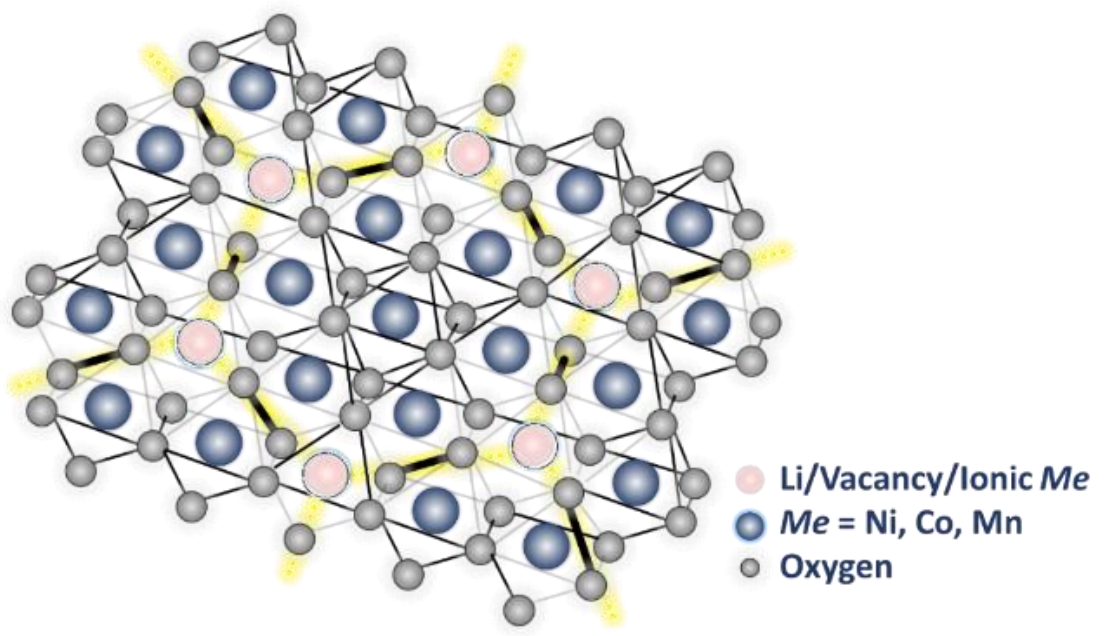

**Figure 26:** Honeycomb ordering in the transition metal layer of layered oxides if not all transition metals (3a sites) undergo covalent interactions with oxygen (6c sites) either due to an excess of Li ions (Li-rich materials), Li vacancies or very ionic *Me*.

The redox processes in LNMOs appear relatively concentrated at one voltage as evident from the sharp dq/dV peaks in **Figure 21** (first three curves). The $Ni^{2+}$ content remains close to the maximum in the samples and thus, LNMOs reveal the highest voltages. With the introduction of Co, the dq/dV peaks become broader, split up into two, and shift to lower voltages. In turn, the $Ni^{2+}$ content becomes a function of the overall Ni content, and the voltages significantly decrease with x in $LiNi_{1-x}$*Me*$_xO_2$ (*Me* = Co, Mn). $Ni^{2+}$ is the only ionic configuration in the rhombohedral host structures and thus, crucially affects the nature of the valence states. This also becomes obvious by comparing the $Ni^{2+}$ content with the *a,b* lattice parameters (**Figure 25a** and **b**), which are also a measure of the covalence of the *Me*-O bonds (**Paragraph 2.1**) [15,198,199]: While the LNMOs show with ~2.89 Å the highest lattice parameters independent of the overall Ni content, the *a*, *b* lattice parameters become a function of 1-x in $LiNi_{1-x}$*Me*$_xO_2$ for the NCMs (2.85 Å - 2.89 Å) and the LNCOs (2.83 Å - 2.88 Å). The shorter the bonds, the higher the degree of covalence because shared electrons reduce the distance between atoms. The trends in covalence follow the $Ni^{2+}$ content confirming a causal relationship. The peak splitting at low and medium S~~~OCs in the dq/dV plot of Co-containing layered oxides (**Figure 21**) is attributed to Li/Li vacancy ordering in and across the Li planes.[34,35,200–206]

### 3.3.3. Capacity limitations

#### 3.3.3.1. The capacity as a function of the $Ni^{2+}$ content

The capacity of Ni-containing layered oxides is determined by the amount of redox active sites, *i.e.* the amount of $Ni^{2+}$ in the rhombohedral *Me*-O host structures (**Paragraph 3**). Upon charge and discharge electron density is shifted from/to O towards/from $Ni^{2+}$ leading to the formation of covalent bonds (= covalent $Ni^{3+}$). A simple stoichiometric consideration, however, which would argue with a one electron exchange reaction per $Ni^{2+}$, is not possible because the number of electrons donated from the oxygen lattice upon charge cannot be quantified, so far. Nevertheless, the $Ni^{2+}$ content in $LiNi_{1-x}$*Me*$_xO_2$ (*Me* = Co, Mn) as a function of x follows the determined capacities of the materials (**Figure 25a** and **c**) confirming that $Ni^{2+}$ is the redox active center which limits the capacity. From this consideration it can be estimated that ~ 1.3 electrons can be extracted per $Ni^{2+}$. *E.g.*, in LNO (M = 96.6 g/mol) 33% of Ni is $Ni^{2+}$ and the material reveals a 1$^{st}$ discharge capacity of 198 mAh/g. 198 mAh/g, in turn, corresponds to 68856.48 C/mol and ~ 2 electrons per $Ni^{2+}$.

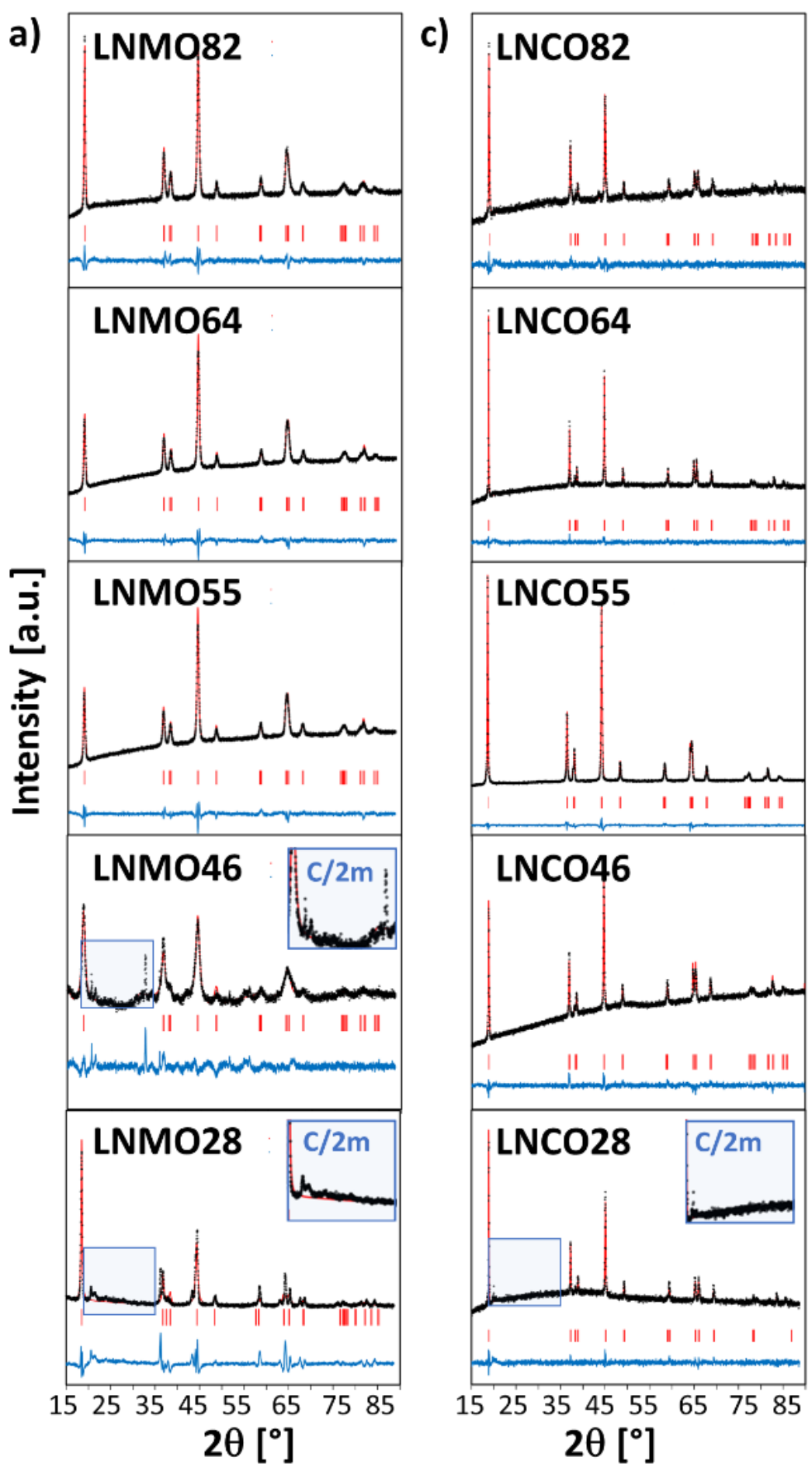

**Figure 27:** Powder diffraction patterns and the corresponding refinement with a $R\bar{3}m$ space group of (a), the LNMOs (LNMO82: $LiNi_{0.8}Mn_{0.2}O_2$, LNMO64: $LiNi_{0.6}Mn_{0.4}O_2$, LNMO55: $LiNi_{0.5}Mn_{0.5}O_2$, LNMO46: $LiNi_{0.4}Mn_{0.6}O_2$, LNMO28: $LiNi_{0.2}Mn_{0.8}O_2$), and (b), the LNCOs (LNCO82: $LiNi_{0.8}Co_{0.2}O_2$, LNCO64: $LiNi_{0.6}Co_{0.4}O_2$, LNCO55: $LiNi_{0.5}Co_{0.5}O_2$, LNCO46: $LiNi_{0.4}Co_{0.6}O_2$, LNCO28: $LiNi_{0.2}Co_{0.8}O_2$).

#### 3.3.3.2. Capacity limitations in Mn containing layered oxides

The amount of $Ni^{2+}$ in LNMOs (**Figure 25a**) remains close to 33% but does not increase further if the Mn content reaches values > 33%. Mn reveals a 4+ oxidations state and thus should enforce the formation of a stoichiometric amount of $Ni^{2+}$. Instead of higher $Ni^{2+}$ values, however, excess Li ions in the transition metal layer are found in LNMOs for x in $LiNi_{1-x}Mn_xO_2$ > 0.5, **Figure 27a**. The excess Li ions form a honeycomb structure in the *Me* layers (**Figure 26**) which lead to C/2m superstructure reflections marked by the insets in **Figure 27a**. The findings demonstrate that structural compromises such as the incorporation of excess Li ions finally lead to fulfilling the charge neutrality condition. 33% of $Ni^{2+}$ correspond to a maximum capacity of 193 mAh/g achieved for LNMO82 ($LiNi_{0.8}Mn_{0.2}O_2$, **Figure 25c**). This means, that ~ 2 electrons can be extracted per $Ni^{2+}$ upon charge.

#### 3.3.3.3. Capacity limitations in Co containing layered oxides

Co-containing layered oxides can be synthesized phase pure down to a Co-content of 60%, **Figure 27b**. Even in LNCO28 ($LiNi_{0.2}Co_{0.8}O_2$) the superstructure reflections of excess Li ions in the transition metal planes are hardly visible. This demonstrates that Co and Ni are much more decoupled compared to Ni and Mn, *i.e.* Co does not oxidize or reduce Ni. Consequently, the amount of $Ni^{2+}$ scales with 1-x in $LiNi_{1-x}Co_xO_2$ and decreases from 0.28 (x = 0.2) to 0.17 (x = 0.6), **Figure 25a**. At the same time, the capacity decreases from 184 mAh/g to 148 mAh/g, **Figure 25c**. This again means, that ~ 2 electrons can be extracted per $Ni^{2+}$. Note that up to this point the charge neutrality condition is not fulfilled if only the formal oxidation states of the elements in $LiNi_{1-x}Co_xO_2$ are considered. Here it must be assumed again that the presence of ionic $Ni^{2+}$ introduces MLCT interactions (**Paragraph 3.3.2.3**) leaving O with an oxidation state > -2 (closer to zero).

### 3.3.3.4. Advances in analytics

Analytical hints towards the honeycomb ordering of $Ni^{2+}$ ions in layered oxides (*e.g.*, $LiNiO_2$) can be obtained from high resolution powder diffractions data, **Figure 28**. The data was measured at beamline I11 using a multianalyzing crystal (MAC) detector offering outstanding signal/noise ratios and a very high energy resolution.[207] These measurements reveal C/2m superstructure reflections (inset in **Figure 28**) which are independent of the synthesis route and not visible in standard Cu Kα diffraction measurements: To exclude impurities as the origin different educts ($NiSO_4$ or $NiCO_3$; LiOH, $Li_2CO_3$ or $LiNO_3$) and synthesis methods (sol gel, all solid state, $OH^-$ precipitation) [199,208,209] have been used but the reflections, which cannot be assigned to common impurities such as NiO, $NiO_x$, $Li_2CO_3$, or $Li_xO_y$, remain.

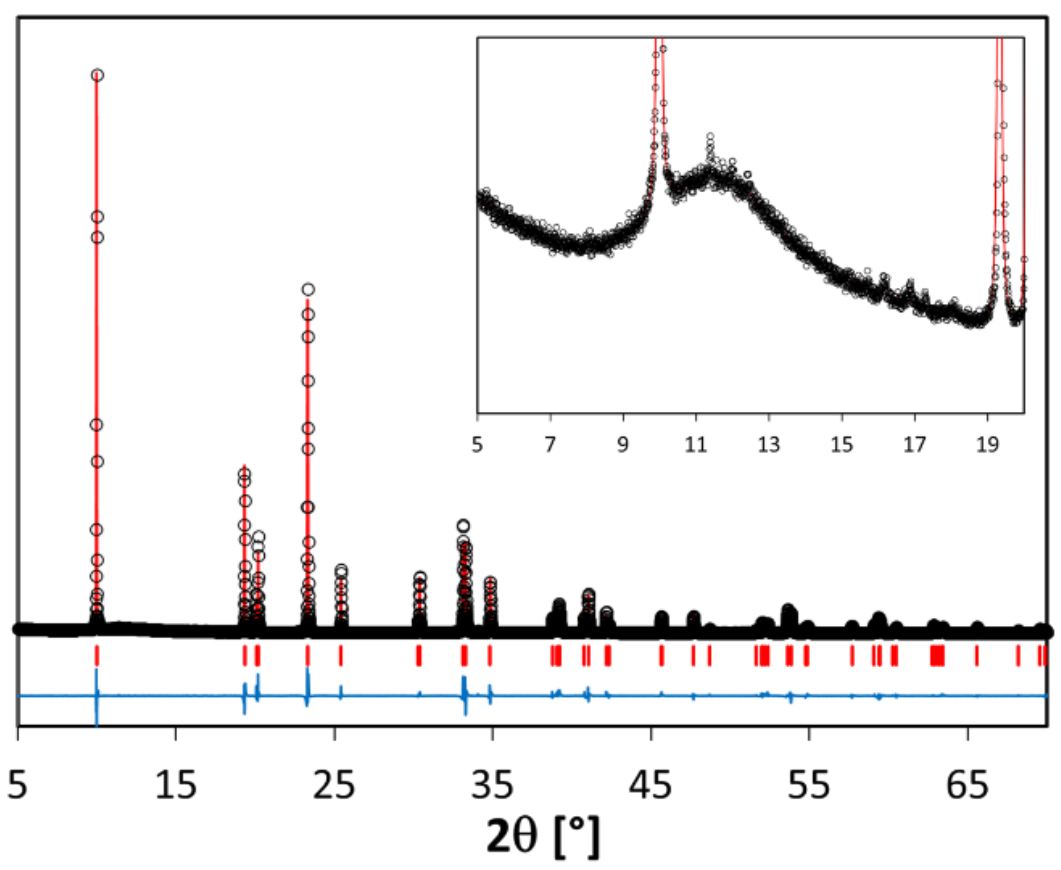

**Figure 28**: Powder diffraction data (λ = 0.825251) of $LiNiO_2$, refined the space group $R\bar{3}m$. The data was measured at beamline I11 (DLS, UK) using the high-resolution MAC detector.

Moreover, varying the temperature does also not lead to changes in the superstructure reflections as revealed with neutron diffraction at the WISH beamline (**Figure 29**) a high-resolution cold-neutron powder diffraction instrument to study magnetism in covalent systems. These findings exclude magnetic ordering as the origin of the additional reflections. Advanced inelastic scattering, x-ray fluorescence spectroscopy in grazing incidence, advanced Raman spectroscopy and other, advanced characterization methods are mandatory to shed more light on transition metal ordering.

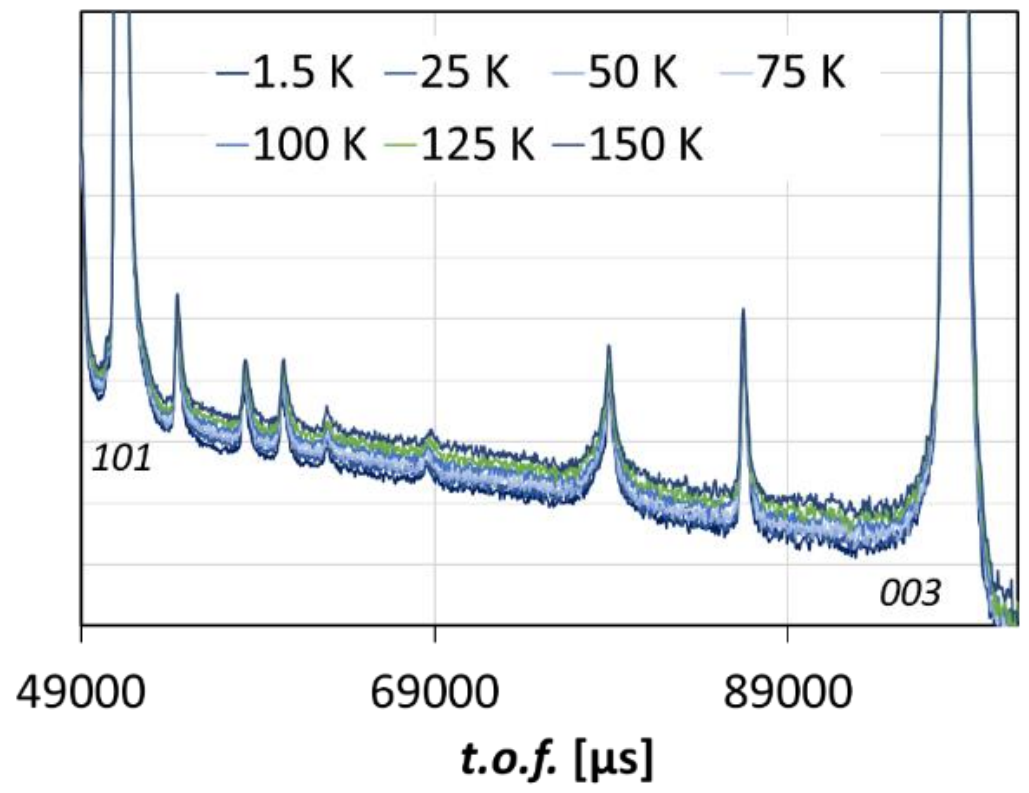

**Figure 29:** Neutron diffraction data measured at the WISH beamline (ISIS, UK) using a time-of-flight detector. The data was collected at 1.5 K, 25 K, 50 K, 75 K, 100 K, 125 K and 150 K.

# 4. Charge compensation at high states of charges

## 4.1. Anionic redox in Li-rich layered oxides

In Li-rich layered oxides (LLO, $Li_{1+x}Me_{1-x}O_2$ with *Me* = Ni, Co, Mn, the LLO described in the present work is $Li_{1.15}Ni_{0.2}Co_{0.1}Mn_{0.55}$), Li ions replace transition metals and form a honeycomb structure in the transition metal layer, **Figure 26**.[20,148,211–214] Upon charge, these Li ions are deintercalated and leave *Me* vacancies in the host structure.[213] Due to the missing *Me* 3d, 4s and 4p states only four σ-type interactions between *Me* and O are possible, **Figure 5**Error! Reference source not found.**b**.[92,215–217] The two remaining, orphaned O 2p states undergo weak π-type interactions (*Me* $t_{2g}$-O 2*p*) parallel to existing σ-bonds, **Figure 5**.[92,114] These π-states show a dominant O 2*p* character and become redox active at high SOCs.[57,61,92,114–116] This charge compensation occurs after $Ni^{2+}$ oxidation, leading to delocalized holes L̲ at O 2*p* sites through orbital hybridization.[15,29,35,39,106,112] A schematic depiction of charge compensation at SOCs > 75% is given in **Figure 30** which is reproduced from [92,215–217].

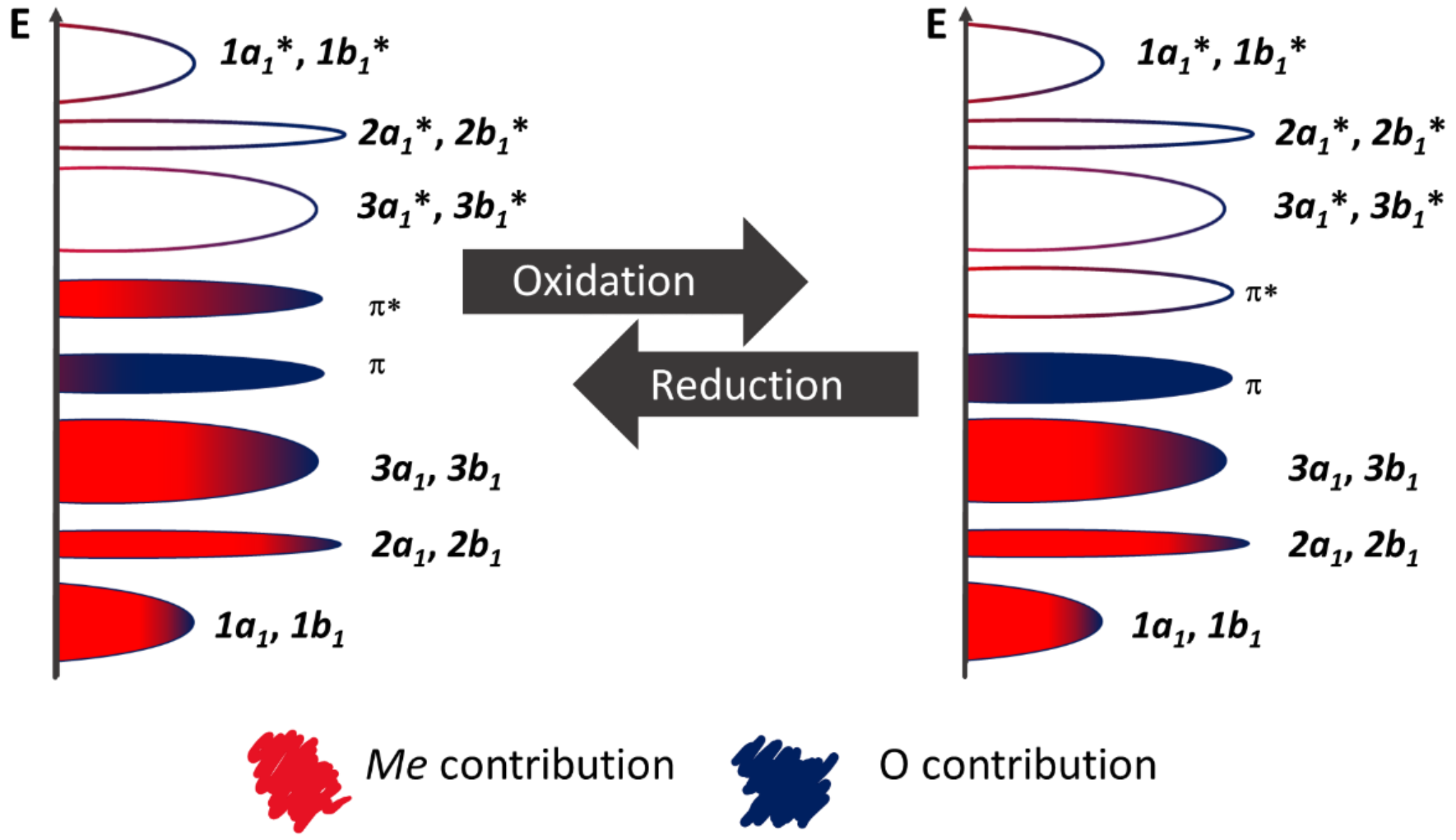

**Figure 30:** Depiction of the DOS of Li-rich layered oxides at ~ 75% SOC (left) and at 100% SOC (right), deduced from the MO scheme in **Figure 5b** and reproduced from [92].

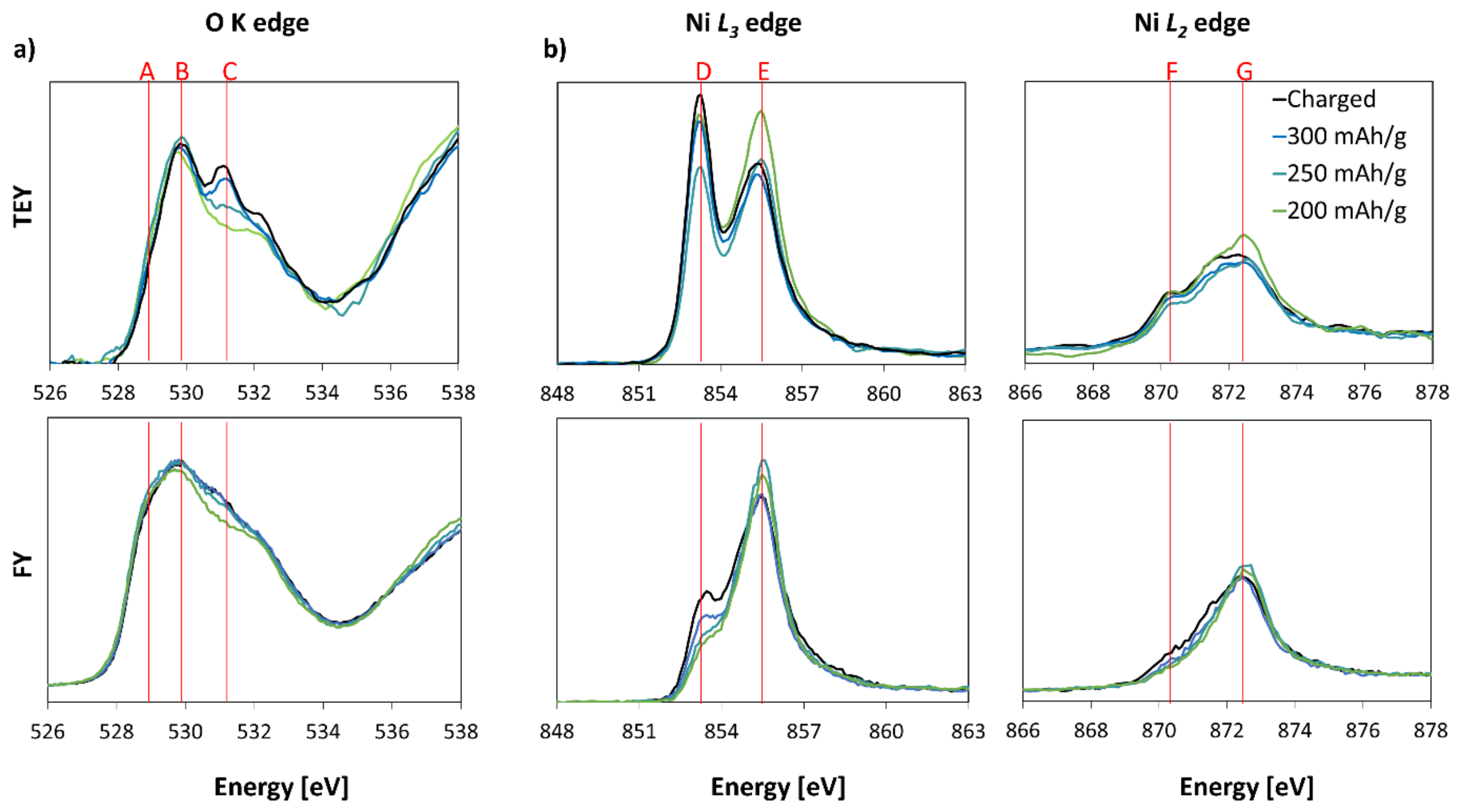

**Figure 31:** O K (a) and Ni $L_{2,3}$ (b) edge of LLO ($Li_{1.15}Ni_{0.2}Co_{0.1}Mn_{0.55}$), measured in total electron yield (TEY, top) and fluorescence yield mode (FY, bottom) at 200 mAh/g, 250 mAh/g, 300 mAh/g and in the charged state. The data was collected upon the 1st C/20 charge, *i.e.*, upon the so-called activation of layered oxides. The spectra are reproduced from reference [61].

According to [92,114], the bonding character of the π-states (*Me* $t_{2g}$ – O 2*p* interactions) increases while electrons are taken out of the antibonding π-states upon charge. This shifts the *Me* $t_{2g}$-O 2*p* peak at ~529 eV (peaks B in **Figure 31a** and **c**) to slightly higher voltages and an additional peak in the O K spectra at 532 eV is coming up, **Figure 31a** (peak C).[57,61,115,116,217] Peak C in the O K NEXAFS spectra (**Figure 31a**) is attributed to excitations into holes in the π* band, **Figure 30**.[53,57,59,61,92,218–220] Groups led by Jean-Marie Tarascon and Peter Bruce have shown that the oxidation continuous until the formation of molecular oxygen within the *Me*-O host structure.[57,59,217,221] Thereby, electron density is redirected from *Me*-O towards O-O bonds, **Figure 32**.[51,56,92,217,222,223] This is confirmed by changes in the Ni $L_{2,3}$ spectra which show increasing $Ni^{2+}$ peaks (ionic configuration, peak D and F, **Figure 31b**) and decreasing $Ni^{3+}$ peaks (peak E and G, **Figure 31b**). This so-called anionic redox described above is more pronounced on the surface (TEY spectra, top panels in **Figure 31**) than in the bulk (FY spectra, bottom panels in **Figure 31**).

Amongst others, resonant inelastic x-ray scattering (RIXS) proofs the presence of O=O bonds (= molecular oxygen) due to $O_2$ vibrations visible in the 0–2 eV region of the inelastic spectra.[58–60] Herein, a clear distinction between $O_2$, $O^{2-}$ and $O_2^{2-}$ dimers is possible because the O–O bond length (= strength) leads to unique vibrational frequencies.[58,59] NEXAFS and RIXS at the O K edge require ultra-high vacuum (UHV). Thus, the electrodes can only be measured under dry (*ex situ*) conditions which leads to the conclusion that molecular oxygen must be trapped in the layered structure.[59,60]

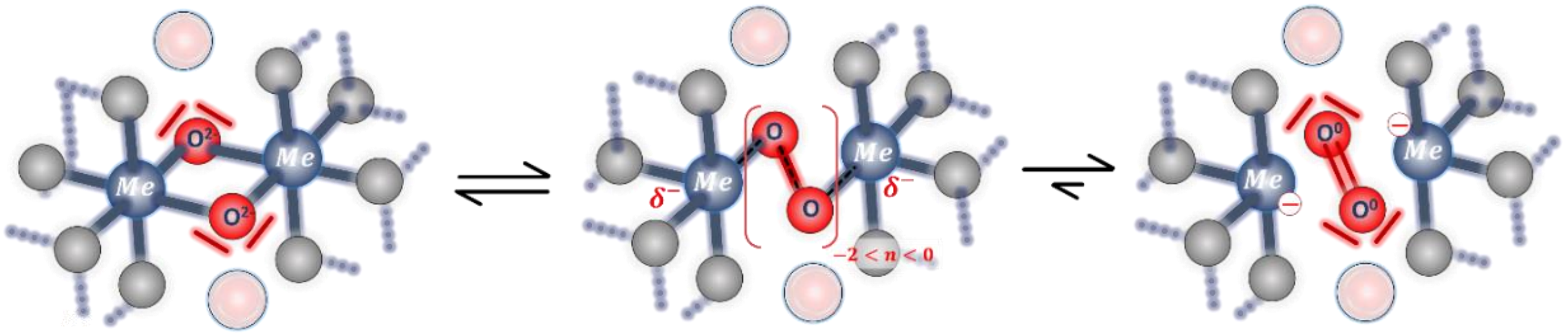

**Figure 32:** Schematic depiction of the oxidation of Li- or Na-rich layered oxides and the associated dislocation of non-bonding electrons as reported in [51,56,92,217,222,223].

DEMS and OEMS measurements upon the initial charge of LLO show that the formation of molecular oxygen in the structure is finally accompanied by gas evolution reactions, **Figure 33** (2nd panel).[17,61,115,138,224–227] Thereby, oxygen does not only escape from the surface near regions of the particles ($O_2$ formation), it might also react with electrolyte and form gases like CO and $CO_2$.[138]

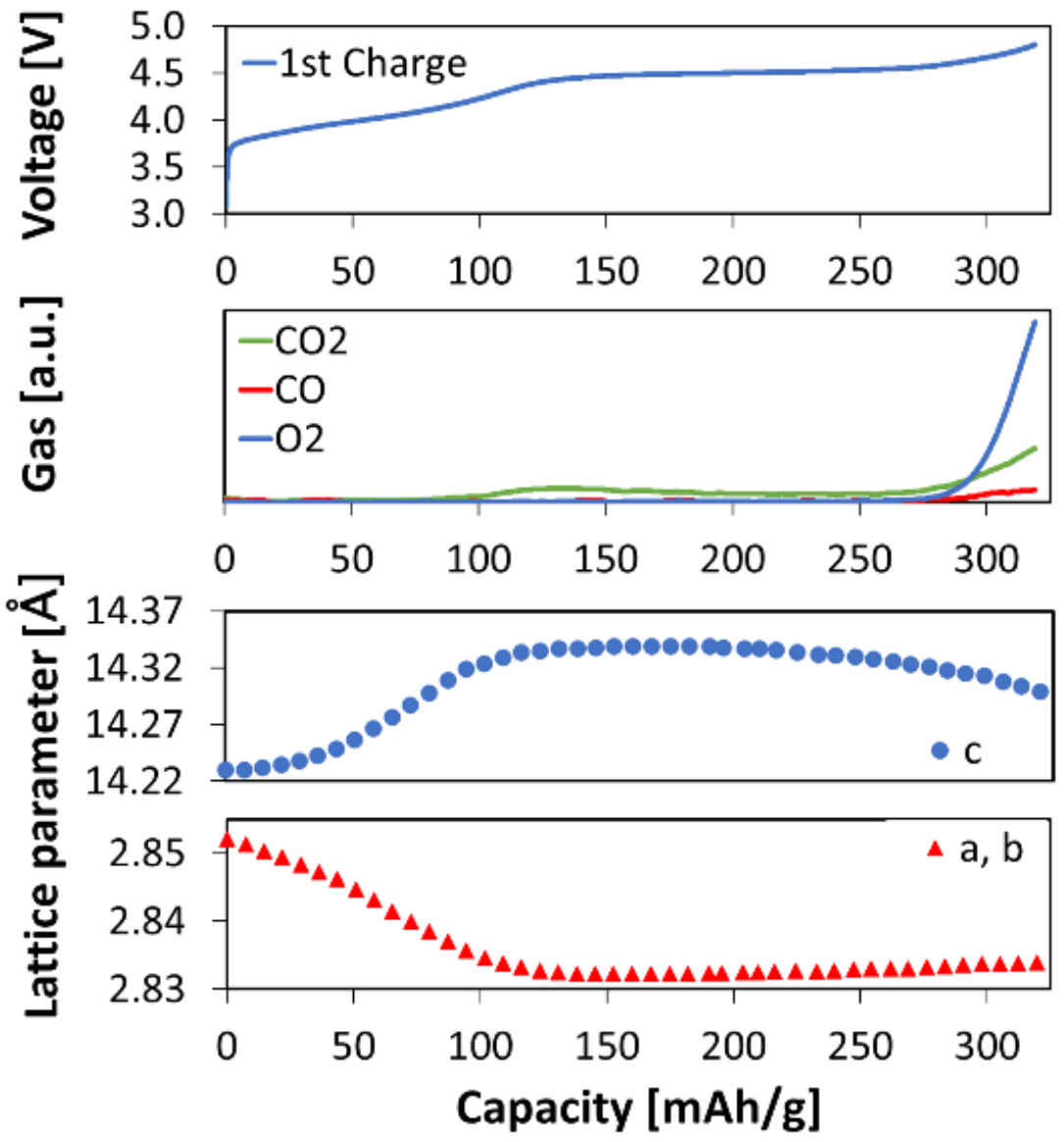

**Figure 33:** Characteristic voltage curve (1st panel), gas evolution (2nd panel), and changes in the lattice parameters (3rd and 4th panel) observed upon the 1st charge of LLO ($Li_{1.15}Ni_{0.2}Co_{0.1}Mn_{0.55}$). The data is reprinted from reference [61].

This also agrees with the reduction of covalent $Ni^{3+}$ to ionic $Ni^{2+}$ (**Figure 31b**) – Upon oxygen release the surface morphology of the layered structure transforms from a rhombohedral *via* a spinel into a rock salt structure, **Eq. 3**.[3,17,138,148,224]

$$\mathrm{Li}_x\mathrm{MeO}_2 \rightarrow \frac{x+1}{3}\ \mathrm{Li}_{3-\frac{3}{x+1}} Me_{\frac{3}{x+1}}\mathrm{O}_4 + \frac{1-2x}{3}\ \mathrm{O}_2 \quad \text{(spinel)}$$
$$\rightarrow (x+1)\ \mathrm{Li}_{1-\frac{1}{x+1}} Me_{\frac{1}{x+1}}\mathrm{O} + \frac{1-x}{2}\ \mathrm{O}_2 \quad \text{(rock salt)} \qquad \textbf{Eq. 3}$$

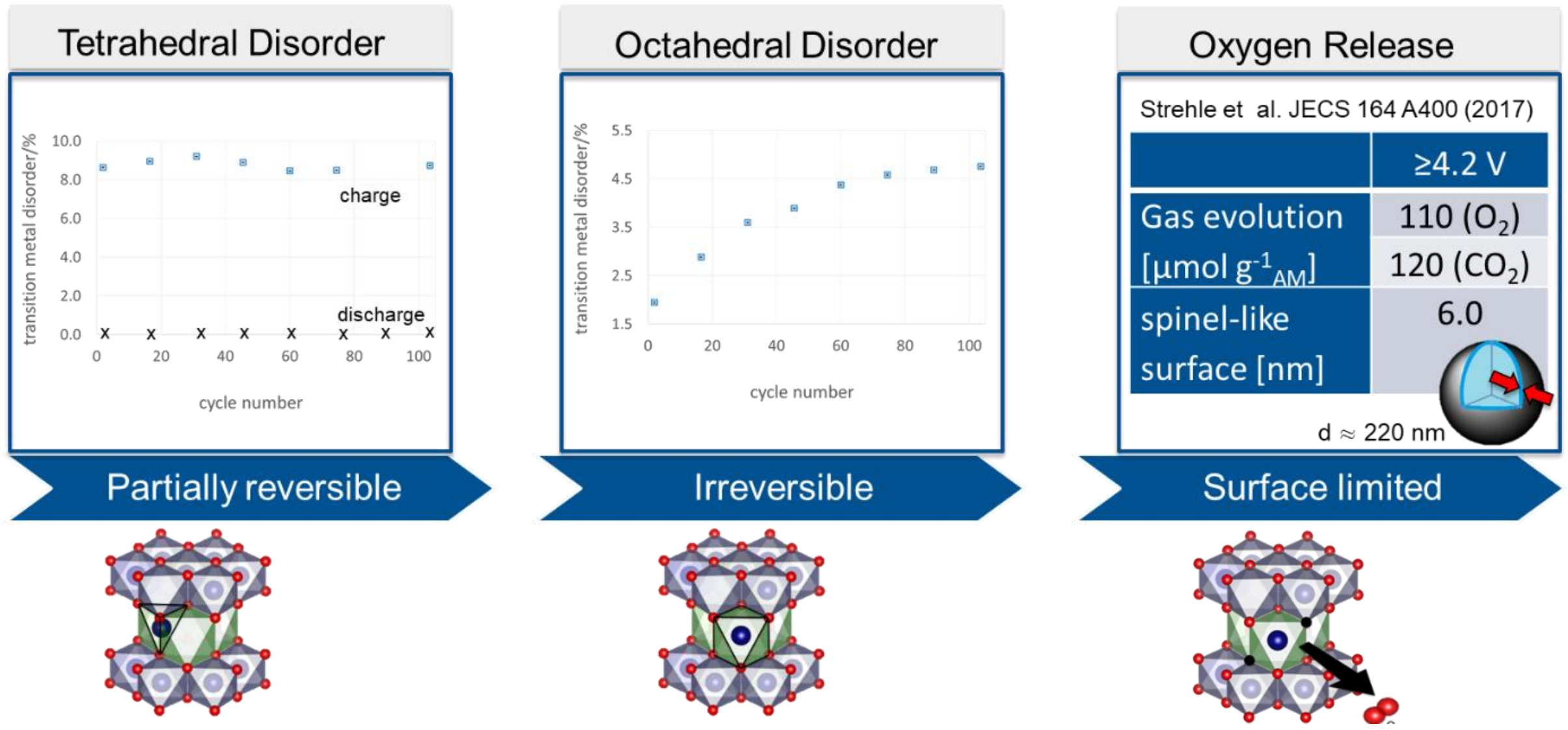

**Figure 34:** Occupation of tetrahedral (a) and octahedral sites (b) by transition metals upon > 100 cycles of LLO. c) gives the amount of gas molecules in µmol/g evolving upon the first charge of LLO and an estimate of the surface layer thickness which is affected by oxygen release. The data is reproduced from the references [138,148].

With the onset of gas evolution reactions, the lattice parameters collapse (**Figure 33**) because in the oxygen depleted rock salt structure the transition metals occupy randomly *Me* and Li sites.[148] At high SOCs, transition metals can migrate through interstitial tetrahedral sites into the Li layer, a partially irreversible process which finally leads to oxygen release from, **Figure 34**.

The anionic redox process upon the first charge (at very high SOCs) enables, beside oxygen release and surface morphology changes, the reduction of Mn and Co (maybe as well Ni) to 2+ species in the subsequent discharge [47,61] which delivers outstanding capacities. But this process is only partially reversible and leads to transition metal dissolution into the electrolyte.[23,24,61,228]

## 4.2. Anionic redox in Ni-rich layered oxides

As discussed for Li-rich layered oxides (**Paragraph 4.1**), an additional redox process is observed in Ni-rich NCMs (x in $LiNi_{1-x-y}Co_xMn_yO_2$ with 1-x-y ≥ 0.8) at high SOCs (≥ $150\ mAh/g$), **Figure 21**. *Operando* powder diffraction reveals a breakdown of the lattice parameters affiliated with the new redox process, **Figure 1** and **Figure 16**. Thereby, the *Me*-O distance increases while the O-O bond shortens significantly (**Figure 16**), showing a redirection of electron density from the *Me*-O towards the O-O bonds, **Figure 33Figure 14**. Moreover, gas evolution reactions are observed as revealed with online electrochemical and differential mass spectrometry (OEMS and DEMS).[15,17,19,21,22,34,119,135,228–230] The NEXAFS data of Ni-rich layered oxides (NCM811 served as reference material) at > 150 mAh/g confirms the change in the charge compensation mechanism, **Figure 35**. From 150 mAh/g to 175 mAh/g an additional, relatively sharp peak at 532 eV in the O K edge (peak C, **Figure 35a**) appears which cannot be assigned to a *Me*-O hybrid state (see reference spectra in **Figure 11**). With the appearance of peak C in the O K edges at 175 mAh/g the $Ni^{2+}$ peaks D and F (**Figure 35b**, **d**) run through a minimum and increases again at higher SOCs. The energetic position of peak C and its relative narrow peak shape agrees with the O=O dimer peak observed upon the so called "activation" or "anionic redox" in Li-, or more recently discussed, Na- rich layered oxides (see **Paragraph 4.1**).[51,53,55–57,61,92,219–221,223,231] Thus it is anticipated that molecular oxygen is formed in NCM811 at SOCs > 150 mAh/g, as well.

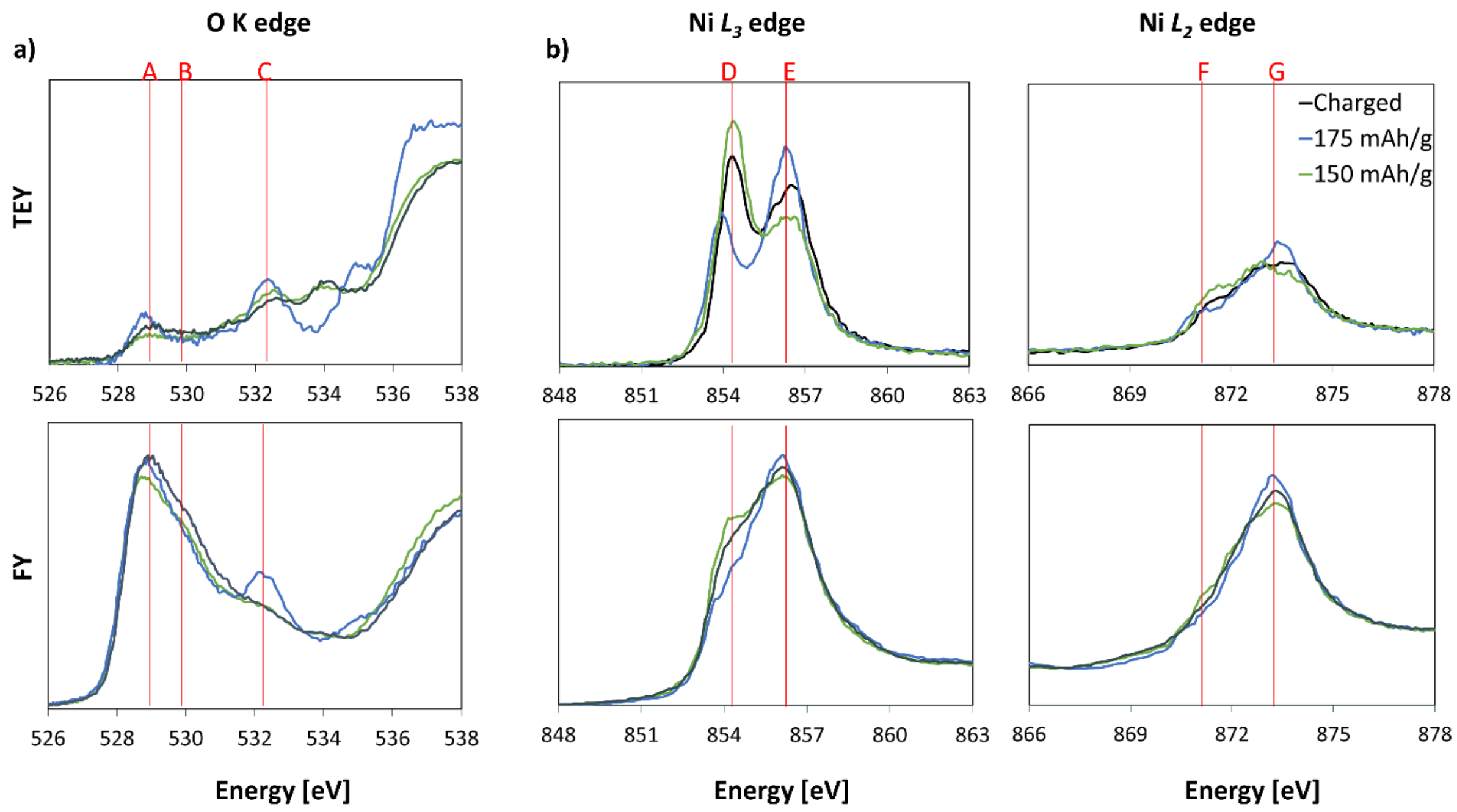

**Figure 35:** O K (a) and Ni $L_{2,3}$ (b) edge of NCM811, measured in total electron yield (TEY, top) and fluorescence yield mode (FY, bottom) at 150 mAh/g, 175 mAh/g and in the charged state. The data was collected upon the 3rd C/20 discharge of NCM811.

At low and medium SOCs $Ni^{2+}$ is completely oxidized to covalent $Ni^{3+}$, **Figure 12**.[15] The π-states are still completely occupied and thus there is no corresponding peak C in the O K spectra at 532 eV. Once all $Ni^{2+}$ is converted into covalent $Ni^{3+}$, electrons are extracted from the π-states and peak C in **Figure 35a** appears (see also **Figure 30**).[51,56,92,217,222,223] From **Figure 21** (red circles) it is evident, that mainly Ni-rich materials show an anionic redox. In these materials, the formal oxidation state derived from charge neutrality conditions does not equal the oxidation states derived from core level spectroscopy (NEXAFS, XPS) and CTM calculations.[15,34]

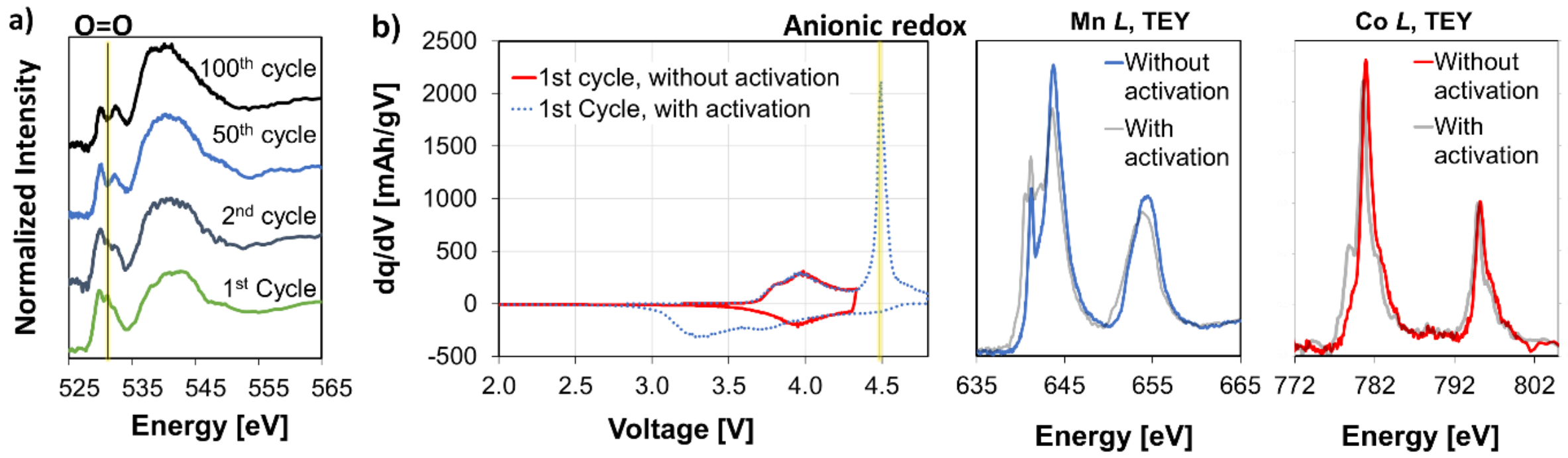

**Figure 37:** a) O K edge of LLO ($Li_{1.15}Ni_{0.2}Co_{0.1}Mn_{0.55}$), measured in total electron yield mode and in the charged state of LLO upon the 1st, the 2nd, the 50th, and the 100th cycle. b) shows the dq/dV plot of LLO, the Mn *L* as well as the Co *L* edge with and without activation (peak at 4.5 V). The Co *L* and Mn *L* edges were measured after the 1st cycle of LLO. The figures are reprinted from reference [61].

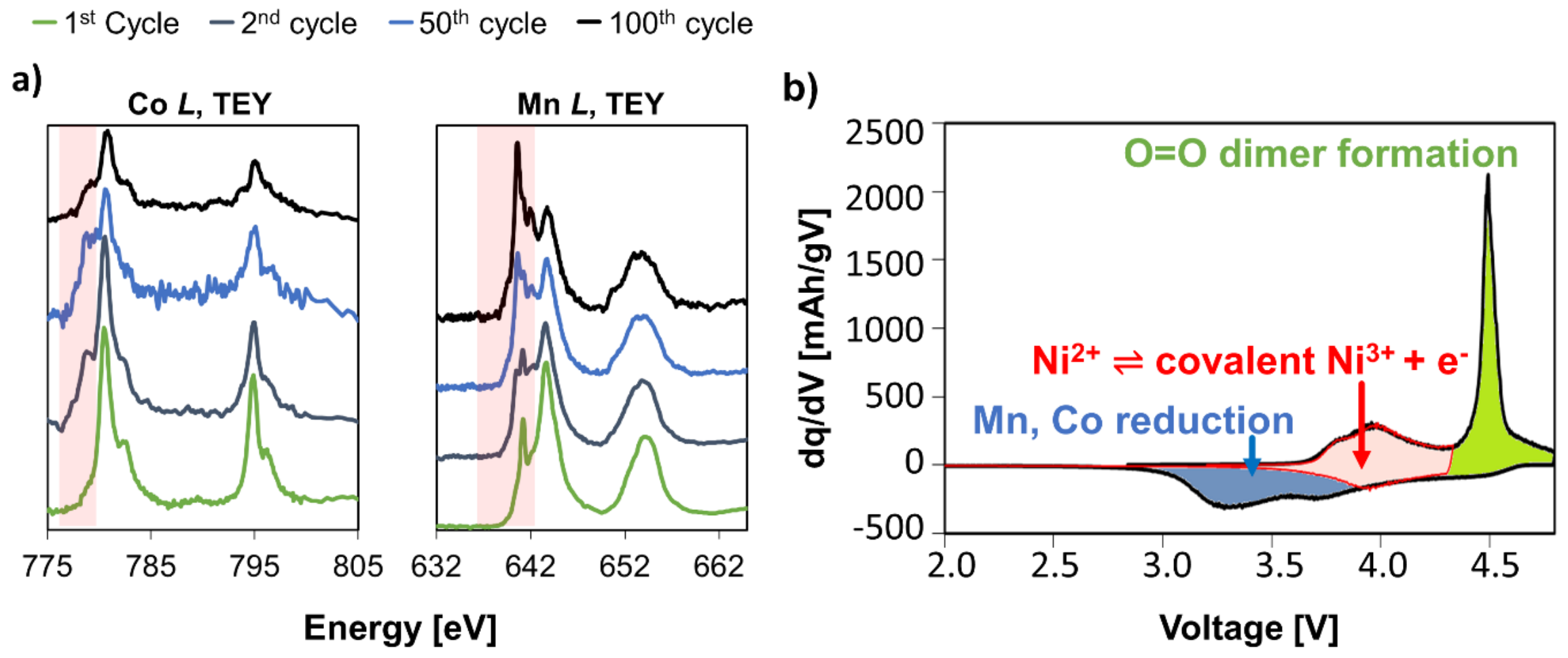

**Figure 36:** a) Co L and Mn L edge of LLO ($Li_{1.15}Ni_{0.2}Co_{0.1}Mn_{0.55}$), measured in the discharged state after the 1st, the 2nd, the 50th and the 100th cycle. b) Dq/dV plot of LLO in the first cycle with an assignment of the different redox processes observed in the NEXAFS spectra to individual peaks.

## 4.3. Irreversible nature of the anionic redox

The irreversible nature of the O=O dimer formation was discussed for Li-rich layered oxides in [61] based on NEXAFS data obtained after extensive cycling. Herein, it was shown that the O=O peak at 532 eV in the O K spectra does not occur again after just a few cycles (**Figure 37a**) while oxygen release and gas evolution reactions are accompanying the anionic redox (**Figure 33**). Nevertheless, the formation of O=O dimers within the *Me*-O host structure activates Mn and Co reduction upon discharge (**Figure 37b**), redox reactions which proceed or even intensify upon long term cycling. The steady increase of reduced Mn and Co upon long term cycling (**Figure 36a**) shows, that these additional redox processes are also partially irreversible in nature but they contribute almost half to the outstanding capacities of Li-rich layered oxides (**Figure 36b**, relative peak area). [61]

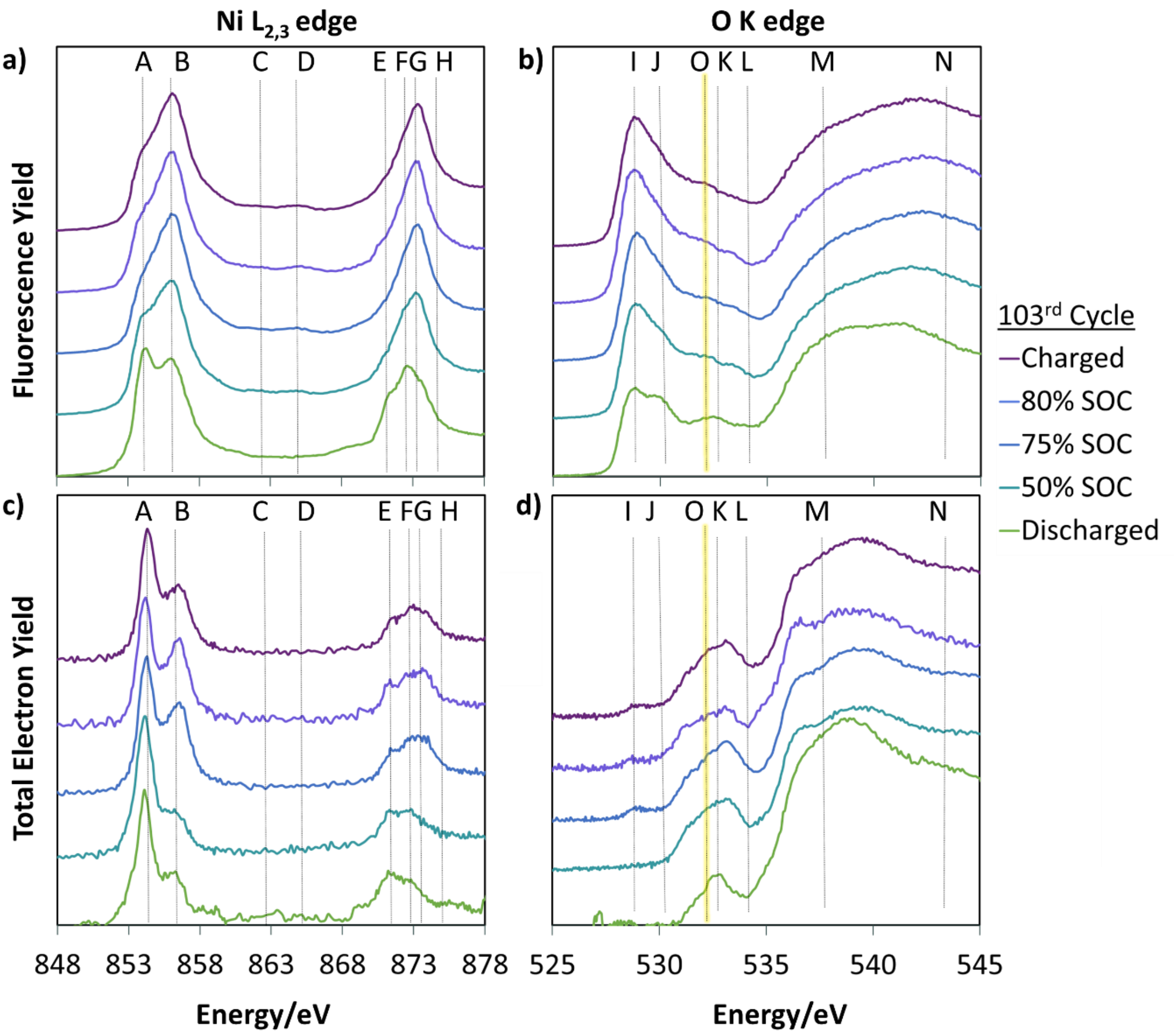

**Figure 38**: Ni $L_{2,3}$ (a, c) and O K (b, d) NEXAFS data of NCM811 upon the 103rd discharge, measured in FY (a, b) and TEY (c, d).

In the case of conventional layered oxides, the irreversibility of this process becomes evident from NEXAFS data, as well (**Figure 38**). Upon the 103rd cycle of NCM811, for example, the O=O dimer peak O at 532 eV is hardly visible any longer, **Figure 38b** and **d** (yellow line). Although this is the case in the more bulk like FY spectra and in the surface sensitive TEY spectra, the reduction of O near the grain boundaries finally leads to oxygen release [17,138,226,227], surface morphology changes (from a rhombohedral *via* a spinel-like to a NiO rock salt structure [3,17,138], **Eq. *3***), and transition metal dissolution into the electrolyte.[23,24,228] This is confirmed by the changes in the Ni $L_{2,3}$ edge, which do not reveal a reversion of the Ni redox process at any point upon the 103rd cycle: Peak A and E, marking ionic $Ni^{2+}$, continuously decrease, peak B and F, features of covalent $Ni^{3+}$, continuously increase from the discharged to the charged state of NCM811, **Figure 38a** and **c**. Moreover, the comparison of reveal spectra of pristine NCM811, discharged NCM811 after the 1st, the 3rd and the 103rd cycle reveals a change in the electronic structure (**Figure 39**). The Hubbard peaks K and L move closer to each other and resemble to one peak on the surface with an increasing number of cycles, **Figure 39a** and **b**. This single peak at ~ 533 eV is characteristic for NiO (dashed black lines, **Figure 39**).[34] Moreover, while the pristine material exhibit Ni-O hybrid states even on the surface (peaks I and J), these Ni-O-hybrid states decrease (**Figure 39a**) or even vanish near the grain boundaries after 100 cycles (**Figure 39b**). The more NiO-like character of the O K spectra with an increasing number of cycles is also confirmed by the increasing O K peaks at 538 eV and 540 eV, both prominent features in NiO (for reference see **Figure 39**). Upon long term cycling, the redox process at lower SOCs (< 75%) remains quantitatively the same, although at 75% SOC the degree of Ni-O hybridization is lower if compared to the 3rd cycle (**Figure 39**), *i.e.*, the peaks A and B in the O K spectra become lower in intensity. The redox process remains incomplete and therefore, less charges are (or less capacity is) exchanged.

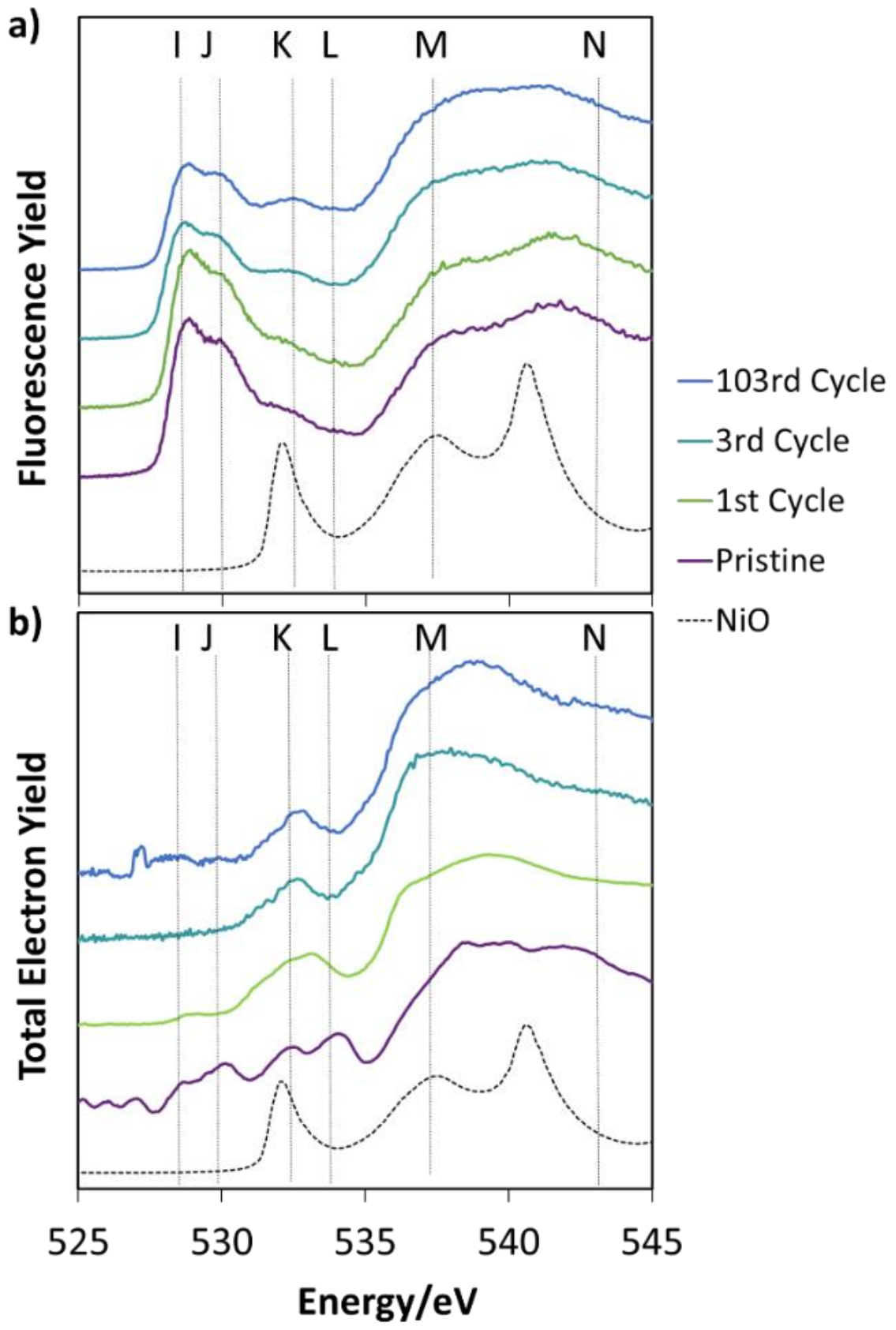

**Figure 39:** FY (a) and TEY (b) O K spectra of pristine NCM811, discharged NCM811 after the 1st, the 3rd and the 103rd cycle in comparison to NiO.

# 5. Summary and Conclusions

The coordination of O in layered oxides (space group $R\bar{3}m$) and structural related materials determines the underlying redox processes as active materials in electrochemical devices: In an ideal *Me*-O host structure, O is coordinated by three transition metals which form six *Me*-O hybrid states with the six O 2*p* orbitals. The lack of covalent interactions caused by *Me* vacancies [92,114] or very ionic *Me* configurations, however, leaves orphaned O 2*p* states which form weak π-type interactions with *Me* $t_{2g}$ states. Initially, the *Me*-O hybrid states (σ-type interactions between *Me* 3*d*, 4*s*, 4*p* and O 2*p* orbitals) form the highest occupied states and are emptied upon charge/filled upon discharge at low and medium SOCs (**Figure 5a**). Note that the formation/breaking of these hybrid states (= switch between a more "Hubburd type" to a more "charge transfer type" semiconductor, **Figure 3**) itself represents the charge compensation mechanism in layered oxides at low and medium SOCs.

In Ni-containing layered oxides, the redox configuration is ionic $Ni^{2+}$ which forms covalent Ni-O states upon oxidation (**Figure 14**). Consequently, the amount of $Ni^{2+}$ limits the capacity of the cathode materials which is, in turn, in maximum 1/3 of the *Me* 3a sites of the rhombohedral *Me*-O host structures, **Figure 25**. $Ni^{2+}$ is a very ionic configuration (> 95% Ni $3d^8$ O $2p^6$ ⇌ < 5% Ni $3d^9$ O $2p^5$)[15,34] which does hardly undergo covalent interactions with its oxygen neighbors, suggesting that it can be treated like excess Li ions in Li-rich layered oxides (**Figure 26**). In these materials, Li ions (or after Li deintercalation *Me* vacancies) replace 1/3 of the transition metals in the *Me* layers. In Mn containing layered oxides (*e.g.* $LiNi_xMn_{1-x}O_2$), the total amount of *Me* 3a sites multiplied by 1/3 yields the amount of $Ni^{2+}$ since Mn and Ni couple, *i.e.* while Mn reveals a 4+ oxidation state a stoichiometric amount of $Ni^{2+}$ is formed. At high Mn contents, however, Li-rich structures are formed instead of phase pure, rhombohedral layered oxides. In Co-containing layered oxides such as $LiNi_xCo_{1-x}O_2$, the amount of $Ni^{2+}$ scales with 1/3 of the overall amount of Ni (Co and Ni do not couple). Thus, the $Ni^{2+}$ content and the capacity decrease linearly with the Ni content.

At higher SOCs in Li- and Ni-rich layered oxides, O=O dimer formation, the so-called anionic redox, is observed. In this case excess Li ions, *Me* vacancies and/or ionic $Ni^{2+}$ leave orphaned O 2*p* orbitals which undergo weak π-type interactions with Me $t_{2g}$ states (**Figure 5b**). At the point where no ionic $Ni^{2+}$ states are left (no Ni-O hybrid states can be formed), these orbitals become redox active. Due to the dominant O character of the π-states this process is called "anionic redox", although it is affiliated with a strong Ni reduction, as well. Upon this redox reaction electron density is redirected from the Ni-O towards the O-O bonds as depicted in **Figure 31**. The process is highly irreversible, *e.g.*, it is affiliated with oxygen release, and it levels off after a few cycles. Moreover, the presence of an anionic redox process at high SOCs enables the reductions of Mn and Co at low SOCs, as deduced for LLO (**Figure 37**). The additional reduction/oxidation of these elements leads to higher capacities but also to transition metal dissolution, surface morphology changes and further side reactions on the anode surface.

# References

1. Lebens-Higgins, Z. W. *et al.* Revisiting the charge compensation mechanisms in LiNi 0.8 Co 0.2−y Al y O 2 systems. *Mater. Horizons* **6**, 2112–2123 (2019).
2. Xu, J., Lin, F., Doeff, M. M. & Tong, W. A review of Ni-based layered oxides for rechargeable Li-ion batteries. *J. Mater. Chem. A* **5**, 874–901 (2017).
3. Radin, M. D. *et al.* Narrowing the Gap between Theoretical and Practical Capacities in Li-Ion Layered Oxide Cathode Materials. *Adv. Energy Mater.* **7**, 1602888 (2017).
4. Myung, S.-T. *et al.* Nickel-Rich Layered Cathode Materials for Automotive Lithium-Ion Batteries: Achievements and Perspectives. *ACS Energy Lett.* **2**, 196–223 (2017).
5. De Biasi, L. *et al.* Between Scylla and Charybdis: Balancing among Structural Stability and Energy Density of Layered NCM Cathode Materials for Advanced Lithium-Ion Batteries. *J. Phys. Chem. C* **121**, 26163–26171 (2017).
6. Butt, A. *et al.* Recent Advances in Enhanced Performance of Ni-Rich Cathode Materials for Li-Ion Batteries: A Review. *Energy Technol.* **10**, 2100775 (2022).
7. Jiang, M., Danilov, D. L., Eichel, R. & Notten, P. H. L. A Review of Degradation Mechanisms and Recent Achievements for Ni-Rich Cathode-Based Li-Ion Batteries. *Adv. Energy Mater.* **11**, 2103005 (2021).
8. Li, W., Asl, H. Y., Xie, Q. & Manthiram, A. Collapse of LiNi 1– x – y Co x Mn y O 2 Lattice at Deep Charge Irrespective of Nickel Content in Lithium-Ion Batteries. *J. Am. Chem. Soc.* **141**, 5097–5101 (2019).
9. Kondrakov, A. O. *et al.* Charge-Transfer-Induced Lattice Collapse in Ni-Rich NCM Cathode Materials during Delithiation. *J. Phys. Chem. C* **121**, 24381–24388 (2017).
10. Schweidler, S. *et al.* Investigation into Mechanical Degradation and Fatigue of High-Ni NCM Cathode Material: A Long-Term Cycling Study of Full Cells. *ACS Appl. Energy Mater.* **2**, 7375–7384 (2019).
11. Kondrakov, A. O. *et al.* Anisotropic Lattice Strain and Mechanical Degradation of High- and Low-Nickel NCM Cathode Materials for Li-Ion Batteries. *J. Phys. Chem. C* **121**, 3286–3294 (2017).
12. Kleiner, K. *et al.* Fatigue of LiNi0.8Co0.15Al0.05O2 in commercial Li ion batteries. *J. Power Sources* **273**, 70–82 (2015).
13. K. Nikolowski. In situ Strukturuntersuchungen an Li(Ni , Co)O2 als Kathodenmaterial für Lithiumionenbatterien. (PhD dissertation, Technische Universität Darmstadt, Germany, 2007).
14. Märker, K., Reeves, P. J., Xu, C., Griffith, K. J. & Grey, C. P. Evolution of Structure and Lithium Dynamics in LiNi 0.8 Mn 0.1 Co 0.1 O 2 (NMC811) Cathodes during Electrochemical Cycling. *Chem. Mater.* **31**, 2545–2554 (2019).
15. Kleiner, K. *et al.* On the Origin of Reversible and Irreversible Reactions in LiNi x Co (1−x)/2 Mn (1−x)/2 O 2. *J. Electrochem. Soc.* **168**, 120533 (2021).
16. Schweidler, S., de Biasi, L., Hartmann, P., Brezesinski, T. & Janek, J. Kinetic Limitations in Cycled Nickel-Rich NCM Cathodes and Their Effect on the Phase Transformation Behavior. *ACS Appl. Energy Mater.* **3**, 2821–2827 (2020).
17. Jung, R., Metzger, M., Maglia, F., Stinner, C. & Gasteiger, H. A. Oxygen Release and Its Effect on the Cycling Stability of LiNi x Mn y Co z O 2 (NMC) Cathode Materials for Li-Ion Batteries. *J. Electrochem. Soc.* **164**, A1361–A1377 (2017).
18. Michalak, B. *et al.* Gas Evolution in LiNi0.5Mn1.5O4/Graphite Cells Studied In Operando by a Combination of Differential Electrochemical Mass Spectrometry, Neutron Imaging, and Pressure Measurements. *Anal. Chem.* **88**, 2877–2883 (2016).
19. Wandt, J., Jakes, P., Granwehr, J., Gasteiger, H. A. & Eichel, R.-A. Singlet Oxygen Formation during the Charging Process of an Aprotic Lithium-Oxygen Battery. *Angew. Chemie Int. Ed.* **55**, 6892–6895 (2016).
20. Duan, Y. *et al.* Insights into Li/Ni ordering and surface reconstruction during synthesis of Ni-rich layered oxides. *J. Mater. Chem. A* **7**, 513–519 (2019).
21. Streich, D. *et al.* Operando Monitoring of Early Ni-mediated Surface Reconstruction in Layered Lithiated Ni–Co–Mn Oxides. *J. Phys. Chem. C* **121**, 13481–13486 (2017).
22. Yoon, W. S. *et al.* In situ soft XAS study on nickel-based layered cathode material at elevated temperatures: a novel approach to study thermal stability. *Sci Rep* **4**, 6827 (2014).
23. Zhan, C., Wu, T., Lu, J. & Amine, K. Dissolution, migration, and deposition of transition metal ions in Li-ion batteries exemplified by Mn-based cathodes – a critical review. *Energy Environ. Sci.* **11**, 243–257 (2018).
24. Evertz, M., Kasnatscheew, J., Winter, M. & Nowak, S. Investigation of various layered lithium ion battery cathode materials by plasma- and X-ray-based element analytical techniques. *Anal. Bioanal. Chem.* **411**, 277–285 (2019).
25. Liu, H. *et al.* Intergranular Cracking as a Major Cause of Long-Term Capacity Fading of Layered Cathodes. *Nano Lett.* **17**, 3452–3457 (2017).
26. Yoon, T. *et al.* Failure mechanisms of LiNi 0.5Mn 1.5O 4 electrode at elevated temperature. *J. Power Sources* **215**, 312–316 (2012).
27. Becker, D. *et al.* Surface Modification of Ni-Rich LiNi0.8Co0.1Mn0.1O2 Cathode Material by Tungsten Oxide Coating for Improved Electrochemical Performance in Lithium-Ion Batteries. *ACS Appl. Mater. Interfaces* **11**, 18404–18414 (2019).
28. Lang, M. *et al.* Post mortem analysis of fatigue mechanisms in LiNi0.8Co0.15Al0.05O2 – LiNi0.5Co0.2Mn0.3O2 – LiMn2O4/graphite lithium ion batteries. *J. Power Sources* **326**, 397–409 (2016).
29. Koyama, Y., Mizoguchi, T., Ikeno, H. & Tanaka, I. Electronic Structure of Lithium Nickel Oxides by Electron Energy Loss Spectroscopy. *J. Phys. Chem. B* **109**, 10749–10755 (2005).
30. Koyama, Y., Tanaka, I., Adachi, H., Makimura, Y. & Ohzuku, T. Crystal and electronic structures of superstructural Li1-x[Co1/3Ni1/3Mn1/3]O2 (0 ≤ x ≤ 1). *J. Power Sources* **119**–**121**, 644–648 (2003).
31. Tarascon, J. M. *et al.* In Situ Structural and Electrochemical Study of Ni1−xCoxO2 Metastable Oxides Prepared by Soft Chemistry. *J. Solid State Chem.* **147**, 410–420 (1999).
32. Yoon, W. S. *et al.* Investigation of the charge compensation mechanism on the electrochemically Li-ion deintercalated Li1-xCo1/3Ni1/3Mn 1/3O2 electrode system by combination of soft and hard X-ray absorption spectroscopy. *J. Am. Chem. Soc.* **127**, 17479–17487 (2005).
33. Kleiner, K. & Ying, B. Charge Compensation in Layered Oxide 3D Band Engineering - Path to Next Gen. Cathode Materials? in *IMLB 2022* 1 (IMLB 2022, 2022).
34. Kleiner, K. *et al.* Unraveling the Degradation Process of LiNi0.8Co0.15Al 0.05O2 Electrodes in Commercial Lithium Ion Batteries by Electronic Structure Investigations. *ACS Appl. Mater. Interfaces* **7**, 19589–19600 (2015).
35. Montoro, L. A. & Rosolen, J. M. The role of structural and electronic alterations on the lithium diffusion in LixCo0.5Ni0.5O2. *Electrochim. Acta* **49**, 3243–3249 (2004).
36. de Biasi, L. *et al.* Chemical, Structural, and Electronic Aspects of Formation and Degradation Behavior on Different Length Scales of Ni-Rich NCM and Li-Rich HE-NCM Cathode Materials in Li-Ion Batteries. *Adv.*

*Mater.* **31**, 1900985 (2019).
37. Yoon, W.-S., Chung, K. Y., McBreen, J. & Yang, X.-Q. A comparative study on structural changes of LiCo1/3Ni1/3Mn1/3O2 and LiNi0.8Co0.15Al0.05O2 during first charge using in situ XRD. *Electrochem. commun.* **8**, 1257–1262 (2006).
38. Yoon, W.-S., Chung, K. Y., McBreen, J., Fischer, D. a. & Yang, X.-Q. Electronic structural changes of the electrochemically Li-ion deintercalated LiNi0.8Co0.15Al0.05O2 cathode material investigated by X-ray absorption spectroscopy. *J. Power Sources* **174**, 1015–1020 (2007).
39. Kim, M. G. *et al.* Ni and oxygen K-edge XAS investigation into the chemical bonding for lithiation of LiyNi1−xAlxO2 cathode material. *Electrochim. Acta* **50**, 501–504 (2004).
40. Aydinol, M. K., Kohan, A. F., Ceder, G., Cho, K. & Joannopoulos, J. Ab initio study of lithium intercalation in metal oxides and metal dichalcogenides. *Phys. Rev. B* **56**, 1354–1365 (1997).
41. Goodenough, J. B. & Kim, Y. Challenges for Rechargeable Li Batteries. *Chem. Mater.* **22**, 587–603 (2010).
42. Kleiner, K. & Ehrenberg, H. Challenges Considering the Degradation of Cell Components in Commercial Lithium-Ion Cells: A Review and Evaluation of Present Systems. *Top. Curr. Chem.* **375**, 45 (2017).
43. Li, S. *et al.* Anionic redox reaction and structural evolution of Ni-rich layered oxide cathode material. *Nano Energy* **98**, 107335 (2022).
44. Thackeray, M. M. *et al.* Li2MnO3-stabilized LiMO2 (M = Mn, Ni, Co) electrodes for lithium-ion batteries. *J. Mater. Chem.* **17**, 3112 (2007).
45. Jain, G. R., Yang, J. S., Balasubramanian, M. & Xu, J. J. Synthesis, electrochemistry, and structural studies of lithium intercalation of a nanocrystalline Li2MnO3-like compound. *Chem. Mater.* **17**, 3850–3860 (2005).
46. Kim, J.-S. *et al.* Electrochemical and Structural Properties of xLi2M'O3·(1−x)LiMn0.5Ni0.5O2 Electrodes for Lithium Batteries (M' = Ti, Mn, Zr; 0 ≤ x ≤ 0.3). *Chem. Mater.* **16**, 1996–2006 (2004).
47. Croy, J. R. *et al.* Examining Hysteresis in Composite xLi2MnO3·(1–x)LiMO2 Cathode Structures. *J. Phys. Chem. C* **117**, 6525–6536 (2013).
48. Lee, G.-H. *et al.* Reversible Anionic Redox Activities in Conventional LiNi1/3Co1/3Mn1/3O2 Cathodes. *Angew. Chem. Int. Ed.* **59**, 2–10 (2020).
49. Qiao, Y. *et al.* Reversible anionic redox activity in Na 3 RuO 4 cathodes: a prototype Na-rich layered oxide. *Energy Environ. Sci.* **11**, 299–305 (2018).
50. Zhang, X. *et al.* Manganese-Based Na-Rich Materials Boost Anionic Redox in High-Performance Layered Cathodes for Sodium-Ion Batteries. *Adv. Mater.* **31**, 1807770 (2019).
51. Saubanère, M. *et al.* The intriguing question of anionic redox in high-energy density cathodes for Li-ion batteries. *Energy Environ. Sci.* **9**, 984–991 (2016).
52. Wu, J., Yang, Y. & Yang, W. Advances in soft X-ray RIXS for studying redox reaction states in batteries. *Dalt. Trans.* **49**, 13519–13527 (2020).
53. Lee, W. *et al.* Anionic Redox Chemistry as a Clue for Understanding the Structural Behavior in Layered Cathode Materials. *Small* **16**, 1905875 (2020).
54. Assat, G. & Tarascon, J.-M. Fundamental understanding and practical challenges of anionic redox activity in Li-ion batteries. *Nat. Energy* **3**, 373–386 (2018).
55. Sathiya, M. *et al.* Reversible anionic redox chemistry in high-capacity layered-oxide electrodes. *Nat. Mater.* **12**, 827–835 (2013).
56. Xu, H., Guo, S. & Zhou, H. Review on anionic redox in sodium-ion batteries. *J. Mater. Chem. A* **7**, 23662–23678 (2019).
57. McCalla, E. *et al.* Visualization of O-O peroxo-like dimers in high-capacity layered oxides for Li-ion batteries. *Science (80-. ).* **350**, 1516–1521 (2015).
58. Århammar, C. *et al.* Unveiling the complex electronic structure of amorphous metal oxides. *Proc. Natl. Acad. Sci.* **108**, 6355–6360 (2011).
59. House, R. A. *et al.* The role of O2 in O-redox cathodes for Li-ion batteries. *Nat. Energy* **6**, 781–789 (2021).
60. Sharpe, R. *et al.* Redox Chemistry and the Role of Trapped Molecular O 2 in Li-Rich Disordered Rocksalt Oxyfluoride Cathodes. *J. Am. Chem. Soc.* **142**, 21799–21809 (2020).
61. Merz, M., Ying, B., Nagel, P., Schuppler, S. & Kleiner, K. Reversible and Irreversible Redox Processes in Li-Rich Layered Oxides. *Chem. Mater.* **33**, 9534–9545 (2021).
62. Liu, X. *et al.* Origin and regulation of oxygen redox instability in high-voltage battery cathodes. *Nat. Energy* **7**, 808–817 (2022).
63. Li, N. *et al.* Correlating the phase evolution and anionic redox in Co-Free Ni-Rich layered oxide cathodes. *Nano Energy* **78**, 105365 (2020).
64. Hwang, S. *et al.* Investigating Local Degradation and Thermal Stability of Charged Nickel-Based Cathode Materials through Real-Time Electron Microscopy. *ACS Appl. Mater. Interfaces* (2014) doi:10.1021/am503278f.
65. Cuisinier, M. *et al.* Quantitative MAS NMR characterization of the LiMn(1/2)Ni(1/2)O(2) electrode/electrolyte interphase. *Solid State Nucl. Magn. Reson.* **42**, 51–61 (2012).
66. Su, Y. *et al.* High-Rate Structure-Gradient Ni-Rich Cathode Material for Lithium-Ion Batteries. *ACS Appl. Mater. Interfaces* **11**, 36697–36704 (2019).
67. Liu, T. *et al.* Rational design of mechanically robust Ni-rich cathode materials via concentration gradient strategy. *Nat. Commun.* **12**, 6024 (2021).
68. Liao, J.-Y., Oh, S.-M. & Manthiram, A. Core/Double-Shell Type Gradient Ni-Rich LiNi 0.76 Co 0.10 Mn 0.14 O 2 with High Capacity and Long Cycle Life for Lithium-Ion Batteries. *ACS Appl. Mater. Interfaces* **8**, 24543–24549 (2016).
69. Zhecheva, E. & Stoyanova, R. Stabilization of the layered crystal structure of LiNiO2 by Co-substitution. *Solid State Ionics* **66**, 143–149 (1993).
70. Eilers-Rethwisch, M. *et al.* Comparative study of Sn-doped Li[Ni0.6Mn0.2Co0.2-xSnx]O2 cathode active materials (x = 0-0.5) for lithium ion batteries regarding electrochemical performance and structural stability. *J. Power Sources* **397**, 68–78 (2018).
71. Kong, F. *et al.* Multivalent Li-Site Doping of Mn Oxides for Li-Ion Batteries. *J. Phys. Chem. C* **119**, 21904–21912 (2015).
72. Kong, F. *et al.* Conflicting Roles of Anion Doping on the Electrochemical Performance of Li-Ion Battery Cathode Materials. *Chem. Mater.* **28**, 6942–6952 (2016).
73. Cao, H., Xia, B., Xu, N. & Zhang, C. Structural and electrochemical characteristics of Co and Al co-doped lithium nickelate cathode materials for lithium-ion batteries. *J. Alloys Compd.* **376**, 282–286 (2004).
74. Zhou, F., Zhao, X. & Dahn, J. R. Synthesis, Electrochemical Properties, and Thermal Stability of Al-Doped LiNi[sub 1⁄3]Mn[sub 1⁄3]Co[sub (1⁄3−z)]Al[sub z]O[sub 2] Positive Electrode Materials. *J. Electrochem. Soc.* **156**, A343 (2009).
75. Gao, S. *et al.* Boron Doping and LiBO 2 Coating Synergistically Enhance the High-Rate Performance of LiNi 0.6 Co 0.1 Mn 0.3 O 2 Cathode Materials. *ACS Sustain. Chem. Eng.* **9**, 5322–5333 (2021).
76. Ying, B., Sutar, P., Nagel, P., Schuppler, S. & Kleiner, K. (Digital Presentation) Investigations into the Capacity Degradation Due to an Electronic Structural Change in Homogenous Boron-Substituted Ni-Rich Layered Oxides. *ECS Meet. Abstr.* **MA2022-01**, 519–520 (2022).
77. Chowdari, B. V. R., Rao, G. V. S. & Chow, S. Y.

Cathodic behavior of Co , Ti , Mg / -doped LiNiO 2. 0–7 (2001).

78. Rosina, K. J. *et al.* Structure of Aluminum Fluoride Coated Li[Li1/9Ni1/3Mn5/9]O2 Cathodes for Secondary Lithium-ion Batteries. *J. Mater. Chem.* **22**, 20602–20610 (2012).

79. Sun, Y. K. *et al.* The role of AlF 3 coatings in improving electrochemical cycling of Li-enriched nickel-manganese oxide electrodes for Li-ion batteries. *Adv. Mater.* **24**, 1192–1196 (2012).

80. Wu, Y., Ming, J., Zhuo, L., Yu, Y. & Zhao, F. Simultaneous surface coating and chemical activation of the Li-rich solid solution lithium rechargeable cathode and its improved performance. *Electrochim. Acta* **113**, 54–62 (2013).

81. Reissig, F. *et al.* The Role of Protective Surface Coatings on the Thermal Stability of Delithiated Ni-Rich Layered Oxide Cathode Materials. *Batteries* **9**, 245 (2023).

82. Song, B., Li, W., Oh, S.-M. & Manthiram, A. Long-Life Nickel-Rich Layered Oxide Cathodes with a Uniform Li 2 ZrO 3 Surface Coating for Lithium-Ion Batteries. *ACS Appl. Mater. Interfaces* **9**, 9718–9725 (2017).

83. Aurbach, D. *et al.* On the correlation among surface chemistry, 3D structure, morphology, electrochemical and impedance behavior of various lithiated carbon electrodes. *J. Power Sources* **97**–**98**, 92–96 (2001).

84. Wang, L. *et al.* Grain Morphology and Microstructure Control in High-Stable Ni-Rich Layered Oxide Cathodes. *Adv. Funct. Mater.* **33**, (2023).

85. Mo, W. *et al.* Tuning the surface of LiNi0.8Co0.1Mn0.1O2 primary particle with lithium boron oxide toward stable cycling. *Chem. Eng. J.* **400**, 125820 (2020).

86. de Groot, F. & Kotani, A. *Core Level Spectroscopy of Solids*. (CRC Press, 2008).

87. Haverkort, M. W., Zwierzycki, M. & Andersen, O. K. Multiplet ligand-field theory using Wannier orbitals. *Phys. Rev. B - Condens. Matter Mater. Phys.* **85**, 165113–165133 (2012).

88. Ballhausen, C. J. *An introduction to ligand field theory*. (McGraw Hill book company, 1962).

89. Fujimori, A., Minami, F. & Sugano, S. Multielectron satellites and spin polarization in photoemission from Ni compounds. *Phys. Rev. B* **29**, 5225–5227 (1984).

90. S. Sugano, Y. Tanabe, and H. K. *Multiplets of Transition-Metal Ions in Crystals*. (Academic Press. Inc., 1070).

91. P. A. Cox. *Transition Metal Oxides: An Introduction to Their Electronic Structure and Properties (The International Series of Monographs on Chemistry)*. (Oxford University Press, USA, 2010).

92. Okubo, M. & Yamada, A. Molecular Orbital Principles of Oxygen-Redox Battery Electrodes. *ACS Appl. Mater. Interfaces* **9**, 36463–36472 (2017).

93. Zimmermann, R. *et al.* Electronic structure of 3d-transition-metal oxides: on-site Coulomb repulsion versus covalency. *J. Phys. Condens. Matter* **11**, 1657–1682 (1999).

94. Anisimov, V. I., Zaanen, J. & Andersen, O. K. Band theory and Mott insulators: Hubbard U instead of Stoner I. *Phys. Rev. B* **44**, 943–954 (1991).

95. Olalde-Velasco, P., Jiménez-Mier, J., Denlinger, J. D., Hussain, Z. & Yang, W. L. Direct probe of Mott-Hubbard to charge-transfer insulator transition and electronic structure evolution in transition-metal systems. *Phys. Rev. B* **83**, 241102 (2011).

96. Zaanen, J. & Sawatzky, G. A. Systematics in band gaps and optical spectra of 3D transition metal compounds. *J. Solid State Chem.* **88**, 8–27 (1990).

97. Zaanen, J., Sawatzky, G. A. & Allen, J. W. Band gaps and electronic structure of transition-metal compounds. *Phys. Rev. Lett.* **55**, 418–421 (1985).

98. Fink, J. *et al.* 2p absorption spectra of the 3d elements. *Phys. Rev. B* **32**, 4899–4904 (1985).

99. van der Laan, G., Zaanen, J., Sawatzky, G. A., Karnatak, R. & Esteva, J.-M. Comparison of x-ray absorption with x-ray photoemission of nickel dihalides and NiO. *Phys. Rev. B* **33**, 4253–4263 (1986).

100. Galakhov, V. R. *et al.* Electronic structure of LiNiO2, LiFeO2 and LiCrO2: X-ray photoelectron and X-ray emission study. *Solid State Commun.* **95**, 347–351 (1995).

101. Wang, R.-P. *et al.* Charge-Transfer Analysis of 2p3d Resonant Inelastic X-ray Scattering of Cobalt Sulfide and Halides. *J. Phys. Chem. C* **121**, 24919–24928 (2017).

102. Biesinger, M. C., Lau, L. W. M., Gerson, A. R. & Smart, R. S. C. The role of the Auger parameter in XPS studies of nickel metal, halides and oxides. *Phys. Chem. Chem. Phys.* **14**, 2434 (2012).

103. Wilson, J. A. Systematics of the breakdown of Mott insulation in binary transition metal compounds. *Adv. Phys.* **21**, 143–198 (1972).

104. de Groot, F. M. F. X-ray absorption and dichroism of transition metals and their compounds. *J. Electron Spectros. Relat. Phenomena* **67**, 529–622 (1994).

105. Nanba, Y. *et al.* Redox Potential Paradox in Na x MO 2 for Sodium-Ion Battery Cathodes. *Chem. Mater.* **28**, 1058–1065 (2016).

106. Park, H. E., Hong, C. H. & Yoon, W. Y. The effect of internal resistance on dendritic growth on lithium metal electrodes in the lithium secondary batteries. *J. Power Sources* **178**, 765–768 (2008).

107. Kim, J. S. *et al.* Electrochemical and structural properties of xLi2MnO3 (1-x)LiM0.5Ni0.5O2 electrodes for lithium batteries (M = Ti, Mn, Zr; O < x < 0.3). *Chem. Mater.* **16**, 1996–2006 (2004).

108. Thole, B. T., Cowan, R. D., Sawatzky, G. A., Fink, J. & Fuggle, J. C. New probe for the ground-state electronic structure of narrow-band and impurity systems. *Phys. Rev. B* **31**, 6856–6858 (1985).

109. Montoro, L. A., Abbate, M., Almeida, E. C. & Rosolen, J. M. Electronic structure of the transition metal ions in LiCoO2, LiNiO2 and LiCo0.5Ni0.5O2. *Chem. Phys. Lett.* **309**, 14–18 (1999).

110. Montoro, L. A., Abbate, M. & Rosolen, J. M. Electronic Structure of Transition Metal Ions in Deintercalated and Reintercalated LiCo0.5Ni0.5O2. *J. Electrochem. Soc.* **147**, 1651–1657 (2000).

111. Ikeno, H., de Groot, F. M. F., Stavitski, E. & Tanaka, I. Multiplet calculations of L(2,3) x-ray absorption near-edge structures for 3d transition-metal compounds. *J. Phys. Condens. Matter* **21**, 104208 (2009).

112. Ikeno, H., Tanaka, I., Koyama, Y., Mizoguchi, T. & Ogasawara, K. First-principles multielectron calculations of Ni L2,3 NEXAFS and ELNES for LiNiO2 and related compounds. *Phys. Rev. B* **72**, 075123 (2005).

113. de Groot, F. M. F., Fuggle, J. C., Thole, B. T. & Sawatzky, G. A. 2 p x-ray absorption of 3 d transition-metal compounds: An atomic multiplet description including the crystal field. *Phys. Rev. B* **42**, 5459–5468 (1990).

114. Seo, D.-H. *et al.* The structural and chemical origin of the oxygen redox activity in layered and cation-disordered Li-excess cathode materials. *Nat. Chem.* **8**, 692–697 (2016).

115. Luo, K. *et al.* Charge-compensation in 3d-transition-metal-oxide intercalation cathodes through the generation of localized electron holes on oxygen. *Nat. Chem.* **8**, 684–691 (2016).

116. Luo, K. *et al.* Anion Redox Chemistry in the Cobalt Free 3d Transition Metal Oxide Intercalation Electrode Li[Li 0.2 Ni 0.2 Mn 0.6 ]O 2. *J. Am. Chem. Soc.* **138**, 11211–11218 (2016).

117. Yang, W. & Devereaux, T. P. Anionic and cationic redox and interfaces in batteries: Advances from soft X-ray absorption spectroscopy to resonant inelastic scattering. *J. Power Sources* **389**, 188–197 (2018).
118. Yoon, W.-S. *et al.* Investigation of the Local Structure of the LiNi0.5Mn0.5O2 Cathode Material during Electrochemical Cycling by X-Ray Absorption and NMR Spectroscopy. *Electrochem. Solid-State Lett.* **5**, A263 (2002).
119. Mizushima, K., Jones, P. C., Wiseman, P. J. & Goodenough, J. B. B. LiCoO2: A new cathode material for batteries of high energy density. *Mat. Res. Bull.* **15**, 783–789 (1980).
120. Oswald, S. & Gasteiger, H. A. The Structural Stability Limit of Layered Lithium Transition Metal Oxides Due to Oxygen Release at High State of Charge and Its Dependence on the Nickel Content. *J. Electrochem. Soc.* **170**, 030506 (2023).
121. Bak, S.-M. *et al.* Structural Changes and Thermal Stability of Charged LiNi x Mn y Co z O 2 Cathode Materials Studied by Combined In Situ Time-Resolved XRD and Mass Spectroscopy. *ACS Appl. Mater. Interfaces* **6**, 22594–22601 (2014).
122. Moses, A. W., Flores, H. G. G., Kim, J.-G. & Langell, M. A. Surface properties of LiCoO2, LiNiO2 and LiNi1−xCoxO2. *Appl. Surf. Sci.* **253**, 4782–4791 (2007).
123. Friedrich, F. *et al.* Editors' Choice—Capacity Fading Mechanisms of NCM-811 Cathodes in Lithium-Ion Batteries Studied by X-ray Diffraction and Other Diagnostics. *J. Electrochem. Soc.* **166**, A3760–A3774 (2019).
124. Xu, C. *et al.* Bulk fatigue induced by surface reconstruction in layered Ni-rich oxide cathodes for Liion batteries. *ChemRxiv. Prepr.* **10.26434/c**, (2020).
125. Liu, X. *et al.* Prelithiated Li-Enriched Gradient Interphase toward Practical High-Energy NMC–Silicon Full Cell. *ACS Energy Lett.* **6**, 320–328 (2021).
126. Oishi, M. *et al.* Charge compensation mechanisms in Li1.16Ni0.15Co0.19Mn0.50O2 positive electrode material for Li-ion batteries analyzed by a combination of hard and soft X-ray absorption near edge structure. *J. Power Sources* **222**, 45–51 (2013).
127. Yoon, W. *et al.* Oxygen Contribution on Li-Ion Intercalation - Deintercalation in LiCoO 2 Investigated by O K-Edge and Co L-Edge X-ray Absorption Spectroscopy. *J. Phys. Chem. B* **106**, 2526–2532 (2002).
128. Yoon, W.-S. *et al.* Combined NMR and XAS Study on Local Environments and Electronic Structures of Electrochemically Li-Ion Deintercalated Li1−xCo 1/3Ni1/3Mn1/3O2 Electrode System. *Electrochem. Solid-State Lett.* **7**, A53 (2004).
129. Manthiram, A. A reflection on lithium-ion battery cathode chemistry. *Nat. Commun.* **11**, 1550 (2020).
130. Manthiram, A. An Outlook on Lithium Ion Battery Technology. *ACS Cent. Sci.* **3**, 1063–1069 (2017).
131. Julien, C., Mauger, A., Zaghib, K. & Groult, H. Comparative Issues of Cathode Materials for Li-Ion Batteries. *Inorganics* **2**, 132–154 (2014).
132. Wagman, D. D. *et al.* Erratum: The NBS tables of chemical thermodynamic properties. Selected values for inorganic and C1 and C2 organic substances in SI units [J. Phys. Chem. Ref. Data 11 , Suppl. 2 (1982)]. *J. Phys. Chem. Ref. Data* **18**, 1807–1812 (1989).
133. Manthiram, A. Materials Challenges and Opportunities of Lithium Ion Batteries. *J. Phys. Chem. Lett.* **2**, 176–184 (2011).
134. Balasubramanian, M., Sun, X., Yang, X. . & McBreen, J. In situ X-ray diffraction and X-ray absorption studies of high-rate lithium-ion batteries. *J. Power Sources* **92**, 1–8 (2001).
135. Laszczynski, N., Solchenbach, S., Gasteiger, H. A. & Lucht, B. L. Understanding Electrolyte Decomposition of Graphite/NCM811 Cells at Elevated Operating Voltage. *J. Electrochem. Soc.* **166**, A1853–A1859 (2019).
136. Kawamura, T., Okada, S. & Yamaki, J. Decomposition reaction of LiPF6-based electrolytes for lithium ion cells. *J. Power Sources* **156**, 547–554 (2006).
137. Yang, X. Q., Sun, X. & Mcbreen, J. New findings on the phase transitions in Li 1 y x NiO 2 : in situ synchrotron X-ray diffraction studies. **1**, 227–232 (1999).
138. Strehle, B. *et al.* The Role of Oxygen Release from Li- and Mn-Rich Layered Oxides during the First Cycles Investigated by On-Line Electrochemical Mass Spectrometry. *J. Electrochem. Soc.* **164**, A400–A406 (2017).
139. Reinert, F. *et al.* Electron and hole doping in NiO. *Phys. B Condens. Matter* **97**, 83–93 (1995).
140. Hu, Z. *et al.* Hole distribution between the Ni 3d and O 2p orbitals in Nd2-xSrxNiO4-δ. *Phys. Rev. B* **61**, 3739–3744 (2000).
141. Amine, K. *et al.* Factors responsible for impedance rise in high power lithium ion batteries. *J. Power Sources* **97**–**98**, 684–687 (2001).
142. Julien, C. M. Lithium intercalated compounds. *Mater. Sci. Eng. R Reports* **40**, 47–102 (2003).
143. Hüfner, S. Electronic structure of NiO and related 3d-transition-metal compounds. *Adv. Phys.* **43**, 183–356 (1994).
144. van Elp, J., Eskes, H., Kuiper, P. & Sawatzky, G. A. Electronic structure of Li-doped NiO. *Phys. Rev. B* **45**, 1612–1622 (1992).
145. Kuiper, P., Kruizinga, G., Ghijsen, J., Sawatzky, G. & Verweij, H. Character of holes in LixNi1-xO and their magnetic behavior. *Phys. Rev. Lett.* **62**, 221–224 (1989).
146. van Elp, J., Searle, B. G., Sawatzky, G. A. & Sacchi, M. Ligand hole induced symmetry mixing of d8 states in LixNi1−xO, as observed in Ni 2p x-ray absorption spectroscopy. *Solid State Commun.* **80**, 67–71 (1991).
147. Jetybayeva, A. *et al.* Unraveling the State of Charge-Dependent Electronic and Ionic Structure–Property Relationships in NCM622 Cells by Multiscale Characterization. *ACS Appl. Energy Mater.* **5**, 1731–1742 (2022).
148. Kleiner, K. *et al.* Origin of High Capacity and Poor Cycling Stability of Li-Rich Layered Oxides: A Long-Duration in Situ Synchrotron Powder Diffraction Study. *Chem. Mater.* **30**, 3656–3667 (2018).
149. Green, R. J., Haverkort, M. W. & Sawatzky, G. A. Bond disproportionation and dynamical charge fluctuations in the perovskite rare-earth nickelates. *Phys. Rev. B* **94**, 195127 (2016).
150. Lu, Y., Höppner, M., Gunnarsson, O. & Haverkort, M. W. Efficient real-frequency solver for dynamical mean-field theory. *Phys. Rev. B* **90**, 085102 (2014).
151. Haverkort, M. W. Quanty for core level spectroscopy - excitons, resonances and band excitations in time and frequency domain. *J. Phys. Conf. Ser.* **712**, 012001 (2016).
152. Haverkort, M. W. *et al.* Bands, resonances, edge singularities and excitons in core level spectroscopy investigated within the dynamical mean-field theory. *EPL (Europhysics Lett.* **108**, 57004 (2014).
153. Stavitski, E. & de Groot, F. M. F. The CTM4XAS program for EELS and XAS spectral shape analysis of transition metal L edges. *Micron* **41**, 687–94 (2010).
154. Mitterbauer, C. *et al.* Electron energy-loss near-edge structures of 3d transition metal oxides recorded at high-energy resolution. *Ultramicroscopy* **96**, 469–80 (2003).
155. Tröger, L. *et al.* Full correction of the self-absorption in soft-fluorescence extended x-ray-absorption fine structure. *Phys. Rev. B* **46**, 3283–3289 (1992).
156. Anisimov, V. I., Solovyev, I. V., Korotin, M. A., Czyżyk, M. T. & Sawatzky, G. A. Density-functional theory and NiO photoemission spectra. *Phys. Rev. B* **48**, 16929–16934 (1993).
157. Hepting, M. *et al.* Electronic structure of the parent compound of superconducting infinite-layer nickelates.

*Nat. Mater.* **19**, 381–385 (2020).
158. Dahn, J. R., von Sacken, U. & Michal, C. A. Structure and electrochemistry of Li1yNiO2 and a new Li2NiO2 phase with the Ni (OH)2 structure. *Solid State Ionics* **44**, 87–97 (1990).
159. Chung, J.-H. *et al.* Local structure of LiNiO2 studied by neutron diffraction. *Phys. Rev. B* **71**, 064410 (2005).
160. Goodenough, J. B. Reflections on sixty years of solid state chemistry. *Zeitschrift fur Anorg. und Allg. Chemie* **638**, 1893–1896 (2012).
161. Wang, Y. *et al.* Pillar-beam structures prevent layered cathode materials from destructive phase transitions. *Nat. Commun.* **12**, 13 (2021).
162. Li, H. *et al.* Stabilizing nickel-rich layered oxide cathodes by magnesium doping for rechargeable lithium-ion batteries. *Chem. Sci.* **10**, 1374–1379 (2019).
163. Ikeno, H., Mizoguchi, T. & Tanaka, I. Ab initio charge transfer multiplet calculations on the L2,3 XANES and ELNES of 3 d transition metal oxides. *Phys. Rev. B* **83**, 155107 (2011).
164. Genreith-Schriever, A. R. *et al.* Oxygen hole formation controls stability in LiNiO2 cathodes. *Joule* **7**, 1623–1640 (2023).
165. Eilers-Rethwisch, M., Winter, M. & Schappacher, F. M. Synthesis, electrochemical investigation and structural analysis of doped Li[Ni0.6Mn0.2Co0.2-xMx]O2 (x = 0, 0.05; M = Al, Fe, Sn) cathode materials. *J. Power Sources* **387**, 101–107 (2018).
166. Li, R. *et al.* Structure and electrochemical performance modulation of a LiNi 0.8 Co 0.1 Mn 0.1 O 2 cathode material by anion and cation co-doping for lithium ion batteries. *RSC Adv.* **9**, 36849–36857 (2019).
167. Oswald, S., Pritzl, D., Wetjen, M. & Gasteiger, H. A. Novel Method for Monitoring the Electrochemical Capacitance by In Situ Impedance Spectroscopy as Indicator for Particle Cracking of Nickel-Rich NCMs: Part I. Theory and Validation. *J. Electrochem. Soc.* **167**, 100511 (2020).
168. Qiu, L. *et al.* Recent advance in structure regulation of high-capacity Ni-rich layered oxide cathodes. *EcoMat* **3**, (2021).
169. Yin, S. *et al.* Fundamental and solutions of microcrack in Ni-rich layered oxide cathode materials of lithium-ion batteries. *Nano Energy* **83**, 105854 (2021).
170. Zhang, S. S. Problems and their origins of Ni-rich layered oxide cathode materials. *Energy Storage Mater.* **24**, 247–254 (2020).
171. Friedrich, F. Correlation of structural changes and capacity fading in Ni-rich layered oxides. (Technical University of Munich, 2017).
172. Yabuuchi, N., Yoshii, K., Myung, S.-T., Nakai, I. & Komaba, S. Detailed Studies of a High-Capacity Electrode Material for Rechargeable Batteries, Li2MnO3−LiCo1/3Ni1/3Mn1/3O2. *J. Am. Chem. Soc.* **133**, 4404–4419 (2011).
173. Dogan, F. *et al.* Re-entrant lithium local environments and defect driven electrochemistry of Li- and Mn-Rich Li-Ion battery cathodes. *J. Am. Chem. Soc.* **137**, 2328–2335 (2015).
174. Manthiram, A. & Goodenough, J. B. Lithium insertion into Fe2(SO4)3 frameworks. *J. Power Sources* **26**, 403–408 (1989).
175. Atkins, P. & de Paula, J. *Physical Chemistry*. *Oxford University Press* vol. 9 (Oxford University Press, 2010).
176. Janiak, C., Meyer, H.-J., Gudat, D. & Kurz, P. *Riedel Moderne Anorganische Chemie*. (De Gruyter, 2018). doi:10.1515/9783110441635.
177. Cotton, F. A. & Wilkinson, G. *Anorganische Chemie - eine zusammenfassende Darstellung für Fortgeschrittene*. (1974).
178. Liu, C., Neale, Z. G. & Cao, G. Understanding electrochemical potentials of cathode materials in rechargeable batteries. *Mater. Today* **19**, 109–123 (2016).
179. Wu, M. S., Chiang, P. C. J., Lin, J. C. & Jan, Y. S. Correlation between electrochemical characteristics and thermal stability of advanced lithium-ion batteries in abuse tests - Short-circuit tests. *Electrochim. Acta* **49**, 1803–1812 (2004).
180. Wang, X., Tamaru, M., Okubo, M. & Yamada, A. Electrode Properties of P2–Na 2/3 Mn y Co 1– y O 2 as Cathode Materials for Sodium-Ion Batteries. *J. Phys. Chem. C* **117**, 15545–15551 (2013).
181. Gilbert, B. *et al.* Multiple Scattering Calculations of Bonding and X-ray Absorption Spectroscopy of Manganese Oxides. *J. Phys. Chem. A* **107**, 2839–2847 (2003).
182. Mitra, C. *et al.* Direct observation of electron doping in La0.7Ce0.3O3 using x-ray absorption spectroscopy. *Phys. Rev. B* **67**, 092404 (2003).
183. Asakura, D. *et al.* Mn 2p resonant X-ray emission clarifies the redox reaction and charge-transfer effects in LiMn 2 O 4. *Phys. Chem. Chem. Phys.* **21**, 18363–18369 (2019).
184. Asakura, D. *et al.* Material/element-dependent fluorescence-yield modes on soft X-ray absorption spectroscopy of cathode materials for Li-ion batteries. *AIP Adv.* **6**, 035105 (2016).
185. Anjum, G. *et al.* NEXAFS studies of La0.8Bi0.2Fe1−xMnxO3 (0.0 ⩽ x ⩽ 0.4) multiferroic system using x-ray absorption spectroscopy. *J. Phys. D. Appl. Phys.* **44**, 075403 (2011).
186. Kubobuchi, K. *et al.* Mn L2,3-edge X-ray absorption spectroscopic studies on charge-discharge mechanism of Li2MnO3. *Appl. Phys. Lett.* **104**, 053906 (2014).
187. Ghiasi, M. *et al.* Mn and Co Charge and Spin Evolutions in LaMn1– xCoxO3 Nanoparticles. *J. Phys. Chem. C* **120**, 8167–8174 (2016).
188. Ohzuku, T., Kitagawa, M. & Hitai, T. Electrochemistry of Manganese Dioxide in Lithium Nonaqueous Cell II. X-Ray Diffractional and Electrochemical Characterization on Deep Discharge Products of Electrolytic Manganese Dioxide. *J. Electrochem. Soc.* **137**, 40–46 (1990).
189. Chung, K. Y. & Kim, K. B. Investigations into capacity fading as a result of a Jahn-Teller distortion in 4 V LiMn2O4 thin film electrodes. *Electrochim. Acta* **49**, 3327–3337 (2004).
190. Yamada, A., Tanaka, M., Tanaka, K. & Sekai, K. Jahn–Teller instability in spinel Li–Mn–O. *J. Power Sources* **81–82**, 73–78 (1999).
191. Ndmno, C. *et al.* Jahn-Teller Distortion , Charge Ordering , and Magnetic Transitions in. **38**, 354–359 (2000).
192. Merz, M. *et al.* Spin and orbital states in single-layered La2−xCaxCoO4 studied by doping- and temperature-dependent near-edge x-ray absorption fine structure. *Phys. Rev. B* **84**, 014436 (2011).
193. Merz, M. *et al.* X-ray absorption and magnetic circular dichroism of LaCoO3, La0.7Ce0.3CoO3, and La0.7Sr0.3CoO3 films: Evidence for cobalt-valence-dependent magnetism. *Phys. Rev. B* **82**, 174416 (2010).
194. Thien, J. *et al.* Cationic Ordering and Its Influence on the Magnetic Properties of Co-Rich Cobalt Ferrite Thin Films Prepared by Reactive Solid Phase Epitaxy on Nb-Doped SrTiO3(001). *Materials (Basel).* **15**, 46 (2021).
195. Rodewald, J. *et al.* Formation of ultrathin cobalt ferrite films by interdiffusion of Fe3O4CoO bilayers. *Phys. Rev. B* **100**, 155418 (2019).
196. Fujita, E., Furenlid, L. R. & Renner, M. W. Direct XANES Evidence for Charge Transfer in Co−CO 2 Complexes. *J. Am. Chem. Soc.* **119**, 4549–4550 (1997).
197. Li, Y. *et al.* Implementing Metal-to-Ligand Charge Transfer in Organic Semiconductor for Improved Visible-Near-Infrared Photocatalysis. *Adv. Mater.* **28**, 6959–6965 (2016).
198. Bröker, A. Synthesis of layered oxides for cathode materials in Li ion batteries. (University of Münster, 2022).

199. Ying, B. *et al.* Monitoring the Formation of Nickel-Poor and Nickel-Rich Oxide Cathode Materials for Lithium-Ion Batteries with Synchrotron Radiation. *Chem. Mater.* **35**, 1514–1526 (2023).

iffraction Studies of Lithium Intercalation in LixCo02. *J. Electroanal. Chem.* **139**, 2–8 (1992).

201. Ven, A. Van Der & Ceder, G. Ordering in Lix( Ni 0.5Mn0.5) O 2 and its relation to charge capacity and electrochemical behavior in rechargeable lithium batteries. *Electrochem. commun.* **6**, 1045–1050 (2004).

202. Shao-Horn, Y., Levasseur, S., Weill, F. & Delmas, C. Probing Lithium and Vacancy Ordering in O3 Layered LixCoO2 (x≈0.5). *J. Electrochem. Soc.* **150**, A366 (2003).

203. Van der Ven, A. & Ceder, G. Lithium diffusion mechanisms in layered intercalation compounds. *J. Power Sources* **97**–**98**, 529–531 (2001).

204. Ceder, G. & Van der Ven, A. Phase diagrams of lithium transition metal oxides: investigations from first principles. *Electrochim. Acta* **45**, 131–150 (1999).

205. Van der Ven, A. Lithium Diffusion in Layered LixCoO2. *Electrochem. Solid-State Lett.* **3**, 301 (1999).

206. Jang, Y.-I., Neudecker, B. J. & Dudney, N. J. Lithium Diffusion in LixCoO2 (0.45 < x < 0.7) Intercalation Cathodes. *Electrochem. Solid-State Lett.* **4**, A74 (2001).

207. Thompson, S. P. *et al.* Beamline I11 at Diamond: A new instrument for high resolution powder diffraction. *Rev. Sci. Instrum.* **80**, 1–5 (2009).

208. Grosu, C. Correlation between structure and electrochemistry of LiMO2 cathode materials (M = Ni, Co). (University of Bologna, Technical University of Munich, 2016).

209. Fantin, R. *et al.* Synthesis and Postprocessing of Single-Crystalline LiNi 0.8 Co 0.15 Al 0.05 O 2 for Solid-State Lithium-Ion Batteries with High Capacity and Long Cycling Stability. *Chem. Mater.* **33**, (2021).

210. Hausen, F., Scheer, N., Ying, B. & Kleiner, K. Correlation of the electronic structure and Li-ion mobility with modulus and hardness in LiNi 0.6 Co 0.2 Mn 0.2 O 2 cathodes by combined near edge X-ray absorption finestructure spectroscopy, atomic force microscopy, and nanoindentation. *Electrochem. Sci. Adv.* **accepted**, (2023).

200. Reimers, J. N. & Dahn, J. R. Electrochemical and In Situ X-Ray D

211. Bréger, J. *et al.* High-resolution X-ray diffraction, DIFFaX, NMR and first principles study of disorder in the Li 2MnO 3-Li[Ni 1/2Mn 1/2]O 2 solid solution. *J. Solid State Chem.* **178**, 2575–2585 (2005).

212. Meng, Y. S. *et al.* Cation Ordering in Layered O3 Li[Ni x Li 1/3 - 2 x /3 Mn 2/ 3 - x /3 ]O 2 (0 ≤ x ≤ 1 / 2 ) Compounds. *Chem. Mater.* **17**, 2386–2394 (2005).

213. Liu, H. *et al.* Operando Lithium Dynamics in the Li-Rich Layered Oxide Cathode Material via Neutron Diffraction. *Adv. Energy Mater.* **6**, 1502143 (2016).

214. Jansen, A. . *et al.* Development of a high-power lithium-ion battery. *J. Power Sources* **81**–**82**, 902–905 (1999).

215. Hong, J. *et al.* Metal–oxygen decoordination stabilizes anion redox in Li-rich oxides. *Nat. Mater.* **18**, 256–265 (2019).

216. Arroyo y de Dompablo, M. E., Van der Ven, A. & Ceder, G. First-principles calculations of lithium ordering and phase stability on NiO. *Phys. Rev. B* **66**, 064112 (2002).

217. Saubanere, M., McCalla, E., Tarascon, J.-M. & Doublet, M.-L. The intriguing question of anionic redox in high-energy density cathodes for Li-ion batteries. *Energy Environ. Sci.* **9**, 984–991 (2016).

218. Robertson, A. D. & Bruce, P. G. Mechanism of Electrochemical Activity in Li2MnO3. *Chem. Mater.* **15**, 1984–1992 (2003).

219. Ren, H. *et al.* Unraveling Anionic Redox for Sodium Layered Oxide Cathodes: Breakthroughs and Perspectives. *Adv. Mater.* **34**, 2106171 (2022).

220. House, R. A. *et al.* Superstructure control of first-cycle voltage hysteresis in oxygen-redox cathodes. *Nature* **577**, 502–508 (2020).

221. House, R. A. *et al.* Covalency does not suppress O2 formation in 4d and 5d Li-rich O-redox cathodes. *Nat. Commun.* **12**, 2975 (2021).

222. Grimaud, A., Hong, W. T., Shao-Horn, Y. & Tarascon, J.-M. Anionic redox processes for electrochemical devices. *Nat. Mater.* **15**, 121–126 (2016).

223. Ben Yahia, M., Vergnet, J., Saubanère, M. & Doublet, M.-L. Unified picture of anionic

redox in Li/Na-ion batteries. *Nat. Mater.* **18**, 496–502 (2019).
224. Hong, J. *et al.* Critical role of oxygen evolved from layered Li-Excess metal oxides in lithium rechargeable batteries. *Chem. Mater.* **24**, 2692–2697 (2012).
225. Armstrong, A. R. *et al.* Article Demonstrating Oxygen Loss and Associated Structural Reorganization in the Lithium Battery Cathode Li [ Ni Li Mn ] O Demonstrating Oxygen Loss and Associated Structural Reorganization in the Lithium Battery Cathode. 8694–8698 (2006) doi:10.1021/ja062027.
226. Lanz, P., Sommer, H., Schulz-Dobrick, M. & Novák, P. Oxygen release from high-energy xLi2MnO3·(1−x)LiMO2 (M=Mn, Ni, Co): Electrochemical, differential electrochemical mass spectrometric, in situ pressure, and in situ temperature characterization. *Electrochim. Acta* **93**, 114–119 (2013).
227. Kim, S. Y. *et al.* Inhibiting Oxygen Release from Li-rich, Mn-rich Layered Oxides at the Surface with a Solution Processable Oxygen Scavenger Polymer. *Adv. Energy Mater.* **11**, 2100552 (2021).
228. Evertz, M. *et al.* Unraveling transition metal dissolution of Li1.04Ni1/3Co1/3Mn1/3O2 (NCM 111) in lithium ion full cells by using the total reflection X-ray fluorescence technique. *J. Power Sources* **329**, 364–371 (2016).
229. Schnadt, J. *et al.* The new ambient-pressure X-ray photoelectron spectroscopy instrument at MAX-lab. *J. Synchrotron Radiat.* **19**, 701–704 (2012).
230. Metzger, M., Strehle, B., Solchenbach, S. & Gasteiger, H. A. Origin of H 2 Evolution in LIBs: H 2 O Reduction vs. Electrolyte Oxidation. *J. Electrochem. Soc.* **163**, A798–A809 (2016).
231. Boivin, E. *et al.* The Role of Ni and Co in Suppressing O-Loss in Li-Rich Layered Cathodes. *Adv. Funct. Mater.* **31**, 2003660 (2021).
232. Achkar, A. J. *et al.* Bulk sensitive x-ray absorption spectroscopy free of self-absorption effects. *Phys. Rev. B* **83**, 081106 (2011).
233. Thompson, S. P. *et al.* Fast X-ray powder diffraction on I11 at Diamond. *J. Synchrotron Radiat.* **18**, 637–648 (2011).
234. Rodriguez-Carvajal, J. & Roisnel, T. Extended software/methods development issue. in *Newletter N°20* 35–36 (International Union of Crystallography, 1998).
235. Yin, S.-C., Rho, Y.-H., Swainson, I. & Nazar, L. F. X-ray/Neutron Diffraction and Electrochemical Studies of Lithium De/Re-Intercalation in Li 1 - x Co 1/ 3 Ni 1/3 Mn 1/3 O 2 ( x = 0 → 1). *Chem. Mater.* **18**, 1901–1910 (2006).

# Supporting information

## A review and outlook on anionic and cationic redox in Ni-, Li- and Mn-rich layered oxides Li*Me*$O_2$ (*Me* = Li, Ni, Co, Mn)

Bixian Ying[1], Zhenjie Teng[1], Sarah Day[2], Dan Porter[2], Martin Winter[1,3], Adrian Jonas[4], Katja Frenzel[4], Lena Mathies[4], Burkhard Beckhoff[4], Peter Nagel[5,6], Stefan Schuppler[5,6], Michael Merz[5,6], Felix Pfeiffer[3], Matthias Weiling[3], Masoud Baghernejad[3], Karin Kleiner[1,*]

[1] Münster Electrochemical Energy Technology (MEET), University of Münster (WWU) Corrensstraße 46, 48149 Münster, Germany

[2] Harwell Science and Innovation Campus, Diamond Light Source, Didcot, Oxfordshire OX11 0DE, U.K.

[3] Helmholtz-Institute Münster, Forschungszentrum Jülich GmbH, 48149 Muenster, Germany

[4] Physikalisch-Technische Bundesanstalt, Abbestr. 2-12, 10587 Berlin, Germany

[5] Institute for Quantum Materials and Technologies, Karlsruhe Institute of Technology (IQMT, KIT), Hermann-von-Helmholtz-Platz 1, 76344 Eggenstein-Leopoldshafen, Germany

[6] Karlsruhe Nano Micro Facility (KNMFi), Karlsruhe Institute of Technology (KIT), Campus North, Hermann-von-Helmholtz-Platz 1, 76344 Eggenstein-Leopoldshafen, Germany

* Correspondence to: karin.kleiner@wwu.de

## S1. Experimental

*Material Synthesis:* Lithium transition metal oxides are synthesized via an $OH^-$ precipitation of $Me(OH)_2$ (with *Me* = Ni, Co, Mn) by pumping a *Me*-sulfate solution (stoichiometric amount of Ni-, Co-, Mn-sulfate) into a $NH_4OH$/NaOH solution while keeping the pH at 11 by adding NaOH(*aq.*) manually. The final rhombohedral structures ($R\bar{3}m$) are obtained from the calcination of the dried hydroxide precursor with LiOH in a molar ratio of 1:1.03.[15]

*Electrode Preparation:* 80 wt-% of the cathode material (powder), 10 wt-% Imerys Super C65, 10 wt-% PvDF (Solef, Polyvinylidene difluoride, PvdF) and NMP (N-methylpyrrolidone) as processing solvent (6.5 mg) was mixed in a planetary mixer (12 min, 1800 RPM) for electrode preparation. The ink was casted on an Al-foil (200 µm wet film thickness) and dried at 80 °C. 12 mm in diameter samples were punched out of the electrode sheets, pressed to achieve a tab density of ~30%, again dried in a Büchi oven (120°C at $10^{-3}$ mbar) including an air free transfer into the glovebox, where it was assembled to a battery cell in an argon filled glovebox using coin cells with two Celgard separators (15 mm in diameter, glass micro fiber 691, VWR, Germany), 60 µL LP57 (1 M $LiPF_6$ in ethylene carbonate (EC):ethyl methyl carbonate (EMC), 3:7 by weight, <20 ppm of $H_2O$, BASF SE, Germany) and lithium foil as counter electrodes (15 mm in diameter, 99.9% Rockwood Lithium, United States). The underlying capacity used to calculate the C-rate for all layered oxides was 180 mAh/g. The applied voltage cut-offs upon galvanostatic cycling were chosen to be 3.0 V and 4.3 V. GITT measurements are performed upon charge and discharge in 3-electrode Swagelok cells using 41 x 30 min C/20 current pulses followed by a 10 h rest (or a rest until the voltage drop becomes < 0.001 V/20 s) to reach conditions close to equilibrium. At the end of charge the cells are charged to 4.3 V with C/20 and hold at this voltage for 5 h.

*NEXAFS measurements:* The electrodes were cycled with C/20 at 25°C to the charged state, to a remaining SOC of 80%, 75%, 50%, and to the discharged state. The cells were disassembled in an argon filled glovebox. All samples were vacuum dried in a Büchi oven ($10^{-3}$ mbar, 25 °C) prior to the measurements without contact to air to avoid outgassing in the UHV chamber. An argon filled transfer case was used for the sample transfer from the glovebox to the measuring chamber of the beamline. NEXAFS measurements were performed at IQMT's soft X-ray beamline WERA at the Karlsruhe synchrotron light source KARA (Germany). NEXAFS measurements at the Ni $L_{2,3}$, Co $L_{2,3}$, Mn $L_{2,3}$, O $K$ and F $K$ edge were carried out in total electron and fluorescence yield (FY). For the Co $L_{2,3}$ edge measurements, the FY detection window (width 0.24 keV) was shifted up in energy from its nominal centre position to minimize crosstalk of the signal from the preceding F $K$ edge.[15] In the case of the Mn $L_{2,3}$ edge, inverse partial FY detection (IPFY) [232] using the O K fluorescence was employed. The photon-energy resolution in the spectra was set to 0.2-0.4 eV. Energy calibration (using a NiO reference), dark current subtraction, division by $I_0$, background subtraction, data normalization and absorption correction was performed as described in.[192,193] Data evaluation of the *Me* $L_{2,3}$ edges was done using charge-transfer multiplet calculations (CTM4XAS, Crispy and Quanty [86,87,150,153]). The calculated spectra were fitted to the measured data with our Excel macro, applying a least squares method to minimize the difference in area between the data and the fit by shifting the calculated spectra along the energy axis and varying the intensity.

*Synchrotron x-ray powder diffraction:* Time-resolved operando synchrotron X-ray powder diffraction (SXPD) of the first cycle of NCM811 was performed at beamline I11 (Diamond Light Source, UK) using the position sensitive detector (PSD) of the beamline.[233] The energy of the x-ray beam was tuned to 25 keV and the calibrated wavelength was 0.489951(10) Å. The battery cells were mounted on a xyz-stage and each cell was adjusted to the centre of diffraction. The 1D data was refined using the software package Fullprof (2θ range: 5 - 65°).[234]

Due to preferential orientations of Al (current collector of the NCM electrode) and Li (counter electrode), the phases were included in the refinements using the Le Bail method. The structural input parameters for the refinements of NCM811 (in the pristine state) were obtained from crystallographic data files.[235] Upon charge anisotropic microstrain is increasing, which was implemented as described in [148].

*AFM/ESM*: The electrode (active material + binder + conductive carbon casted on Al) was fixed between two steel plates using a clamp. The samples were embedded in epoxy resin (EpoFix), cut and polished so that a cross section of the electrode could be grounded via the steel plates. For AFM-ESM measurements slow scan velocities of roughly 4µm/s were chosen to ensure reliable tip–sample interactions. The amplitude results were normalized with respect to the drive amplitude and noise level for each individual sample, and the corrected amplitude was then referred to as the "ESM amplitude" in the figures. For measuring the polished NCM material, a commercially available atomic force microscope (Bruker Dimension ICON, Santa Barbara, USA) operating inside a glovebox (MBraun, Stratham, USA, O2 < 0.1 ppm, H2O < 0.1 ppm) was used. The contact resonance frequency and amplitude were tracked using a phase-locked loop (HF2LI, Zurich Instruments, Switzerland). The applied drive amplitude and normal load were 6 V and 133 nN, respectively. For each sample, multiple NCM particles were analysed. The in-contact resonance frequency between the samples and tip was observed to be between 300 and 350 kHz. Both topographical and ESM signals were obtained simultaneously.

*XPS*: An Al Kα source ($E_{photon}$ = 1486.6 eV) with a 10 mA filament current and a filament voltage source of 12 kV was used for the measurements. The analyzed area was 300 µm × 700 µm. A charge neutralizer was used to suppress charging effects. Energy calibration was done with the software CasaXPS using the C 1s peak at 284.8 eV.

Results and Discussion

## S2. Evaluation of XPS spectra

The natural linewidth of the Al Kα source (x-ray photoelectron spectroscopy, XPS) is ~ 0.8 eV, *i.e.* twice as much as in the case of the NEXAFS spectra. Without further monochromatization this makes a finestructure analysis difficult and reference measurements as well as supporting charge transfer multiplet (CTM) calculations are necessary to ensure a proper assignment of peaks. While the Ni $L_{2,3}$ spectra reveal small differences (Figure 11b), the Co $L_{2,3}$ (LNCOs, NCMs) and the Mn $L_{2,3}$ (LNMOs, NCMs) do not (Figure S 1) which is in agreement with previous findings.[15,34] Note that background correction and normalization of all spectra was only performed in the case of the Co $L_{2,3}$ edge. The Mn $L_{2,3}$ spectra show, especially in the case of Ni-rich layered oxides ($x > 0.6$ in $LiNi_xMe_{1-x}O_2$), self-absorption and/or other correlation effects, which make the normalization as described in the main text difficult. Thus, the data was only normalized to the main peak intensity. The $L_2/L_3$ ratio differs significantly from 1:2 (core hole state degeneracies) which reinforces that self-absorption effects play a crucial role but the peak position and shape does not differ. Thus it can be concluded, that the electronic configuration must be the same although correlation effects crucially modify the background.

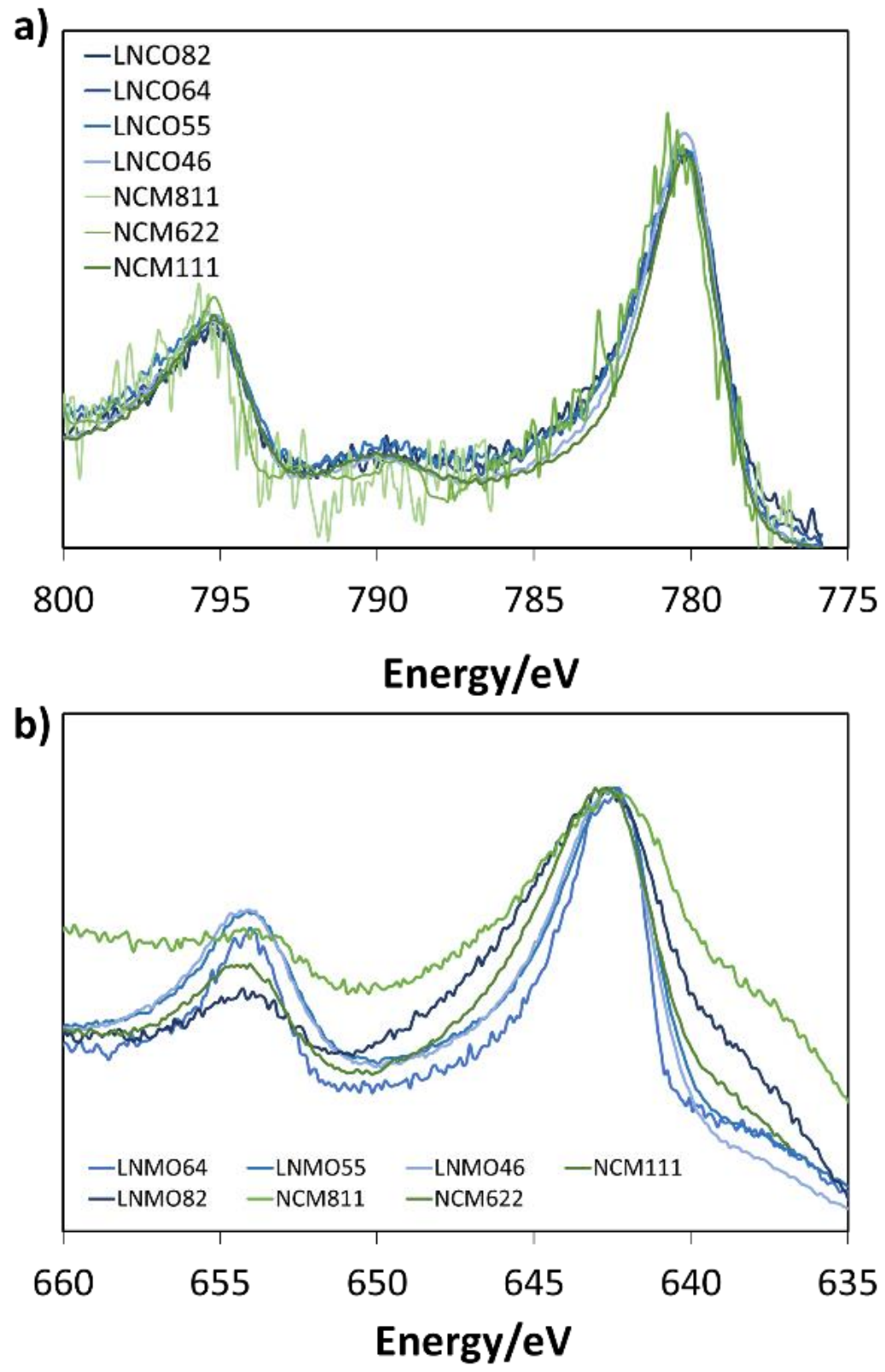

**Figure S 1:** Co $L_{2,3}$ (a) and Mn $L_{2,3}$ (b) XPS spectra of LNCOs (only Co $L_{2,3}$), NCMs LNMOs (only Mn $L_{2,3}$). The Co $L_{2,3}$ edge is normalized as described in the main text. The Mn $L_{2,3}$ edge shows large self-absorption and correlation effects, especially in the case of Ni-rich layered oxides ($x > 0.6$ in $LiNi_xMe_{1-x}O_2$) and thus is only normalized to its maximum peak intensity.

Figure S 2 shows charge transfer multiplet (CTM) calculations [86,87] with a much higher energy resolution than obtained from the experiments ($U_{dd}$ (Coulomb Interactions) – $U_{pd}$ (core-hole potential) = 1eV, Gaussian and Lorentzian broadening = 0.4 eV, Ligand Field energy of $Ni^{2+}/Ni^{3+}$ =2.0 eV/2.5 eV). This is to get an idea, where spectral weight of the $Ni^{2+}$ (> 96% Ni $3d^8$ O $2p^6$ $\rightleftharpoons$ < 4% Ni $3d^9$ O $2p^5$) and $Ni^{3+}$ configuration (> 23% Ni $3d^7$ O $2p^6$ $\rightleftharpoons$ < 77% Ni $3d^8$ O $2p^5$), the configurations present in the Ni $L_{2,3}$ edge of layered oxides according to [15,34,61] and the Ni $L_{2,3}$ NEXAFS spectra in the main text, can be expected. $Ni^{2+}$ is a very ionic

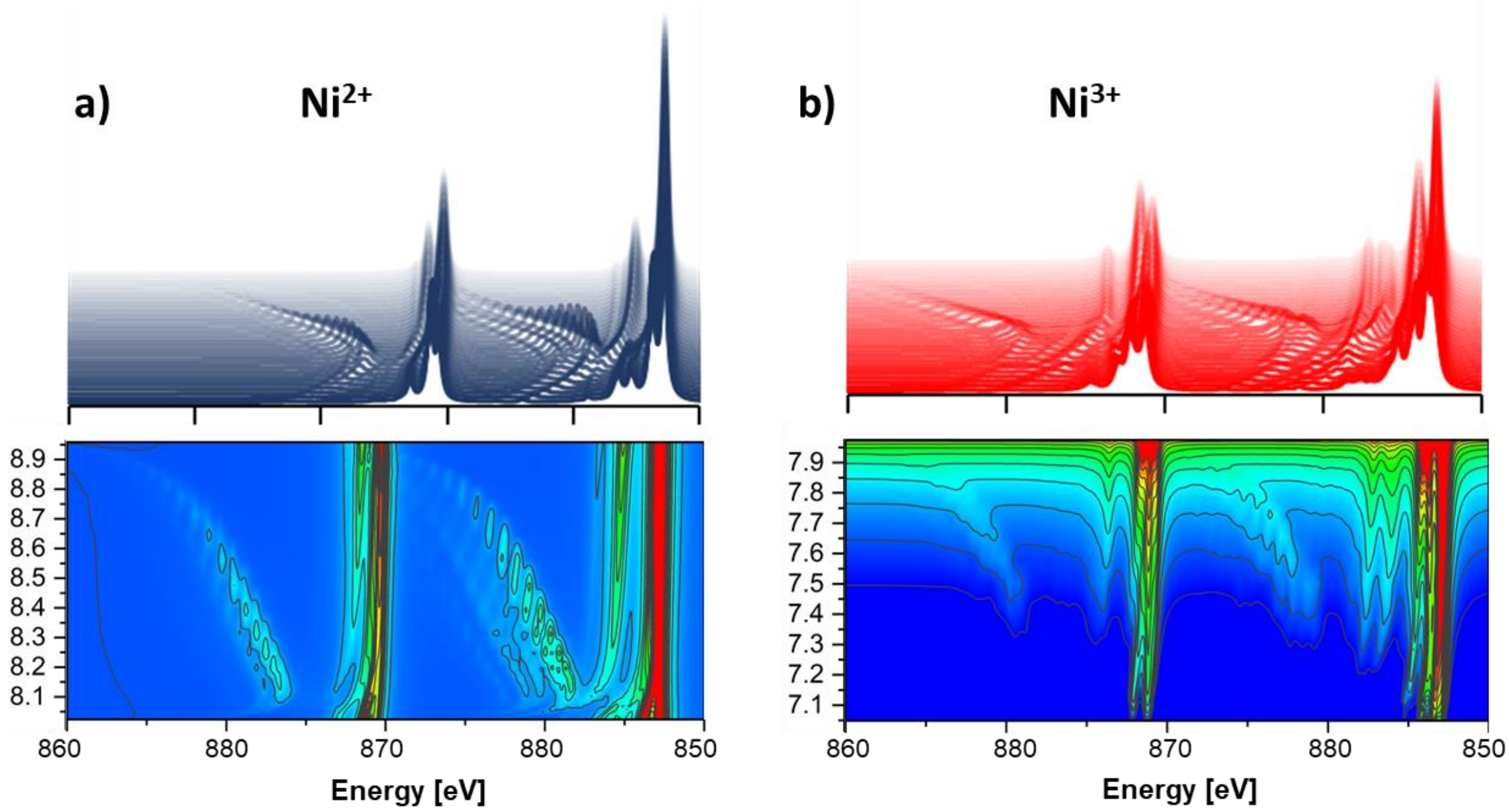

**Figure S 2:** Charge transfer multiplet calculations of the Ni $L_{2,3}$ XPS spectra with the $Ni^{2+}$ configuration (a) and the $Ni^{3+}$ configuration (b).

configuration (mean number of 3d electrons ~8) and satellite peaks range up to ~ 5 eV (Figure S 2a) after the main $L_2$ and $L_3$ edge, while in the covalent $Ni^{3+}$ configuration (mean number of 3d electrons ~7.7) the satellites have a range of > 15 eV (Figure S 2b). The relative energy at which the peaks appear in the calculations must not agree with the energy at which the peaks appear in the real XPS spectra (the spectra can be shifted along the x-axis). As evident from Figure 11, the resolution of the $L_2$ and $L_3$ edge of the measured data does not allow an analysis of the finestructure as it was done in case of the NEXAFS data. Nevertheless, the assignment of broader features to either ionic $Ni^{2+}$ or covalent $Ni^{3+}$ is possible. Therefore, the Ni $L_{2,3}$ edge of NiO as reference is simulated first to obtain the energetic position of peaks and their spectral shape, Figure S 3. This is done using eight gaussian functions (peaks A-G in Figure 11) with different full width at half maxima (FWHM). Two sharp peaks are clearly separated around 855 eV which are assigned to $Ni^{2+}$ (peak A, FWHM = 1.5 eV) and $Ni^{3+}$ (peak B, FWHM = 1.5 eV). The region of the satellite peaks (855 eV – 860 eV) was fitted with two gaussian functions and to account for the different broadening of the $Ni^{2+}$ and $Ni^{3+}$ satellites the FWHM was set to 2 eV ($Ni^{2+}$ satellite peak C) and 3 eV ($Ni^{3+}$ satellite peak D), respectively. The $L_2$ edge was simulated using four gaussian functions, as well, and the FWHM of the gaussian functions was chosen to be the same as for the $L_3$ edge (peak E: $Ni^{2+}$; FWHH = 1.5 eV, peak F: $Ni^{3+}$; FWHH = 1.5 eV, peak G: $Ni^{2+}$ satellite; FWHH = 2 eV, peak H: $Ni^{3+}$ satellite; FWHH = 3 eV).

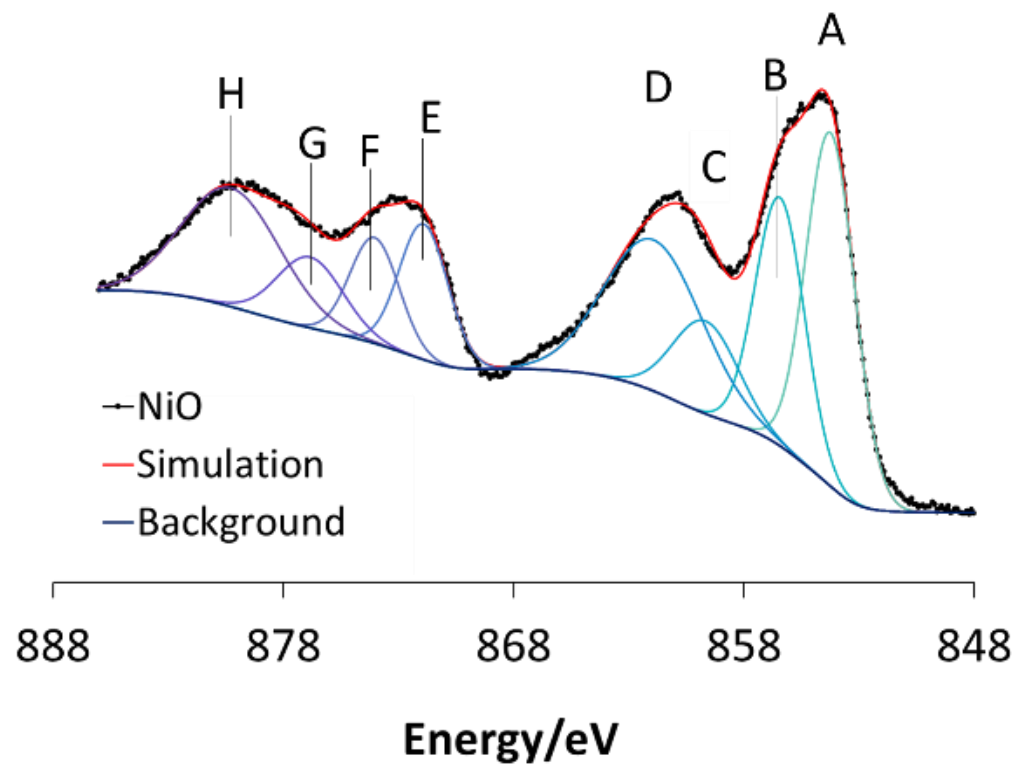

**Figure S 3:** Ni $L_{2,3}$ XPS spectra of NiO and its simulation using eight gaussian functions A-G. A and E represent the $Ni^{2+}$ peaks, B and F the $Ni^{3+}$ peaks, and, C, D G and H account for the satellite peaks. While the Ni peaks are relatively sharp (FWHM 0 1.5 eV) the satellite peaks are broader ($N^{2+}$satealites D and H: FWHM = 2 eV, $Ni^{3+}$satelittes D and H: FWHM = 3 eV).

For the simulation of the Ni $L_2$ XPS spectra of the LMNOs, the NCMs and LNCOs (Figure S 4) the energetic position and the broadening of the peaks from the NiO spectra are constrained. The fixed values are given in Table S 1.

**Table S 1:** Energetic positions and FWHM of the Gaussian functions used to simulate the Ni $L_{2,3}$ edge of NiO.

| Peak | Energy/eV | FWHM/eV |
|---|---|---|
| **Peak A, $Ni^{2+}$** | 854.23 | 1.5 |
| **Peak B, $Ni^{3+}$** | 855.82 | 1.5 |
| **Peak C, $Ni^{2+}$ satellite** | 859.65 | 2.0 |
| **Peak D, $Ni^{3+}$ satellite** | 861.90 | 3.0 |
| **Peak E, $Ni^{2+}$** | 871.90 | 1.5 |
| **Ni3+, Peak F** | 873.83 | 1.5 |
| **Broad, Ni2+, Peak G** | 877.43 | 2.0 |
| **Broad, Ni3+, Peak G** | 880.27 | 3.0 |

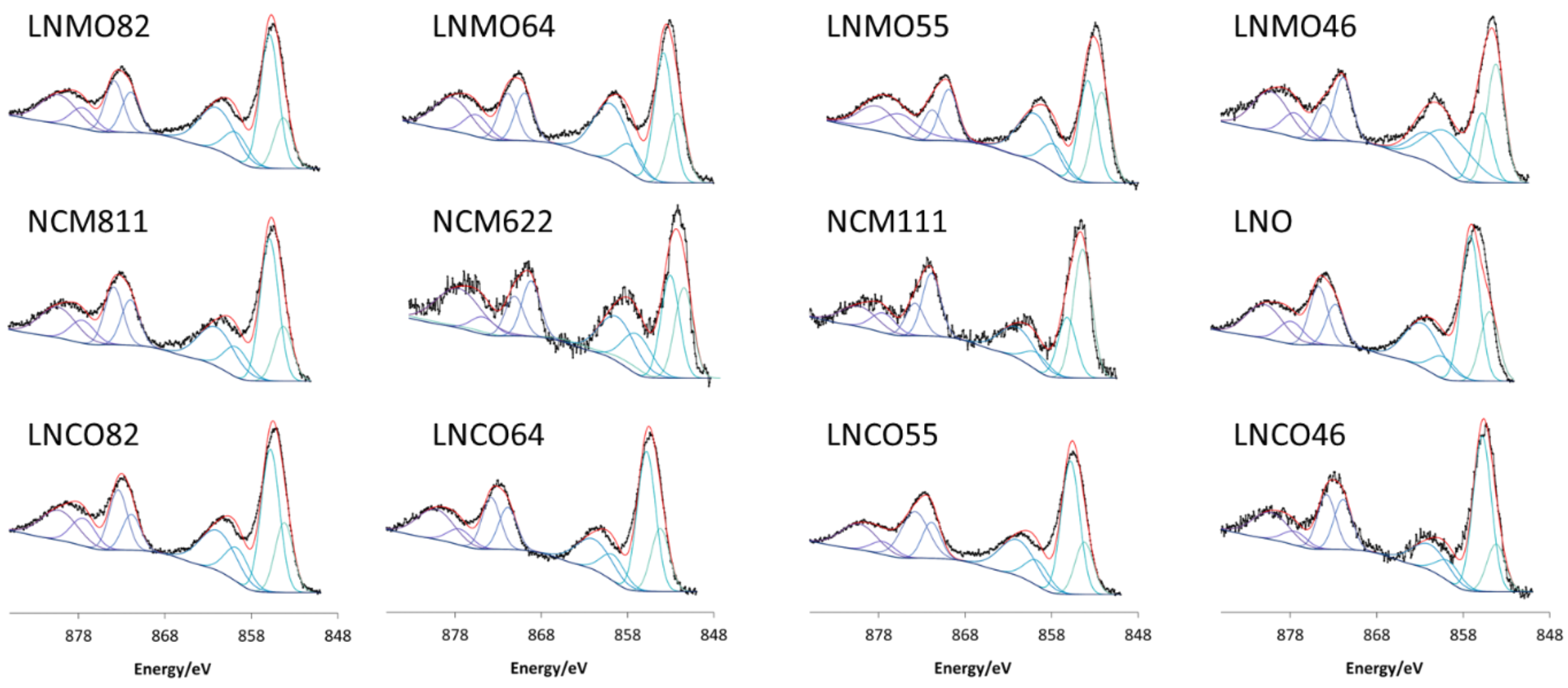

**Figure S 4:** Ni $L_{2,3}$ XPS spectra and simulation of the LNMOs (top), the NCMs (middle), LNO (middle right) and the LNCOs (bottom). The spectra are fitted with eight gaussian functions whereby the energy and the FWHM was kept fix. The energies and FWHM was taken from the NiO reference fit.